%% file: paper.tex
\newif\ifnb     
\nbtrue

\newif\ifcom    
\comtrue

\newif\ifarx    
\arxtrue

\ifarx
  \documentclass[sigconf,nonacm]{acmart}
\else
  \documentclass[sigconf,natbib=true,anonymous=true]{acmart}
\fi

\PassOptionsToPackage{table}{xcolor}     

\usepackage{graphicx}
\usepackage{glossaries}                 
\usepackage{enumitem}                   
\usepackage{makecell}                   
\usepackage{pifont}                     
\usepackage{cleveref}                   
\usepackage{color-edits}                
\usepackage{multirow}                   
\usepackage[most]{tcolorbox}	 		
\usepackage{xcolor}                     
\usepackage{algorithm}                  
\usepackage{algpseudocode}              
\usepackage{placeins}      				
\usepackage{dblfloatfix}   				
\usepackage{soul}

\usepackage[bottom]{footmisc}
\newtcbox{\gray}[1][]{%
  on line,
  colback=gray!30,   
  colframe=gray!30,  
  boxrule=0pt,       
  arc=3pt,           
  boxsep=0pt,        
  left=1pt,
  right=1pt,
  top=1pt,
  bottom=1pt,
  #1
}

\newcommand{\footnoteref}[1]{\textsuperscript{\ref{#1}}}

\setacronymstyle{long-short}
\newacronym{enns}{ENNS}{exact nearest neighbor search}
\newacronym{anns}{ANNS}{approximate nearest neighbor search}
\newacronym{fanns}{FANNS}{filtered approximate nearest neighbor search}
\newacronym{afanns}{AFANNS}{approximately filtered approximate nearest neighbor search}
\newacronym{em}{EM}{exact match}
\newacronym{r}{R}{range}
\newacronym{emis}{EMIS}{exact match in set}
\newacronym{mem}{MEM}{multiple exact match}
\newacronym{mr}{MR}{multiple range}
\newacronym{memis}{MEMIS}{multiple exact match in set}
\newacronym{c}{C}{combined}
\ifarx
\newacronym{lsh}{LSH}{\textbf{locality-sensitive hashing}}
\newacronym{ivf}{IVF}{\textbf{inverted file}}
\newacronym{pq}{PQ}{\textbf{product quantization}}
\newacronym{ivfpq}{IVF-PQ}{\textbf{inverted file with product quantization}}
\newacronym{nsw}{NSW}{\textbf{navigable small world}}
\newacronym{hnsw}{HNSW}{\textbf{hierarchical navigable small world}}
\else
\newacronym{lsh}{LSH}{{locality-sensitive hashing}}
\newacronym{ivf}{IVF}{{inverted file}}
\newacronym{pq}{PQ}{{product quantization}}
\newacronym{ivfpq}{IVF-PQ}{{inverted file with product quantization}}
\newacronym{nsw}{NSW}{{navigable small world}}
\newacronym{hnsw}{HNSW}{{hierarchical navigable small world}}
\fi
\newacronym{sig}{SIG}{\textit{segmented inclusive graph}}                   
\newacronym{hsig}{HSIG}{\textit{hierarchical segmented inclusive graph}}        
\newacronym{kgraph}{KGraph}{\textit{K-nearest-neighbor graph}}      
\newacronym{knn}{KNN}{k nearest neighbors}
\newacronym{aft}{AFT}{\textit{attribute frequency tree}}
\newacronym{vgpq}{VGPQ}{\textit{Voronoi graph product quantization}}
\newacronym{lng}{LNG}{\textit{label navigating graph}}
\newacronym{ung}{UNG}{\textit{unified navigating graph}}
\newacronym{dac}{DAC}{\textit{directed acyclic graph}}
\newacronym{bst}{BST}{\textit{binary search tree}}
\newacronym{llm}{LLM}{large language model}
\newacronym{rag}{RAG}{retrieval-augmented generation}
\newacronym{rlm}{RLM}{reasoning language model}
\newacronym{qps}{QPS}{queries per second}
\newacronym{mteb}{MTEB}{Massive Text Embedding Benchmark}

\addauthor[Patrick]{pi}{teal}
\renewcommand{\pi}[1]{\ifcom\picomment{#1}\fi}
\addauthor[Paragraph Summary]{ps}{orange}

\newcommand{\cmark}{\ding{52}}%
\newcommand{\xmark}{\ding{56}}%
\newcommand{\rone}{\raisebox{-0.125em}{\includegraphics[height=0.75em]{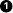}}}
\newcommand{\rtwo}{\raisebox{-0.125em}{\includegraphics[height=0.75em]{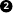}}}
\newcommand{\rthree}{\raisebox{-0.125em}{\includegraphics[height=0.75em]{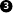}}}
\newcommand{\rfour}{\raisebox{-0.125em}{\includegraphics[height=0.75em]{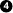}}}
\newcommand{\rfive}{\raisebox{-0.125em}{\includegraphics[height=0.75em]{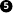}}}
\newcommand{\rsix}{\raisebox{-0.125em}{\includegraphics[height=0.75em]{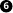}}}
\newcommand{\rseven}{\raisebox{-0.125em}{\includegraphics[height=0.75em]{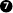}}}
\newcommand{\reight}{\raisebox{-0.125em}{\includegraphics[height=0.75em]{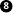}}}
\newcommand{\rnine}{\raisebox{-0.125em}{\includegraphics[height=0.75em]{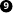}}}
\newcommand{\rten}{\raisebox{-0.125em}{\includegraphics[height=0.75em]{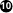}}}
\newcommand{\releven}{\raisebox{-0.125em}{\includegraphics[height=0.75em]{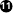}}}

\newcommand{\sone}{\raisebox{-0.125em}{\includegraphics[height=0.75em]{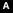}}}
\newcommand{\stwo}{\raisebox{-0.125em}{\includegraphics[height=0.75em]{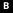}}}
\newcommand{\sthree}{\raisebox{-0.125em}{\includegraphics[height=0.75em]{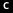}}}
\newcommand{\sfour}{\raisebox{-0.125em}{\includegraphics[height=0.75em]{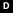}}}
\newcommand{\sfive}{\raisebox{-0.125em}{\includegraphics[height=0.75em]{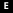}}}

\newcommand{\dataseturl}{\ifnb https://hf.co/datasets/SPCL/arxiv-for-fanns-large \else The URL to the dataset on Hugging Face is omitted for blind review.\fi}
\newcommand{\dataseturlmedium}{\ifnb https://hf.co/datasets/SPCL/arxiv-for-fanns-medium \else Dataset omitted for blind review \fi}
\newcommand{\dataseturlsmall}{\ifnb https://hf.co/datasets/SPCL/arxiv-for-fanns-small \else Dataset omitted for blind review \fi}
\newcommand{\codeurl}{\ifnb https://github.com/spcl/fanns-benchmark \else Repository omitted for blind review (the code cannot be easily anonymized).\vspace{-4em}\fi}  

\sethlcolor{gray!30}

\setcopyright{acmlicensed}
\copyrightyear{2026}
\acmYear{2026}
\acmDOI{XXXXXXX.XXXXXXX}
\acmConference[SIGIR '26]{the 49th International Conference on Research and Development in Information Retrieval}{July 20--24,
  2026}{Melbourne, Australia}

\acmISBN{978-1-4503-XXXX-X/2018/06}

\acmSubmissionID{208}

\begin{document}

\title{Benchmarking Filtered Approximate Nearest Neighbor Search Algorithms on Transformer-based Embedding Vectors}

\settopmatter{authorsperrow=6}

\ifnb
\author{\large Patrick Iff}
\affiliation{
  \institution{ETH Zurich}
  \country{Switzerland}
}
\email{iffp@inf.ethz.ch}

\author{\large Paul Brügger}
\affiliation{
  \institution{ETH Zurich}
  \country{Switzerland}
}
\ifarx
\fi

\author{\large Marcin Chrapek}
\affiliation{
  \institution{ETH Zurich}
  \country{Switzerland}
}
\ifarx
\fi

\author{\large David Kochergin}
\affiliation{
  \institution{ETH Zurich}
  \country{Switzerland}
}

\author{\large Maciej Besta}
\affiliation{
  \institution{ETH Zurich}
  \country{Switzerland}
}
\ifarx
\fi

\author{\large Torsten Hoefler}
\affiliation{
  \institution{ETH Zurich}
  \country{Switzerland}
}
\email{htor@inf.ethz.ch}
\else
\author{Anonymous Authors}
\fi

\renewcommand{\shortauthors}{\ifnb Iff et al. \else Anonymous et al. \fi}

\begin{abstract}
	\input{abstract}
\end{abstract}


\maketitle

\vspace{-1.0em}
\begin{center}
\textbf{Code:} \codeurl\\
\textbf{Dataset:} \dataseturl
\end{center}
\vspace{-0.75em}

\input{01_intro}

\input{02_background}
\input{03_survey}
\input{04_dataset}

\input{05_benchmark}

\input{06_related_work}
\input{07_conclusion}

\ifnb
    \input{08_acknowledgements}
\fi

\bibliographystyle{ACM-Reference-Format}
\bibliography{bibliography}

\end{document}

%% file: abstract.tex
Advances in embedding models for text, image, audio, and video drive progress across multiple domains, including retrieval-\hspace{0pt}augmented generation, recommendation systems, and others.
Many of these applications require an efficient method to retrieve items that are close to a given query in the embedding space while satisfying a filter condition based on the item's attributes, a problem known as filtered approximate nearest neighbor search (FANNS).
By performing an in-depth literature analysis on FANNS, we identify a key gap in the research landscape: publicly available datasets with embedding vectors from state-of-the-art transformer-based text embedding models that contain abundant real-world attributes covering a broad spectrum of attribute types and value distributions.
To fill this gap, we introduce the \texttt{arxiv-for-fanns} dataset of transformer-based embedding vectors for the abstracts of over 2.7 million arXiv papers, enriched with 11 real-world attributes such as authors and categories.
We benchmark eleven different FANNS methods on our new dataset to evaluate their performance across different filter types, numbers of retrieved neighbors, dataset scales, and query selectivities.
We distill our findings into eight key observations that guide users in selecting the most suitable FANNS method for their specific use cases.

%% file: 01_intro.tex
\section{Introduction}
\label{sec:intro}

Advances in embedding models for text \cite{nvembed, lens, stella}, image \cite{googlenet, seresnext}, video \cite{vid-embed-1}, audio \cite{aud-embed-1,aud-embed-2}, and other modalities \cite{biggraph} have significantly enhanced semantic search, where items are mapped to a high-dimensional vector space and similarity is measured by the distance between embedding vectors.
Efficient similarity search algorithms are essential to navigate this space.
Given the scale of modern datasets, \gls{enns} algorithms are too slow, necessitating the use of \gls{anns} methods \cite{lsh-1, nsw, hnsw, diskann, ivf, pq, idec, aq, bapq, opq}.
Additionally, many applications, including \glspl{llm} with retrieval-augmented generation \cite{rag-survey,rag-1,rag-2}, recommendation systems \cite{rec-sys-1, pase, rec-sys-2}, vehicle and person re-\hspace{0pt}identification \cite{re-ident-1, arkgraph}, and others \cite{face-rec-1, pase, voice-rec-1,ecom-1, adbv}, require retrieving only items that satisfy filtering conditions on item attributes, such as access rights for a document, a video's timestamp, or a product’s price.
These requirements have driven the development of \gls{fanns}.
Both academia and industry have recognized its growing importance, leading to numerous research publications \cite{nhq, hqi, fdann, caps, arkgraph, acorn, bwst, serf, irg, unify, wow, dsg, tfanns, digra} and adoption in database solutions \cite{pinecone,vespa,vectara}.

These efforts require suitable benchmarks to guide the development of \gls{fanns} methods, and while transformer-based text embeddings are widely used in practice, no existing \gls{fanns} benchmarks with transformer-based embedding vectors exist.
Motivated by these developments, we make three main contributions:  

First, to clarify the requirements for \gls{fanns} benchmarks, we present a comprehensive \textbf{taxonomy and survey} of existing \gls{fanns} methods.  
We identify three key dimensions for the classification of \gls{fanns} methods: \textit{filtering approach}, \textit{indexing technique}, and supported \textit{filter types}.  
To cover all \textit{filter types}, a \gls{fanns} benchmark must contain categorical, numerical, and set-valued attributes, and to reflect the strengths and weaknesses of various \textit{filtering approaches}, these attributes should exhibit diverse value distributions.  
The \textit{indexing technique} is largely orthogonal to the benchmark.  

Second, we present the novel \texttt{arxiv-for-fanns} \textbf{dataset} with transformer-based embeddings of paper abstracts from the arXiv dataset~\cite{arxiv-ds} using the model \texttt{stella\_en\_400M\_v5}~\cite{stella-hf, stella}.  
Our dataset has eleven real-world attributes, including categorical (e.g., venue), numerical (e.g., publication year), and set-valued (e.g., authors) ones, and exhibits diverse value distributions.

Third, we perform a comprehensive \textbf{benchmarking} study of eleven FANNS methods on our novel dataset, evaluating their performance across different filter types, varying numbers of retrieved neighbors, different dataset scales, and a wide range of query selectivities.
We distill the results of this benchmarking effort into eight key observations that offer practical guidance for selecting the most appropriate FANNS method for specific application scenarios.
\vspace{-0.75em}

%% file: 02_background.tex
\section{Background}
\label{sec:back}

\vspace{-0.25em}
\subsection{Filtered Approx. Nearest Neighbor Search}
\label{sec:back-fanns}

To introduce the \gls{fanns} concepts that we leverage throughout the paper, we walk the reader through the process of building and using a semantic search engine for research papers with filtering capabilities, which is a practical use case of \gls{fanns} that can be built using our \texttt{arxiv-for-fanns} dataset.
\Cref{fig:example} provides an overview of this example application, with components and actions referenced using letters \sone~to \sfive~and numbers \rone~to \releven, respectively.

\input{fig/fig_example}

\textbf{Index Construction:}
To build such a system, we transform each of the $n$ papers in our database~\sone~into an \textit{item}.
Formally, an \textit{item} $I_i = (v_i, a_{i,1}, \dots, a_{i,m})$ for $i \in \{1, \dots, n\}$ consists of a $d$-dimensional embedding vector $v_i$ and a value for each of the $m$ \textit{attributes} ($a_{i,1}$ to $a_{i,m}$).
In our example, the paper's venue, publication year, and authors define~\rone~the \textit{item's} \textit{attributes}.
The paper's abstract is processed~\rtwo~by a text embedding model~\stwo, such as NV-Embed \cite{nvembed, nvembed-hf}, LENS \cite{lens, lens-hf}, Stella \cite{stella, stella-hf}, GTE \cite{gte, gte-hf}, or BGE \cite{bge, bge-hf}, to generate the \textit{item's} embedding vector $v_i$.
To efficiently answer \gls{fanns} queries for these \textit{items}, we insert~\rthree~all items into a \gls{fanns} index~\sthree.
\vspace{0.25em}

\textbf{Query execution:}
During runtime, users interact with the search engine by entering \rfour~search terms and filter conditions.
We use $p$ to denote the number of user requests submitted during a session.
We transform each user request into a \gls{fanns} \textit{query}, formally defined as $Q_j = (q_j, k_j, f_j)$ for $j \in \{1,\dots,p\}$ and consisting of a $d$-dimensional query vector $q_j$, the number of \textit{items} to return $k_j$, and a filter function $f_j$.
The filter $f_j$ and the number of requested results $k_j$ are set \rfive~based on the user's filter conditions.
The search term is processed \rsix~by the same text embedding model \stwo~used during index construction to generate the query vector $q_j$.
The \textit{query} is submitted \rseven~to a \gls{fanns} algorithm \sfive, which utilizes \reight~the \gls{fanns} index \sthree~to retrieve the IDs of the most relevant papers.
These IDs are forwarded \rnine~to a retriever \sfour, which fetches \rten~the corresponding papers from the database \sone~and returns \releven~them to the user.

\subsection{Approximate Nearest Neighbor Search}
\label{sec:back-anns}

\ifarx
Since many \gls{fanns} algorithms build on \gls{anns} methods without filtering, we first review the most relevant \gls{anns} approaches. They fall into four categories: tree-based, hash-based, graph-based, and quantization-based. For detailed surveys, see Han et al.~\cite{anns-survey-han} and Echihabi et al.~\cite{anns-survey-echihabi}. Benchmarks \cite{ann-benchmark,ann-benchmark-web} show that no algorithm dominates across datasets; performance is dataset-dependent.

\subsubsection{Tree-based methods}
\label{sec:back-anns-tree}

Tree-based methods partition the vector space into regions using search trees, where nodes denote regions or their boundaries.
Deeper levels represent finer partitions.
Examples include \textbf{k-d trees} \cite{kdtree}, \textbf{R-trees} \cite{rtree}, \textbf{ball trees} \cite{balltree}, \textbf{cover trees} \cite{covertree}, \textbf{RP trees} \cite{rp-tree}, \textbf{M-trees} \cite{mtree}, \textbf{K-means trees} \cite{kmtree}, and \textbf{best bin first} \cite{bbf}, as well as many variants \cite{anns-tree-1,anns-tree-2,anns-tree-3,anns-tree-4}.
These methods are effective for low-dimensional vectors but degrade under the curse of dimensionality \cite{curse-of-dim}.

\subsubsection{Hash-based methods}
\label{sec:back-anns-hash}

Hash-based methods use a set of hash functions to map vectors from a continuous $d$-dimensional space into discrete hash buckets.  
These functions are designed so that similar vectors are mapped into the same bucket with high probability.  
When querying a hash-based index, the query vector is hashed, and the vectors in the corresponding bucket are compared to the query vector.  
Hash-based methods include \gls{lsh} \cite{lsh-1,lsh-2}, \textbf{spectral hashing} \cite{sh}, \textbf{iDEC} \cite{idec}, \textbf{deep hashing} \cite{dh}, \textbf{mmLSH} \cite{mmlsh}, \textbf{PM-LSH} \cite{pm-lsh}, \textbf{R2LSH} \cite{r2lsh}, \textbf{EI-LSH} \cite{ei-lsh}, and others \cite{anns-hash-1,anns-hash-2,anns-hash-3,anns-hash-4,anns-hash-5,anns-hash-6,anns-hash-7,anns-hash-8,anns-hash-9}.

\subsubsection{Graph-based methods}
\label{sec:back-anns-graph}

Graph-based methods represent vectors as vertices, connecting those close in the embedding space to form a neighborhood graph \cite{ng,dr-ng}.
Nearest neighbors are found by traversing the graph from one or more entry points, following the shortest distance to the query.
\Gls{nsw} graphs \cite{nsw} add long edges between distant vertices to accelerate traversal.
\Gls{hnsw} graphs \cite{hnsw} extend \gls{nsw} with a hierarchy: upper layers hold long edges, lower layers short ones, and traversal proceeds top-down.
Other graph-based methods include \textbf{DiskANN} \cite{diskann}, \textbf{FreshDiskANN} \cite{freshdann}, \textbf{NSG} \cite{nsg}, \textbf{GRNG} \cite{grng}, and others \cite{anns-graph-1,anns-graph-3,anns-graph-4,anns-graph-5,anns-graph-6}.

\subsubsection{Quantization-based methods}
\label{sec:back-anns-quant}

Quantization-based methods map the $n$ vectors to a set of $c \ll n$ clusters, each represented by a cluster centroid. 
Each vector is assigned to the cluster whose centroid is closest to the vector.
One can either store all vectors assigned to a given cluster, as in \gls{ivf} \cite{ivf}, or approximate all vectors of a cluster by the respective cluster centroid, which is less accurate but requires less storage space and allows for faster query times. 
To find the nearest neighbors of a query vector, we only compare it to the database vectors in the $w \leq c$ clusters whose centroids are closest to the query vector.
\Gls{pq} \cite{pq} and \gls{ivfpq} \cite{pq} are more advanced quantization-based methods that split the vector space into $s$ orthogonal subspaces and quantize each subspace separately.
Other quantization-based \gls{anns} methods include \textbf{additive quantization} \cite{aq}, \textbf{BAPQ} \cite{bapq}, \textbf{LOPQ} \cite{lopq}, \textbf{OPQ} \cite{opq}, \textbf{RaBitQ} \cite{rabitq}, \textbf{SPANN} \cite{spann}, and others \cite{anns-quant-1,anns-quant-2,anns-quant-3,anns-quant-4,anns-quant-5}.
\fi

%% file: fig/fig_example.tex
\begin{figure}[h]
\centering
\captionsetup{justification=centering}
\includegraphics[width=1.0\columnwidth]{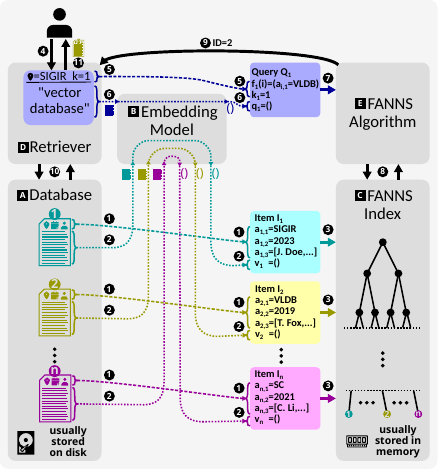}
\vspace{-2.25em}
\caption{\textbf{(\textsection \ref{sec:back-fanns}) An Example Application of FANNS.}}
\label{fig:example}
\vspace{-1em}
\end{figure}

%% file: 03_survey.tex
\input{tab/tab_fanns}

\section{Taxonomy and Survey of FANNS Methods}
\label{sec:fanns}

By analyzing the vast landscape of existing \gls{fanns} methods, we identify three key dimensions for classifying these methods.
First, they differ in the \textit{approach} they take to combine \gls{anns} with attribute filtering, which we discuss in \Cref{sec:fanns-approaches}.
Second, we observe that all \gls{fanns} methods are based on the same four fundamental \textit{indexing techniques} as \gls{anns} methods, namely, hashing, trees, graphs, and quantization, or combinations thereof (see \Cref{sec:back-anns}).
Third, we distinguish between different \textit{filter types} that \gls{fanns} methods support, which we discuss in \Cref{sec:back-filter}.
A fourth, less important dimension is the precise problem definition of \gls{fanns} that is addressed.
Most works guarantee that all returned items pass the filter and only the distance in embedding space is approximated; however, a small number of methods also approximate the filtering step, sometimes returning items that do not match the filter.
We refer to this relaxed problem as \gls{afanns}.

\subsection{Filtering Approaches}
\label{sec:fanns-approaches}

Two common approaches to solving the \gls{fanns} problem are \textbf{pre-filtering} and \textbf{post-filtering}.  
In \textit{pre-filtering}, an attribute-only index such as a B-tree \cite{btree}, B+-tree \cite{bptree}, or qd-tree \cite{qdtree} identifies all items matching the filter. The \gls{knn} of these items are then determined by computing the distance to the query vector or approximating this distance using quantized embedding vectors precomputed during index construction.  
In \textit{post-filtering}, a nearest neighbor search is performed on an unfiltered \gls{anns} index, followed by filtering out non-matching items. This can be done by retrieving $k' > k$ items in the hope that enough pass the filter or by iteratively retrieving more items until $k$ matches are found.  
Many graph-based \gls{anns} indices support a third approach, \textbf{in-filtering}, where attributes are ignored during index construction, and only vertices satisfying the filter are considered during query execution. A fourth approach involves a \textbf{hybrid index} that integrates embedding vectors and attributes in a single index. 

\input{fig/fig_selectivity}

The efficiency of different approaches to solving the \gls{fanns} problem depends on the dataset and filtering condition (see \Cref{fig:selectivity}).
We define the selectivity of a filter $f$ on a dataset $D$ as the fraction of items in $D$ that satisfy the filter $f$, following previous work\footnote{Note that some works \cite{adbv,milvus,irg} use the inverse definition of selectivity.} \cite{hqi,vbase,acorn,unify}.
\textit{Pre-filtering} is most efficient when selectivity is low, meaning only a few items remain for the \gls{enns}.
\textit{Post-filtering} and \textit{in-filtering} are most efficient when selectivity is high, i.e., most items pass the filter.
If selectivity is too low, \gls{anns} may not return enough valid candidates for \textit{post-filtering}, and graph traversal with \textit{in-filtering} may fail due to a sparsely connected or disconnected graph.
A \textit{hybrid index} is most efficient when selectivity is moderate, making none of the three previous approaches optimal.

\subsection{Filter Types}
\label{sec:back-filter}

We identify three fundamental \textit{filter types} based on a single attribute.
In \textbf{\Gls{em} filtering}, the item's attribute value must equal a specified query value.
While \gls{em} filtering is usually applied to a categorical attribute, it can also be used on numerical attributes.
\textbf{\Gls{r} filtering} applies to numerical attributes and considers only items where the attribute value is within a specified query range.
Finally, \textbf{\Gls{emis} filtering} applies to set-valued attributes and considers only items whose attribute set contains a queried value.
\gls{fanns} algorithms that apply the same filter type to multiple attributes implement \textbf{\gls{mem}}, \textbf{\gls{mr}}, and \textbf{\gls{memis}} filters.
We define a \textbf{\gls{c} filter} as a filter that can combine the three fundamental single-attribute filters using arbitrary logical operators and may also support arbitrary SQL-style filter predicates.


\ifarx
\input{03a_methods_explained_long}

\fi

%% file: tab/tab_fanns.tex
\balance
\begin{table*}[t]
\small
\setlength{\tabcolsep}{5pt}
\centering
\captionsetup{justification=centering}
\caption{\textbf{(\textsection \ref{sec:fanns-explanation}) Overview of FANNS methods}. 
$^\dagger$ Does not mention a filtering step during query execution.
*We have not been able to find the source code of this work.
\rone~The index is constructed for a single ordered attribute, in-filtering is proposed to handle secondary attributes.
\rtwo~Any EMIS filter can be used as an EM filter by using a set attribute with only a single value.
}
\vspace{-1.5em}
\begin{tabular}{lllllcccccccl}
\toprule
&&&&&\multicolumn{7}{c}{\textbf{Filters}}&\textbf{Open}\\
\cline{6-12}
\textbf{\makecell[l]{Method}}
&\textbf{\makecell[l]{Year}}
&\textbf{\makecell[l]{Problem}}
&\textbf{\makecell[l]{Approach}}
&\textbf{\makecell[l]{Technique(s)}}
&\textbf{\rotatebox{0}{\makecell[l]{EM}}}
&\textbf{\rotatebox{0}{\makecell[l]{R}}}
&\textbf{\rotatebox{0}{\makecell[l]{EMIS}}}
&\textbf{\rotatebox{0}{\makecell[l]{MEM}}}
&\textbf{\rotatebox{0}{\makecell[l]{MR}}}
&\textbf{\rotatebox{0}{\makecell[l]{MEMIS}}}
&\textbf{\rotatebox{0}{\makecell[l]{C}}}
&\textbf{\rotatebox{0}{\makecell[l]{source}}}\\
\midrule
Rii \cite{rii}					& 2018 & FANNS 	& pre-/in-filtering			& \makecell[l]{quantization}			& \cmark & \cmark & \cmark & \cmark  & \cmark & \cmark  & \cmark & \quad\cmark \cite{rii-code}		\\	
MA-NSW \cite{mansw}				& 2019 & FANNS 	& hybrid index				& \makecell[l]{graph}					& \cmark & \xmark & \xmark & \cmark  & \xmark & \xmark  & \xmark & \quad\xmark*						\\	
PASE \cite{pase}				& 2020 & FANNS	& post-filtering			& \makecell[l]{quantization / graph}	& \cmark & \cmark & \cmark & \cmark  & \cmark & \cmark  & \cmark & \quad\cmark \cite{pase-code} 	\\  
AnalyticDB-V \cite{adbv}		& 2020 & FANNS	& pre-/post-filtering		& \makecell[l]{quantization}			& \cmark & \cmark & \cmark & \cmark  & \cmark & \cmark  & \cmark & \quad\xmark* 					\\	
Milvus \cite{milvus}			& 2021 & FANNS	& pre-/post-filtering		& \makecell[l]{quantization / graph} 	& \xmark & \cmark & \xmark & \xmark  & \cmark & \xmark  & \xmark & \quad\cmark \cite{milvus-code}	\\	
NHQ \cite{nhq,nhq-2}			& 2022 & AFANNS	& hybrid index				& \makecell[l]{graph}					& \cmark & \xmark & \xmark & \cmark  & \xmark & \xmark  & \xmark & \quad\cmark \cite{nhq-code}		\\	
HQANN \cite{hqann}				& 2022 & AFANNS $^\dagger$& hybrid index	& \makecell[l]{graph}					& \cmark & \xmark & \xmark & \cmark  & \xmark & \xmark  & \xmark & \quad\xmark* 					\\	
AIRSHIP \cite{airship}			& 2022 & FANNS 	& in-filtering				& \makecell[l]{graph}					& \cmark & \cmark & \cmark & \cmark  & \cmark & \cmark  & \cmark & \quad\xmark* 					\\	
HQI \cite{hqi}					& 2023 & FANNS 	& hybrid index				& \makecell[l]{tree + quantization}		& \cmark & \cmark & \xmark & \cmark  & \cmark & \xmark  & \cmark & \quad\xmark* 					\\	
VBASE \cite{vbase}				& 2023 & FANNS 	& post-filtering			& \makecell[l]{any}						& \cmark & \cmark & \cmark & \cmark  & \cmark & \cmark  & \cmark & \quad\cmark \cite{vbase-code}	\\
FDANN \cite{fdann}				& 2023 & FANNS 	& hybrid index				& \makecell[l]{graph}					& \rtwo	 & \xmark & \cmark & \xmark  & \xmark & \xmark	& \xmark & \quad\cmark \cite{fdann-code}	\\	
CAPS \cite{caps}				& 2023 & FANNS 	& hybrid index				& \makecell[l]{quantization + tree}		& \cmark & \xmark & \xmark & \cmark  & \xmark & \xmark  & \xmark & \quad\cmark \cite{caps-code} 	\\	
ARKGraph \cite{arkgraph}		& 2023 & FANNS 	& hybrid index				& \makecell[l]{graph}					& \xmark & \cmark & \xmark & \xmark  & \xmark & \xmark  & \xmark & \quad\cmark \cite{arkgraph-code}	\\
ACORN \cite{acorn}				& 2024 & FANNS 	& in-filtering				& \makecell[l]{graph}					& \cmark & \cmark & \cmark & \cmark  & \cmark & \cmark  & \cmark & \quad\cmark \cite{acorn-code}	\\	
$\beta$-WST \cite{bwst}			& 2024 & FANNS 	& hybrid index				& \makecell[l]{tree + graph}			& \xmark & \cmark & \xmark & \xmark  & \xmark & \xmark  & \xmark & \quad\cmark \cite{bwst-code} 	\\
SeRF \cite{serf}				& 2024 & FANNS 	& hybrid index				& \makecell[l]{graph}					& \xmark & \cmark & \xmark & \xmark  & \rone  & \xmark  & \xmark & \quad\cmark \cite{serf-code} 	\\
iRangeGraph \cite{irg}			& 2024 & FANNS 	& hybrid index				& \makecell[l]{tree + graph}			& \xmark & \cmark & \xmark & \xmark  & \rone  & \xmark  & \rone  & \quad\cmark \cite{irg-code}  	\\
UNIFY \cite{unify}				& 2024 & FANNS 	& hybrid index				& \makecell[l]{graph}					& \xmark & \cmark & \xmark & \xmark  & \xmark & \xmark  & \xmark & \quad\cmark \cite{unify-code}	\\
UNG \cite{ung}					& 2024 & FANNS 	& hybrid index 				& \makecell[l]{graph}					& \rtwo	 & \xmark & \cmark & \rtwo	 & \xmark & \cmark  & \xmark & \quad\cmark \cite{ung-code}		\\	
RangePQ \cite{rpq}				& 2025 & FANNS 	& hybrid index 				& \makecell[l]{tree + quantization}		& \xmark & \cmark & \xmark & \xmark	 & \xmark & \xmark  & \xmark & \quad\cmark \cite{rpq-code}		\\	
DIGRA \cite{digra}				& 2025 & AFANNS & hybrid index 				& \makecell[l]{tree + graph}			& \xmark & \cmark & \xmark & \xmark	 & \xmark & \xmark  & \xmark & \quad\cmark \cite{digra-code}	\\	
DSG \cite{dsg}					& 2025 & FANNS 	& hybrid index 				& \makecell[l]{graph}					& \xmark & \cmark & \xmark & \xmark	 & \xmark & \xmark  & \xmark & \quad\cmark \cite{dsg-code}		\\	
TFANNS \cite{tfanns}			& 2025 & FANNS 	& hybrid index 				& \makecell[l]{graph}					& \rtwo	 & \xmark & \cmark & \xmark	 & \xmark & \xmark  & \xmark & \quad\cmark \cite{tfanns-code}	\\	
\bottomrule
\end{tabular}
\label{tab:fanns}
\vspace{-1.5em}
\end{table*}

%% file: fig/fig_selectivity.tex
\begin{figure}[h]
\centering
\captionsetup{justification=centering}
\includegraphics[width=1.0\columnwidth]{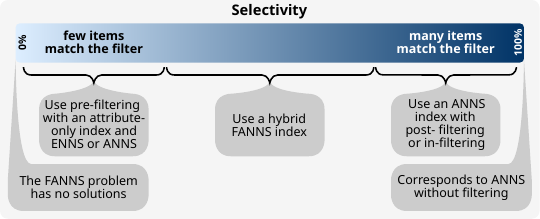}
\vspace{-2.5em}
\caption{\textbf{(\textsection \ref{sec:fanns-approaches}) The best approach for solving FANNS depends on the filter's selectivity within a given dataset.}}
\label{fig:selectivity}
\vspace{-1.0em}
\end{figure}

%% file: 03a_methods_explained_long.tex
\subsection{Overview of FANNS Methods}
\label{sec:fanns-explanation}

In \Cref{tab:fanns}, we organize existing \gls{fanns} methods according to our taxonomy.
We observe that most \gls{fanns} methods are based on graphs or quantization, with trees often used in combination with quantization- or graph-based techniques.
\Cref{fig:technique-classification} groups \gls{fanns} methods according to their internal working principles, and in the following, we describe these methods in more detail.

\input{fig/fig_technique_classification}

\subsubsection{Rii}
\label{sec:fanns-rii}

The Rii~\cite{rii} index is based on \gls{ivfpq}.
It takes a bitmap of matching items as input.
If the selectivity is low, it compares the query vector directly with the efficiently accessible, contiguously stored \gls{pq} codes (pre-filtering).
If the selectivity is high, it first performs the \gls{ivf} step of \gls{ivfpq} to identify the closest clusters.
Within those clusters, a \gls{pq} code is compared to the query vector only if the corresponding item matches the filter (in-filtering).

\subsubsection{MA-NSW}
\label{sec:fanns-mansw}

MA-NSW \cite{mansw} is a graph-based \gls{fanns} index that supports both \gls{em} and \gls{mem} filters. 
It constructs multiple \gls{nsw}-like, graph-based \gls{anns} indices, one for each possible combination of attribute values.
During query execution, MA-NSW queries the \gls{anns} index that matches the query's filter expression to approximately retrieve the \gls{knn} that satisfy the filter conditions.

\subsubsection{PASE}
\label{sec:fanns-pase}

PASE \cite{pase} integrates \gls{anns} into the PostgreSQL \cite{postgres} database by implementing the \gls{ivf} and \gls{hnsw} \gls{anns} indices. 
It employs an iterative post-filtering approach, in which it repeatedly retrieves items from the \gls{anns} index and filters them until $k$ matching items are found.

\subsubsection{AnalyticDB-V}
\label{sec:fanns-adbv}

Alibaba's AnalyticDB-V \cite{adbv} extends the AnalyticDB \cite{adb} SQL database with \gls{fanns} capabilities. 
It introduces its own \gls{anns} index, called \gls{vgpq}, which is similar to \gls{ivfpq}. 
For \gls{fanns} implementation, it supports both pre-filtering and post-filtering.

\subsubsection{Milvus}
\label{sec:fanns-milvus}

Milvus \cite{milvus} extends the FAISS \gls{anns} library \cite{faiss,faiss-github} by incorporating range filtering for one or multiple attributes. 
It supports both pre- and post-filtering, as well as a filtering scheme that partitions the data based on the most frequently filtered attribute and builds a separate \gls{anns} index for each partition.

\subsubsection{NHQ}
\label{sec:fanns-nhq}

NHQ \cite{nhq} proposes a method to extend any proximity graph-based \gls{anns} index into a hybrid \gls{fanns} index. 
Instead of constructing the \gls{anns} index solely based on embedding distance, it introduces a fusion distance that combines embedding distance with attribute dissimilarity. 
While this approach is applicable to any proximity graph-based \gls{anns} index, NHQ also presents construction schemes for two novel proximity graphs. 
Since its query execution scheme operates using the fusion distance without an explicit filtering step, NHQ may return items that do not match the filter, thus addressing the \gls{afanns} rather than the \gls{fanns} problem.

\subsubsection{HQANN}
\label{sec:fanns-hqann}

Like NHQ \cite{nhq}, HQANN \cite{hqann} employs a fusion distance to transform any proximity graph-based \gls{anns} index into a hybrid \gls{fanns} index. 
The key difference to NHQ is that HQANN prioritizes attribute dissimilarity in the fusion distance, whereas NHQ emphasizes embedding distance.

\subsubsection{AIRSHIP}
\label{sec:fanns-airship}

AIRSHIP \cite{airship} \ifarx(see \Cref{fig:alg-airship})~\fi is a graph-based \gls{fanns} index capable of handling arbitrary filter functions. 
It relies on a graph-based \gls{anns} index constructed without considering attributes but adapts query execution to \gls{fanns} through in-filtering and a series of optimizations. 
During query execution, AIRSHIP traverses the entire graph but includes only vertices that satisfy the filter in the results. 
\ifarx It selects a starting point within a cluster of items that match the filter and traverses the graph in multiple directions concurrently.\fi

\ifarx
\input{fig/fig_alg_airship}
\fi

\subsubsection{HQI}
\label{sec:fanns-hqi}

Apple's HQI \cite{hqi} \ifarx(see \Cref{fig:alg-hqi})~\fi is a hybrid \gls{fanns} index.
To integrate attribute filtering with nearest neighbor search, HQI transforms embedding vectors into attributes by applying k-means clustering and storing each item's cluster ID as an attribute. 
It then employs a balanced qd-tree \cite{qdtree}, an attribute-only index, to partition the items. 
Each non-leaf node in the qd-tree contains a predicate over the item attributes (including the cluster ID), with its two child nodes holding items that do or do not satisfy the predicate. 
During query execution, HQI identifies the query vector’s $w$ closest cluster IDs and prunes all branches of the tree that do not contain any of these $w$ cluster IDs or whose attributes do not match the filter. 
\ifarx
An \gls{ivf} index is used within each leaf.
\fi

\ifarx
\input{fig/fig_alg_hqi}
\fi

\subsubsection{VBASE}
\label{sec:fanns-vbase}

VBASE \cite{vbase} extends PostgreSQL \cite{postgres} by integrating \gls{anns} into the database system. 
Its key insight is that most \gls{anns} query execution methods follow a two-stage process: an initial stage where traversal moves from a starting point in the direction of the query vector, and a second stage where the search gradually moves away from the query vector to identify its \gls{knn}. 
Rather than employing a naïve pre-filtering approach, where the \gls{anns} index is queried with a fixed $k' \geq k$ and non-matching items are filtered out, VBASE runs the \gls{anns} query only until the traversal starts moving away from the query vector. 
From that point onward, it iteratively retrieves and filters additional items until the required $k$ matches are found.

\subsubsection{FDANN}
\label{sec:fanns-fdann}

Microsoft's Filtered-DiskANN (FDANN) \cite{fdann} \ifarx(see \Cref{fig:alg-fdann})~\fi implements a graph-based index with \gls{emis} filtering. 
In addition to supporting an \gls{emis} filter with a single query-label, it also allows filtering with multiple query-labels, where at least one of them must be present in the item's set attribute. 
Filtered-DiskANN introduces two methods for constructing \gls{fanns} indices based on Vamana graphs \cite{diskann}: \textit{FilteredVamana} and \textit{StitchedVamana}. 
During query execution, it traverses the graph starting from a set of entry points that contain the query-label. 
Throughout its \gls{nsw}-like graph traversal, it considers only those vertices that satisfy the filter criteria.

\ifarx
\input{fig/fig_alg_fdann}
\fi

\subsubsection{CAPS}
\label{sec:fanns-caps}

In contrast to most \gls{fanns} methods, CAPS \cite{caps} \ifarx(see \Cref{fig:alg-caps})~\fi introduces a quantization- and tree-based index that is easier to parallelize than the more common graph-based indices. 
The CAPS index consists of two levels. The first level organizes items into clusters based on their embedding distance only or a combination of embedding distance and attribute dissimilarity.
At the second level, CAPS constructs an \gls{aft}, which is an attribute-only index. 
Each level of the \gls{aft} splits items into two buckets: one containing items with the most common attribute value (a leaf node) and another with the remaining items (a non-leaf node), unless the maximum depth is reached.
During query execution, CAPS selects the $w$ clusters closest to the query vector and uses the corresponding \glspl{aft} to identify matching items.
\ifarx
\input{fig/fig_alg_caps}
\fi

\subsubsection{ARKGraph}
\label{sec:fanns-arkgraph}

Given a set of items, each with a single \textit{ordered attribute}, ARKGraph \cite{arkgraph} \ifarx(see \Cref{fig:alg-arkgraph})~\fi addresses the more general problem of constructing the \gls{kgraph} \cite{knn-graph} for the subset of items that satisfy a range filter. 
The resulting \gls{kgraph} can be utilized for solving the \gls{fanns} problem or for other data analysis tasks.
A naïve approach would store $O(n^2)$ \glspl{kgraph}, one for each possible query range. 
For any given item $I$ with attribute value $x$, this would require maintaining $O(n^2)$ different lists of \gls{knn}, one for each possible query range $[l,r]$ such that $l \leq x \leq r$. 
ARKGraph's first key insight is that the \gls{knn} of $I$ for any query range $[l,r]$ can be efficiently computed by merging the \gls{knn} of $I$ from the partial query ranges $[l,x)$ and $(x,r]$. 
Thus, instead of storing \gls{knn} lists for all $O(n^2)$ query ranges, it suffices to maintain \gls{knn} lists for only $O(n)$ partial query ranges and reconstruct the \gls{knn} dynamically for any given range.
The second key insight is that an item often shares the same \gls{knn} across multiple similar partial query ranges. 
By grouping query ranges and storing only one \gls{knn} set per group, ARKGraph further reduces index size and storage overhead. 

\ifarx
\input{fig/fig_alg_arkgraph}
\fi

\subsubsection{ACORN}
\label{sec:fanns-acorn}

ACORN \cite{acorn} \ifarx(see \Cref{fig:alg-acorn})~\fi supports arbitrary filter types and filtering across multiple attributes. 
The ACORN index is a denser variant of the HNSW \gls{anns} index \cite{hnsw}, constructed without considering attributes. 
During query execution, ACORN traverses only the subgraph induced by vertices that match the filter. 
This approach differs from AIRSHIP \cite{airship}, which traverses the entire graph but includes only matching vertices in the result set. 
The ACORN index's subgraphs approximate an \gls{hnsw} \cite{hnsw}.

\ifarx
\input{fig/fig_alg_acorn}
\fi

\subsubsection{$\beta$-WST}
\label{sec:fanns-bwst}

$\beta$-WST \cite{bwst} \ifarx(see \Cref{fig:alg-bwst})~\fi is a \gls{fanns} index designed for range filtering on a single attribute.  
A $\beta$-WST index is structured as a segment tree \cite{seg-tree} with a uniform branching factor of $\beta$. 
At layer $l$, the attribute value range is divided into $\beta^l$ segments, with a graph-based \gls{anns} index constructed for each segment.
\ifarx
The default query strategy selects a minimal set of tree nodes that fully cover the query range, queries the corresponding \gls{anns} indices, and merges the results.
Additional strategies include \textit{OptimizedPostfiltering}, \textit{TreeSplit}, and \textit{SuperPostfiltering}.
\fi

\ifarx
\input{fig/fig_alg_bwst}
\fi

\subsubsection{SeRF}
\label{sec:fanns-serf}

SeRF \cite{serf} introduces the \textit{segment graph}, a data structure for range-filtered \gls{fanns} queries with half-bounded query ranges (i.e., where the query range has an upper limit but no lower limit). 
It losslessly compresses $n$ \gls{hnsw} indices \cite{hnsw}, corresponding to the $n$ possible half-bounded query ranges, into a single index. 
This is achieved by incrementally inserting items into an \gls{hnsw} index in attribute-value order. 
Each edge in the resulting graph stores the attribute value at which it was added and, if pruned during construction, the attribute value at which it was removed. 
As a result, the \gls{hnsw} graph contains edges that are valid only within specific attribute value intervals.
During query execution, only edges whose validity interval contains the upper limit of the query range are considered. 
For query ranges with an upper and a lower limit, SeRF introduces the \textit{2D segment graph}\ifarx~(see \Cref{fig:alg-serf})\fi. 
This structure compresses $n$ segment graphs, one for each possible lower limit, into a single graph with $n$ vertices, where edges store validity intervals for the upper and lower limit of the query range.

\ifarx
\input{fig/fig_alg_serf}
\fi

\subsubsection{TFANNS}
\label{sec:fanns-tfanns}

TFANNS \cite{tfanns} proposes three graph-based \gls{fanns} indices for \gls{emis} filtering: The \textit{local-}, \textit{global-}, and \textit{packing method}.

\subsubsection{iRangeGraph}
\label{sec:fanns-irangegraph}

iRangeGraph \cite{irg} \ifarx(see \Cref{fig:alg-irangegraph})~\fi is a graph-based \gls{fanns} index designed for range filtering. 
It is primarily built around a single \textit{ordered attribute} but also supports in-filtering or post-filtering for secondary attributes of arbitrary types.
iRangeGraph constructs a set of graph-based \gls{anns} indices, referred to as \textit{elemental graphs}. 
These elemental graphs are organized in a segment tree \cite{seg-tree} with $O(\log n)$ layers. 
At layer $l$, the entire attribute-value range is divided into $2^l$ segments, and a graph-based \gls{anns} index is built for each segment, ensuring that each item appears once in each of the $\log n$ layers.
Query execution employs an efficient on-the-fly method to dynamically merge a subset of elemental graphs into a single \gls{anns} index containing only items within the query range. 
\ifarx
While its index construction is similar to $\beta$-WST \cite{bwst}, iRangeGraph differs in query execution: instead of querying multiple graph-based indices for different subsets of items and merging the results, it constructs a single graph-based \gls{anns} index dynamically at query time.
\fi

\ifarx
\input{fig/fig_alg_irangegraph}
\fi

\subsubsection{RangePQ}
\label{sec:fanns-rpq}

RangePQ \cite{rpq} answers \gls{r}-filtered \gls{fanns} queries by combining a \gls{pq} index for embedding vectors with a \gls{bst} for attribute values.

\subsubsection{UNIFY}
\label{sec:fanns-unify}

UNIFY \cite{unify} is a graph-based method that supports range filtering for a single attribute.
It introduces two new types of proximity graph, the \gls{sig} and its \gls{hnsw}-like variant the \gls{hsig} \ifarx(see \Cref{fig:alg-unify})\fi.
It divides the range of possible attribute values into $s$ segments, s.t., each segment contains approximately the same number of items.
The high-level idea of a \gls{sig} is to build a proximity graph \cite{551} for each of the $2^s-1$ possible combinations of segments, and then merge these graphs into a single graph that contains the $2^s-1$ smaller graphs as subgraph.
Due to a clever construction algorithm, the \gls{sig} can be built without explicitly constructing all $2^s-1$ smaller graphs.
\ifarx
The \gls{sig} is stored as a segmented adjacency list, where outgoing edges are grouped by their destination segment.
\fi
When querying the \gls{sig}, we only traverse the subgraph that correspond to those segments that intersect with the query range.
The \gls{hsig} is a multi-layered variant of the \gls{sig} which is traversed from top to bottom in order to speed-up the query execution.
\ifarx
To handle queries with low selectivity, UNIFY supports a pre-filtering variant and for queries with high selectivity, it supports unfiltered \gls{anns} with post-filtering.
\fi

\ifarx
\input{fig/fig_alg_unify}
\fi

\subsubsection{UNG}
\label{sec:fanns-ung}

UNG \cite{ung} \ifarx(see \Cref{fig:alg-ung})~\fi is a graph-based \gls{fanns} index that supports both \gls{emis} and \gls{memis} filters. 
It first identifies all distinct label sets in the database and constructs a \gls{lng}, which is a \gls{dac} with one vertex for each distinct label set and a directed edge from label set $a$ to label set $b$ if $a$ is a subset of $b$.
The \gls{ung} index is built by constructing a proximity graph for each vertex in the \gls{lng}, where the vertices of each proximity graph represent the items with the corresponding label set.
To unify these proximity graphs, cross-edges are added between vertices in the proximity graphs of label sets $a$ and $b$ if there is a corresponding edge from $a$ to $b$ in the \gls{lng}.
When querying the \gls{ung}, the standard graph traversal is initiated, starting in all proximity graphs whose label sets are minimal supersets of the query label set.

\ifarx
\input{fig/fig_alg_ung}
\fi

\subsubsection{DIGRA}
\label{sec:fanns-digra}

DIGRA \cite{digra} uses a multi-way tree structure to organize items based on the value of their single ordered attribute.
At each node of this tree, a \gls{nsw} \cite{nsw} graph is used to perform \gls{anns} on its subtree.
Each \gls{r} filtered query can be answered by querying only two of these \gls{nsw} graphs.

\subsubsection{DSG}
\label{sec:fanns-dsg}

The Dynamic Segment Graph (DSG) \cite{dsg} is an extension of SeRF \cite{serf} that is optimized for cases, where the vectors are dynamically inserted into the index in random order, and first sorting them by attribute value is not possible.

%% file: fig/fig_technique_classification.tex
\begin{figure}[h]
\centering
\captionsetup{justification=centering}
\includegraphics[width=1.0\columnwidth]{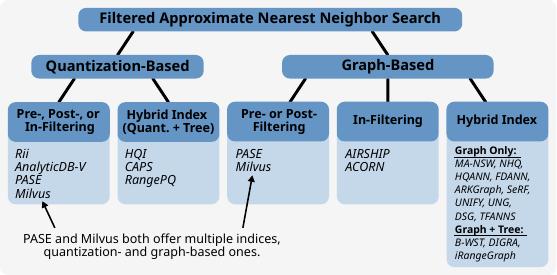}
\vspace{-2.5em}
\caption{\textbf{(\textsection \ref{sec:fanns-explanation}) Grouping of FANNS method based on their internal working principles.}}
\label{fig:technique-classification}
\vspace{-1em}
\end{figure}

%% file: fig/fig_alg_airship.tex
\begin{figure}[H]
\vspace{-0.5em}
\centering
\captionsetup{justification=centering}
\includegraphics[width=1.0\columnwidth]{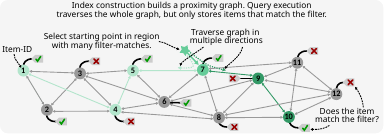}
\vspace{-2.25em}
\caption{\textbf{(\textsection \ref{sec:fanns-airship}) Visualization of the AIRSHIP index.}}
\label{fig:alg-airship}
\vspace{-1.0em}
\end{figure}

%% file: fig/fig_alg_hqi.tex
\begin{figure}[H]
\vspace{-1.0em}
\centering
\captionsetup{justification=centering}
\includegraphics[width=1.0\columnwidth]{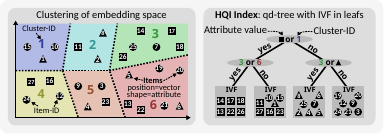}
\vspace{-2.25em}
\caption{\textbf{(\textsection \ref{sec:fanns-hqi}) Visualization of the HQI index.}}
\label{fig:alg-hqi}
\end{figure}

%% file: fig/fig_alg_fdann.tex
\begin{figure}[H]
\vspace{-0.5em}
\centering
\captionsetup{justification=centering}
\includegraphics[width=1.0\columnwidth]{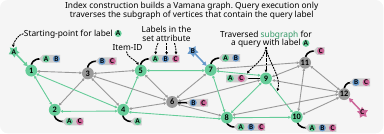}
\vspace{-2em}
\caption{\textbf{(\textsection \ref{sec:fanns-fdann}) The Filtered-DiskANN index visualized.}}
\label{fig:alg-fdann}
\vspace{-0.5em}
\end{figure}

%% file: fig/fig_alg_caps.tex
\begin{figure}[H]
\vspace{-0.5em}
\centering
\captionsetup{justification=centering}
\includegraphics[width=1.0\columnwidth]{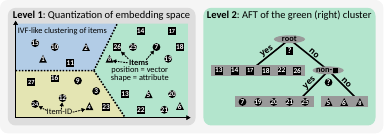}
\vspace{-2em}
\caption{\textbf{(\textsection \ref{sec:fanns-caps}) Visualization of the CAPS index.}}
\label{fig:alg-caps}
\end{figure}

%% file: fig/fig_alg_arkgraph.tex
\begin{figure}[H]
\vspace{-1.0em}
\centering
\captionsetup{justification=centering}
\includegraphics[width=1.0\columnwidth]{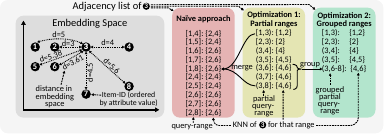}
\vspace{-2.25em}
\caption{\textbf{(\textsection \ref{sec:fanns-arkgraph}) Visualization of the adjacency lists of item 3 in ARKGraph using the three different approaches.}}
\label{fig:alg-arkgraph}
\vspace{-1.0em}
\end{figure}

%% file: fig/fig_alg_acorn.tex
\begin{figure}[H]
\vspace{-1.0em}
\centering
\captionsetup{justification=centering}
\includegraphics[width=1.0\columnwidth]{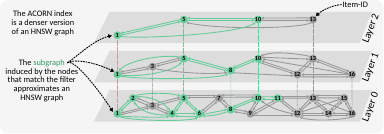}
\vspace{-2.25em}
\caption{\textbf{(\textsection \ref{sec:fanns-acorn}) Visualization of the ACORN index.}}
\label{fig:alg-acorn}
\vspace{-1.0em}
\end{figure}

%% file: fig/fig_alg_bwst.tex
\begin{figure}[H]
\centering
\captionsetup{justification=centering}
\includegraphics[width=1.0\columnwidth]{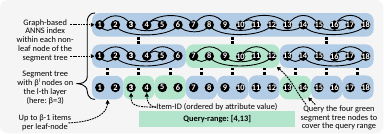}
\vspace{-2.25em}
\caption{\textbf{(\textsection \ref{sec:fanns-bwst}) Visualization of the $\beta$-WST for $\beta = 3$.}}
\label{fig:alg-bwst}
\vspace{-1.0em}
\end{figure}

%% file: fig/fig_alg_serf.tex
\begin{figure}[H]
\centering
\captionsetup{justification=centering}
\includegraphics[width=1.0\columnwidth]{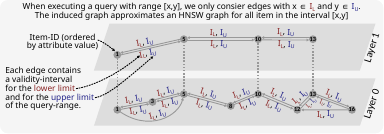}
\vspace{-2em}
\caption{\textbf{(\textsection \ref{sec:fanns-serf}) SeRF's 2D segment graph visualized.}}
\label{fig:alg-serf}
\vspace{-1.0em}
\end{figure}

%% file: fig/fig_alg_irangegraph.tex
\begin{figure}[H]
\centering
\captionsetup{justification=centering}
\includegraphics[width=1.0\columnwidth]{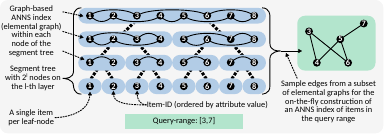}
\vspace{-2.25em}
\caption{\textbf{(\textsection \ref{sec:fanns-irangegraph}) Visualization of the iRangeGraph index.}}
\label{fig:alg-irangegraph}
\vspace{-1.0em}
\end{figure}

%% file: fig/fig_alg_unify.tex
\begin{figure}[H]
\vspace{-1.0em}
\centering
\captionsetup{justification=centering}
\includegraphics[width=1.0\columnwidth]{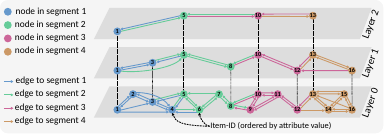}
\vspace{-2.25em}
\caption{\textbf{(\textsection \ref{sec:fanns-unify}) UNIFY's HSIG index visualized.}}
\label{fig:alg-unify}
\vspace{-1.5em}
\end{figure}

%% file: fig/fig_alg_ung.tex
\begin{figure}[H]
\centering
\captionsetup{justification=centering}
\includegraphics[width=1.0\columnwidth]{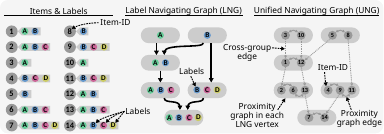}
\vspace{-2.25em}
\caption{\textbf{(\textsection \ref{sec:fanns-ung}) Visualization of the UNG index.}}
\label{fig:alg-ung}
\vspace{-1.0em}
\end{figure}

%% file: 03a_methods_explained_short.tex
\subsection{Overview of FANNS Methods}
\label{sec:fanns-explanation}

In \Cref{tab:fanns}, we organize existing \gls{fanns} methods according to our taxonomy.
We observe that most \gls{fanns} methods are based on graphs or quantization, with trees often used in combination with quantization- or graph-based techniques.
\Cref{fig:technique-classification} groups \gls{fanns} methods according to their internal working principles, and in the following, we describe these methods in more detail.

\input{fig/fig_technique_classification}
\subsubsection{Quantization-based Methods}
\label{sec:fanns-explained-quant}

Quantization-based \gls{fanns} indices use \gls{anns} methods such as \gls{ivf} \cite{ivf} or \gls{ivfpq} \cite{pq} to group nearby embedding vectors into clusters.
By pruning vectors that belong to clusters far from the query vector and by approximating distances between query and database vectors using distances to cluster centroids, these methods enable fast approximate nearest neighbor search.
To support filtering, quantization-based \gls{anns} methods can be combined with pre-, post-, or in-filtering, as done by Rii \cite{rii}, PASE \cite{pase}, AnalyticDB-V \cite{adbv}, and Milvus \cite{milvus}.
Alternatively, filtering can be integrated by constructing a hybrid index that combines quantization with a tree-based attribute index, as done by HQI \cite{hqi}, CAPS \cite{caps}, and RangePQ \cite{rpq}.

Apple's \textbf{HQI} \cite{hqi} uses quantization to transform embedding vectors into attributes by applying k-means clustering and storing each item's cluster ID as an attribute.
It then employs a balanced qd-tree \cite{qdtree}, which is an attribute-only index, to partition items based on both the original attributes and the cluster ID.
Each non-leaf node in the qd-tree contains a predicate over the item attributes, including the cluster ID, and its two child nodes hold items that do or do not satisfy the predicate.
During query execution, HQI identifies the $w$ cluster IDs closest to the query vector and prunes all branches of the tree that do not contain any of these $w$ cluster IDs or whose attributes do not satisfy the filter.

The \textbf{CAPS} index \cite{caps} follows a different approach.
It uses a two-level structure in which, at the first level, items are organized into clusters based on their embedding distances, and at the second level, items within each cluster are organized by attributes using an \gls{aft}, an attribute-only index.
During query execution, CAPS selects the $w$ clusters closest to the query vector and uses the corresponding \glspl{aft} to identify matching items.

\textbf{RangePQ} \cite{rpq}, a method for \gls{r}-filtered \gls{fanns}, also uses quantization to prune items that are far from the query vector.
In addition to pruning by embedding distance, it also removes clusters that do not contain items matching the range filter.

\subsubsection{Graph-based Methods}
\label{sec:fanns-explained-graph}

Graph-based \gls{fanns} indices build proximity graphs \cite{551}, such as \gls{nsw} \cite{nsw} or \gls{hnsw} \cite{hnsw}, over the embedding vectors.
To answer queries, they traverse the graph, moving from an entry point towards the query vector to find its approximate \gls{knn}.
There are multiple ways to integrate filtering into this process.
PASE \cite{pase} and Milvus \cite{milvus} use pre- or post-filtering in combination with standard graph-based \gls{anns} indices.
\textbf{AIRSHIP} \cite{airship} and \textbf{ACORN} \cite{acorn} both use in-filtering during graph traversal.
The main difference between the two is that ACORN traverses only the subgraph induced by vertices matching the filter, whereas AIRSHIP traverses the entire graph but includes only matching vertices in the result set.

Most graph-based \gls{fanns} methods, however,  construct a hybrid index where both embedding vectors and attributes are considered during index construction.
These hybrid indices can again be classified into two groups.
Graph-only indices, such as MA-NSW \cite{mansw}, NHQ \cite{nhq}, HQANN \cite{hqann}, FDANN \cite{fdann}, ARKGraph \cite{arkgraph}, SeRF \cite{serf}, UNIFY \cite{unify}, UNG \cite{ung}, DSG \cite{dsg}, and TFANNS \cite{tfanns}, build a proximity graph-based index that reflects both embedding vectors and attribute values.
The second group consists of two-level indices, such as $\beta$-WST \cite{bwst}, iRangeGraph \cite{irg}, and DIGRA \cite{digra}, which use a tree structure to partition items based on their attribute values and then build a proximity graph-based \gls{anns} index for each partition of the tree.

\textbf{MA-NSW} \cite{mansw} constructs multiple \gls{nsw}-like, graph-based \gls{anns} indices, one for each possible combination of attribute values.
\textbf{NHQ} \cite{nhq} and \textbf{HQANN} \cite{hqann} both employ a fusion distance that combines embedding distance with attribute dissimilarity to transform any proximity graph-based \gls{anns} index into a hybrid \gls{fanns} index.
The key difference is that NHQ prioritizes embedding distance in the fusion distance, whereas HQANN emphasizes attribute dissimilarity.
\textbf{FDANN} \cite{fdann} works similarly to the aforementioned in-filtering methods but uses the attributes to precompute suitable entry points for the graph traversal based on the query's filter.
\textbf{ARKGraph} \cite{arkgraph} addresses the general problem of constructing the \gls{kgraph} \cite{knn-graph} for the subset of items that satisfy a range filter, which can then be used to solve \gls{r}-filtered \gls{fanns} queries.
Using a series of efficient techniques, the $O(n^2)$ possible \glspl{kgraph} for all possible range filters can be compressed into a single index with only minor storage overhead.
\textbf{SeRF} \cite{serf} and \textbf{DSG} \cite{dsg} are indices for \gls{r}-filtered \gls{fanns} queries.
They both construct an \gls{hnsw}-like proximity graph in which each edge is only valid for a specific range of \gls{r} filters.
The more recent DSG \cite{dsg} is an optimized version of SeRF \cite{serf} for dynamic datasets where items are inserted in random order.
\textbf{UNIFY} \cite{unify} introduces two new types of proximity graph: the \gls{sig} and its \gls{hnsw}-like variant, the \gls{hsig}.
In both, the range of possible attribute values is divided into segments and edges are grouped by their destination segment.
During query execution, only edges whose destination segment intersects with the query range are traversed.
\textbf{UNG} \cite{ung} is a two-level graph-based \gls{fanns} index, where at the first level, a \gls{lng} is constructed to represent all distinct label sets in the database and the subset relationships between them.
At the second level, each vertex of the \gls{lng} contains a proximity graph over the items with the corresponding label set.

The second group of graph-based \gls{fanns} indices consists of two-level indices that first use a tree structure to partition items based on their attribute values and then build a proximity graph for each tree partition.
\textbf{$\beta$-WST} \cite{bwst} employs a segment tree \cite{seg-tree} with a uniform branching factor of $\beta$ to divide the attribute value range into segments at each tree layer.
A graph-based \gls{anns} index is constructed for each segment.
During query execution, a minimal set of tree nodes that fully cover the query range is selected, and the corresponding \gls{anns} indices are queried and merged.
\textbf{iRangeGraph} \cite{irg} similarly constructs a set of graph-based \gls{anns} indices, called \textit{elemental graphs}, organized in a segment tree \cite{seg-tree} with $O(\log n)$ layers.
At layer $l$, the entire attribute-value range is divided into $2^l$ segments, and a graph-based \gls{anns} index is built for each segment.
Query execution uses an efficient on-the-fly method to dynamically merge a subset of elemental graphs into a single \gls{anns} index containing only items within the query range.
\textbf{DIGRA} \cite{digra} organizes items using a multi-way tree structure based on the value of a single ordered attribute.
At each tree node, a \gls{nsw} \cite{nsw} graph is used to perform \gls{anns} on its subtree.
The solution for each \gls{r}-filtered query can be approximated by querying only two of these \gls{nsw} graphs.

%% file: 04_dataset.tex
\section{A New Dataset to Benchmark FANNS}
\label{sec:dataset}

Benchmarking \gls{fanns} methods requires suitable datasets. 
Ideally, such datasets should contain real-world attributes (criterion 1), queries with ground truth (criterion 2), and be publicly available (criterion 3).
The most common approach to evaluate \gls{fanns} methods that we observed in scientific literature is to use an \gls{anns} dataset such as SIFT \cite{sift-ds}, GIST \cite{sift-ds}, GloVe \cite{glove-ds}, UQ-V \cite{nhq-code}, Msong \cite{nhq-code}, Audio \cite{nhq-code}, Crawl \cite{nhq-code}, Enron \cite{nhq-code}, BIGANN \cite{bigann-ds}, Deep1B \cite{deep1b-ds}, MSTuring \cite{bigann-ds}, or YandexT2I \cite{bigann-ds} and extend it with synthetic, randomly generated attributes~\cite{mansw,milvus,nhq,hqann,airship,hqi,fdann,caps,bwst,serf,unify,ung,digra,rpq} (failing to meet criterion~1).
Another common, though more labor-intensive, approach is to repurpose a dataset not originally intended for \gls{fanns} such as TripClick \cite{tripclick-ds}, MNIST \cite{mnist-ds}, LAION \cite{laion-ds}, RedCaps \cite{redcaps-ds}, YouTube-Rgb \cite{youtube8m-ds}, Youtube-Audio \cite{youtube8m-ds}, Words \cite{words-ds}, MTG \cite{mtg-ds}, or WIT-Image \cite{wit-ds} by manually creating embedding vectors and queries, or by crawling the web to obtain suitable real-world attributes~\cite{acorn,bwst,serf,irg,unify,ung,digra} (failing to meet criterion~2).
Finally, some papers originating from industry rely on proprietary in-house datasets that are not publicly available~\cite{adbv,hqann,hqi} (violating criterion~3).
The only dataset we are aware of that meets all three criteria is the \texttt{Paper} dataset ~\cite{nhq-code}, containing 200-dimensional embeddings of research paper abstracts.

In addition to the scarcity of suitable datasets that fulfill all three criteria, we observe that text embeddings from state-of-the-art transformer-based models such as NV-Embed \cite{nvembed,nvembed-hf}, LENS \cite{lens,lens-hf}, GTE \cite{gte,gte-hf}, Stella \cite{stella,stella-hf}, and BGE \cite{bge,bge-hf} with thousands of dimensions are not represented in existing \gls{fanns} datasets.
To clarify whether such a dataset is desperately needed or if the performance of \gls{fanns} methods is agnostic to the embedding model used, we need to answer the following two questions:
\begin{enumerate}
	\item Do the characteristics of embedding vectors vary significantly across different embedding models?
	\item Does the performance of \gls{fanns} methods depend on the characteristics of the embedding vectors?
\end{enumerate}

To answer the first question, we compare embeddings from the \texttt{SIFT} \cite{sift-ds} (images), \texttt{Audio} \cite{nhq-code} (audio), \texttt{UQ-V} \cite{nhq-code} (video), and \texttt{Paper} \cite{nhq-code} (text) datasets to transformer-based text embeddings.  
In \Cref{tab:ds-dist-summary} we report the dimensionality of embedding vectors, the mean absolute correlation between dimensions of the embedding space, as well as the mean, standard deviation, and best-fitting distribution (among ten candidates) for vector norms and pairwise distances between vectors in each dataset
\ifarx
. \Cref{fig:ds-norm,fig:ds-distance,fig:ds-correlation} show histograms for the vector norms and pairwise distances, as well as the correlation matrix for inter-dimension correlations of all five datasets.
\fi
These results reveal our first key observation:

\vspace{0.25em}
\gray{\textbf{Observation \rone}:~} The characteristics of embedding vectors vary significantly across different embedding models.
\vspace{0.25em}

\input{tab/tab_dataset_analysis}

\ifarx
\input{fig/fig_dataset_analysis_norm}

\input{fig/fig_dataset_analysis_distance}

\input{fig/fig_dataset_analysis_correlation}

\fi

\input{fig/fig_bench_recall_vs_qps_baselines}		

To answer the second question, we evaluate seven representative \gls{fanns} methods on the \texttt{SIFT} \cite{sift-ds}, \texttt{Audio} \cite{nhq-code}, \texttt{UQ-V} \cite{nhq-code}, and \texttt{Paper} \cite{nhq-code} datasets.
For these experiments, we follow the same methodology as described in \Cref{sec:bench}.
\Cref{fig:bench-recall-baselines} shows the recall vs. QPS trade-off for each method on each dataset and \Cref{fig:bench-tti-mem-baselines} shows the corresponding index construction time, peak memory usage, and index size.
We observe that while some methods such as UNG or FDANN (both stitched and filtered) perform consistently across datasets, many methods exhibit significant performance variations depending on the dataset used.
ACORN is significantly slower on \texttt{Paper} compared to other datasets, NHQ (nsw) fails to achieve a recall of over 0.4 on the \texttt{Audio} dataset, NHQ (kgraph) achieves significantly higher recall on \texttt{SIFT} and \texttt{Paper} than on \texttt{Audio} and \texttt{UQ-V}, and CAPS is way more competitive on \texttt{Paper} and \texttt{Audio} than on \texttt{SIFT} and \texttt{UQ-V}, which leads us to our second observation:

\vspace{0.25em}
\gray{\textbf{Observation \rtwo}:~} The performance of most \gls{fanns} methods depends on the characteristics of the embedding vectors.
\vspace{0.25em}

Our first two observations motivate the need for a new \gls{fanns} dataset that includes embeddings from state-of-the-art transformer-based models.
To this end, we introduce the \texttt{arxiv-for-fanns} dataset, where each item corresponds to a research paper.
Our dataset is available in three scales: \texttt{small} (1k items, 197 MB), \texttt{medium} (100k items, 1.86 GB), and \texttt{large} (over 2.7M items, 46.1 GB).
Each item includes a 4096-dimensional, normalized text embedding of the paper abstract, computed using the \texttt{stella\_en\_400M\_v5} model~\cite{stella-hf,stella}, based on abstracts from the arXiv dataset~\cite{arxiv-ds}.
We select \texttt{stella} because it is the most lightweight among the top 10 models\footnote{As of March 11, 2025}~on the English MTEB leaderboard~\cite{mteb}.
Critically, each item contains 11 real-world attributes \ifarx(see \Cref{tab:arxiv-dataset})~\fi covering categorical, numerical and set-valued attributes with diverse value distributions \ifarx (shown in \Cref{fig:dataset-properties})\fi, fulfilling criterion~1.
To satisfy criterion~2, each dataset variant includes three sets of 10k queries for \gls{em}, \gls{r}, and \gls{emis} filters (see \Cref{sec:back-filter}), along with their corresponding precomputed ground truth.
Our queries filter for the number of sub categories (EM), update date (R), and for the main categories (EMIS), but a user can easily construct his own queries using any combination of the 11 attributes.

\ifarx
\input{tab/tab_arxiv_dataset}

\fi

\ifarx
\input{fig/fig_bench_tti_and_mem_baselines}

\input{fig/fig_dataset_properties}

\fi

Query vectors are generated by embedding search terms synthesized by GPT-4o~\cite{gpt4o}, while filter values are sampled from the dataset attributes.
\Cref{fig:query-properties} illustrates the selectivity distribution (as defined in \Cref{sec:fanns-approaches}) across queries in \texttt{arxiv-for-fanns-large}.
\ifarx To support unfiltered \gls{anns} evaluation, we also provide ground truth results without any filtering.~\fi
All three versions of the dataset are publicly available on Hugging Face\footnote{\dataseturlsmall}\footnote{\dataseturlmedium}\footnote{\dataseturl}, fulfilling criterion~3.

\input{fig/fig_query_properties}

\ifarx
\fi

%% file: tab/tab_dataset_analysis.tex
\begin{table}[h]
\vspace{-1.0em}
\footnotesize
\setlength{\tabcolsep}{5pt}
\centering
\captionsetup{justification=centering}
\caption{\textbf{(\textsection \ref{sec:dataset}) Summary statistics of embedding vectors.}}
\vspace{-1.75em}
\begin{tabular}{llll}
\toprule
\textbf{\makecell[l]{Dataset}} 
&\textbf{\makecell[l]{Quantity}} 
&\textbf{\makecell[l]{Mean ($\pm$ Std)}} 
&\textbf{\makecell[l]{Best Fit}} \\
\midrule
	\multirow{4}{*}{\texttt{SIFT}} 
  & Dimensions & 128 & -- \\
  & Correlation & 0.182 & -- \\
  & Vector Norms & 509 $\pm$ 0.66 & Beta ($a{=}5.5, b{=}8.7$) \\
  & Pairw. Distances & 536 $\pm$ 115  & Beta ($a{=}168, b{=}2.1$) \\
\midrule
	\multirow{4}{*}{\texttt{Audio}} 
  & Dimensions & 192 & -- \\
  & Correlation & 0.155 & -- \\
  & Vector Norms & $7.9{\times}10^{5}\,\pm\,2.0{\times}10^{4}$ & Beta ($a{=}12.2, b{=}10.1$) \\
  & Pairw. Distances & $1.46{\times}10^{5}\,\pm\,3.8{\times}10^{4}$ & Normal \\
\midrule
	\multirow{4}{*}{\texttt{UQ-V}} 
  & Dimensions & 256 & -- \\
  & Correlation & 0.738 & -- \\
  & Vector Norms & 0.225 $\pm$ 0.267 & Lognormal ($s{=}0.81$) \\
  & Pairw. Distances & 0.201 $\pm$ 0.201 & Lognormal ($s{=}0.71$) \\
\midrule
	\multirow{4}{*}{\texttt{Paper}} 
  & Dimensions & 200 & -- \\
  & Correlation & 0.192 & -- \\
  & Vector Norms & 2.13 $\pm$ 0.14 & Lognormal ($s{=}0.31$) \\
  & Pairw. Distances & 1.10 $\pm$ 0.24 & Beta ($a{=}472, b{=}19.6$) \\
\midrule
	\multirow{4}{*}{\makecell{Ours\\(trans-\\former-\\based)}} 
  & Dimensions & 4096 & -- \\
  & Correlation & 0.099 & -- \\
  & Vector Norms & 1.00 $\pm$ $1.3{\times}10^{-7}$ & Normal \\
  & Pairw. Distances & 1.08 $\pm$ 0.14 & Weibull ($k{=}4.6{\times}10^{4}$) \\
\bottomrule
\end{tabular}
\label{tab:ds-dist-summary}
\vspace{-1.0em}
\end{table}

%% file: fig/fig_dataset_analysis_norm.tex
\begin{figure*}[h]
\centering
\captionsetup{justification=centering}
\includegraphics[width=1.0\textwidth]{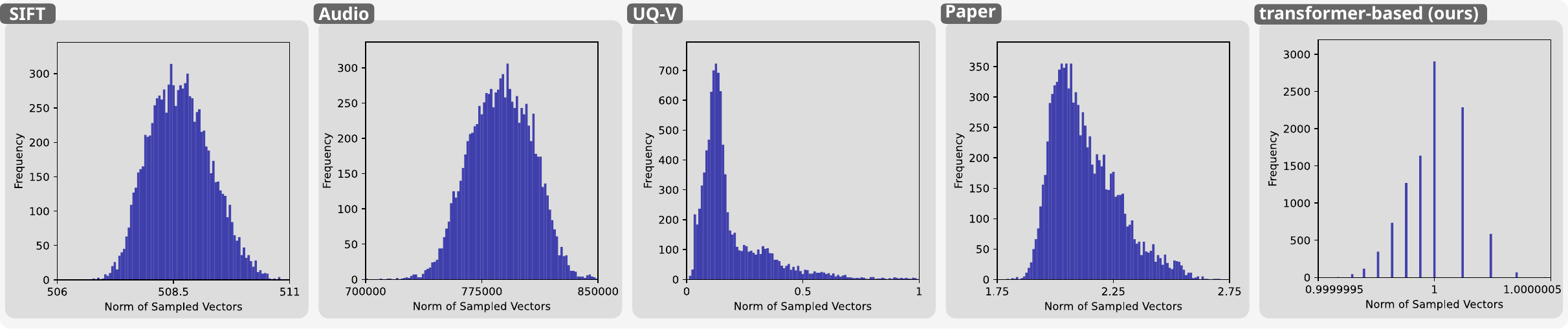}
\vspace{-2.5em}
\caption{\textbf{(\textsection \ref{sec:dataset}) Distribution of vector norm among sampled embedding vectors}.}
\label{fig:ds-norm}
\vspace{-1.0em}
\end{figure*}

%% file: fig/fig_dataset_analysis_distance.tex
\begin{figure*}[h]
\centering
\captionsetup{justification=centering}
\includegraphics[width=1.0\textwidth]{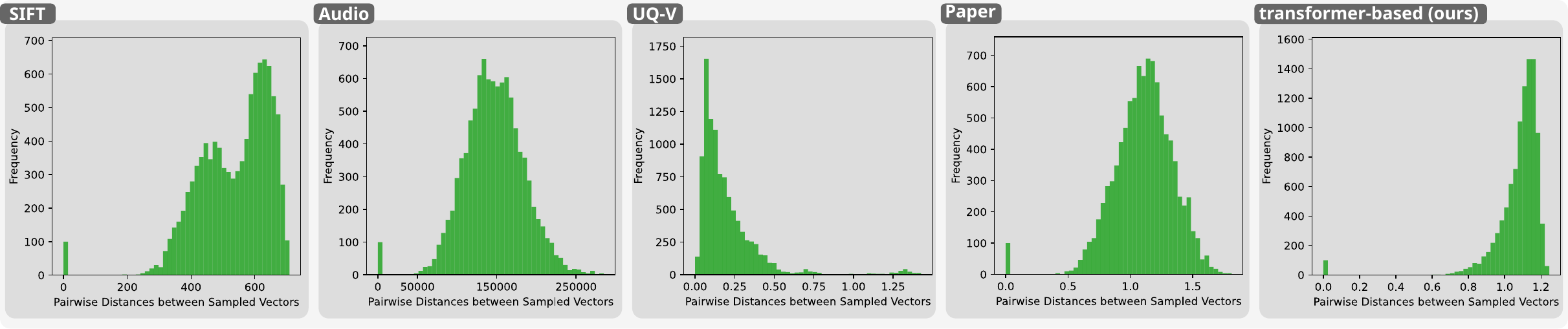}
\vspace{-2.5em}
\caption{\textbf{(\textsection \ref{sec:dataset}) Distribution of pairwise distances among sampled embedding vectors}.}
	\label{fig:ds-distance}
\vspace{-1.0em}
\end{figure*}

%% file: fig/fig_dataset_analysis_correlation.tex
\begin{figure*}[h]
\centering
\captionsetup{justification=centering}
\includegraphics[width=1.0\textwidth]{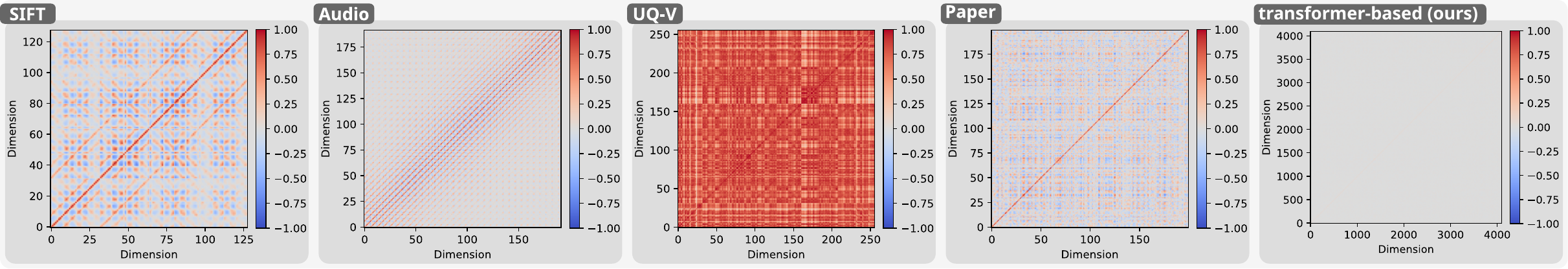}
\vspace{-2.5em}
\caption{\textbf{(\textsection \ref{sec:dataset}) Correlation between different dimensions among sampled embedding vectors}.}
\label{fig:ds-correlation}
\vspace{-1.0em}
\end{figure*}

%% file: fig/fig_bench_recall_vs_qps_baselines.tex
\begin{figure*}[h]
\centering
\captionsetup{justification=centering}
\includegraphics[width=1.0\textwidth]{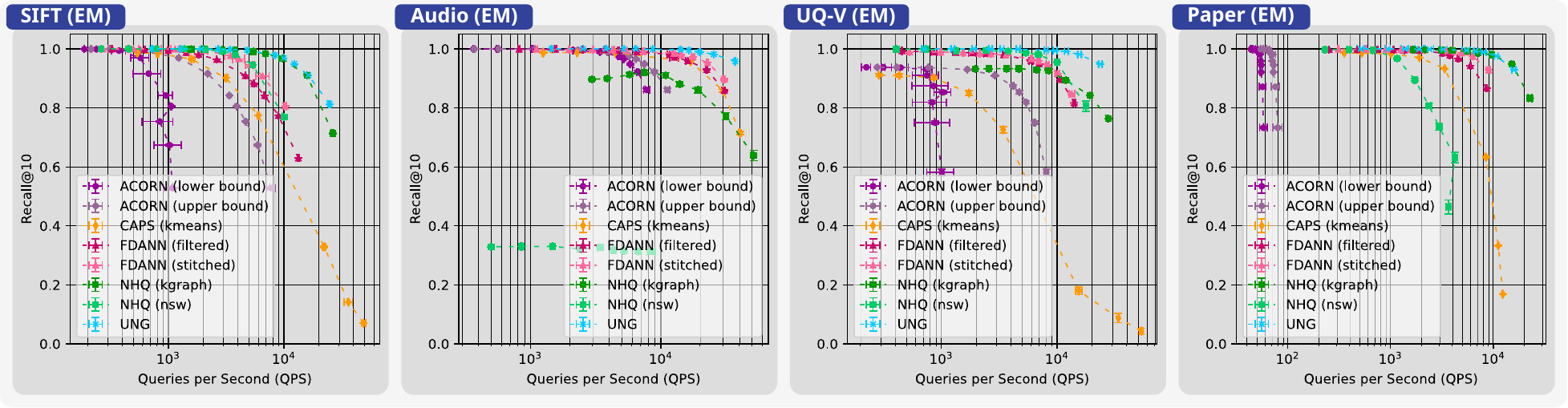}
\vspace{-2.5em}
\caption{\textbf{(\textsection \ref{sec:dataset}) Recall@10 vs. QPS plots} for EM filtering on four widely used datasets of different modalities.}
\label{fig:bench-recall-baselines}
\vspace{-1.5em}
\end{figure*}

%% file: tab/tab_arxiv_dataset.tex
\begin{table}[b]
\footnotesize
\setlength{\tabcolsep}{10pt}
\centering
\captionsetup{justification=centering}
\caption{\textbf{(\textsection \ref{sec:dataset}) Attributes in the novel arXiv dataset}. 
}
\vspace{-1.5em}
\begin{tabular}{lll}
\toprule
\textbf{\makecell[l]{Name}}
&\textbf{\makecell[l]{Attribute Type}}
&\textbf{\makecell[l]{Data type}}
\\
\midrule
submitter	 					& unordered attribute 	& string			\\
has\_comment	 				& unordered attribute 	& boolean			\\
number\_of\_main\_categories 	& ordered attribute	 	& integer			\\
main\_categories 				& set attribute	 		& list of strings 	\\
number\_of\_sub\_categories		& ordered attribute 	& integer			\\
sub\_categories	 				& set attribute 		& list of strings	\\
license 		 				& unordered attribute	& string 			\\
number\_of\_versions		 	& ordered attribute		& integer 			\\
update\_date		 			& ordered attribute		& integer 			\\
number\_of\_authors 			& ordered attribute		& integer 			\\
authors							& set attribute			& list of strings	\\
\bottomrule
\end{tabular}
\label{tab:arxiv-dataset}
\vspace{-0.5em}
\end{table}

%% file: fig/fig_bench_tti_and_mem_baselines.tex
\begin{figure*}[t]
\centering
\captionsetup{justification=centering}
\includegraphics[width=1.0\textwidth]{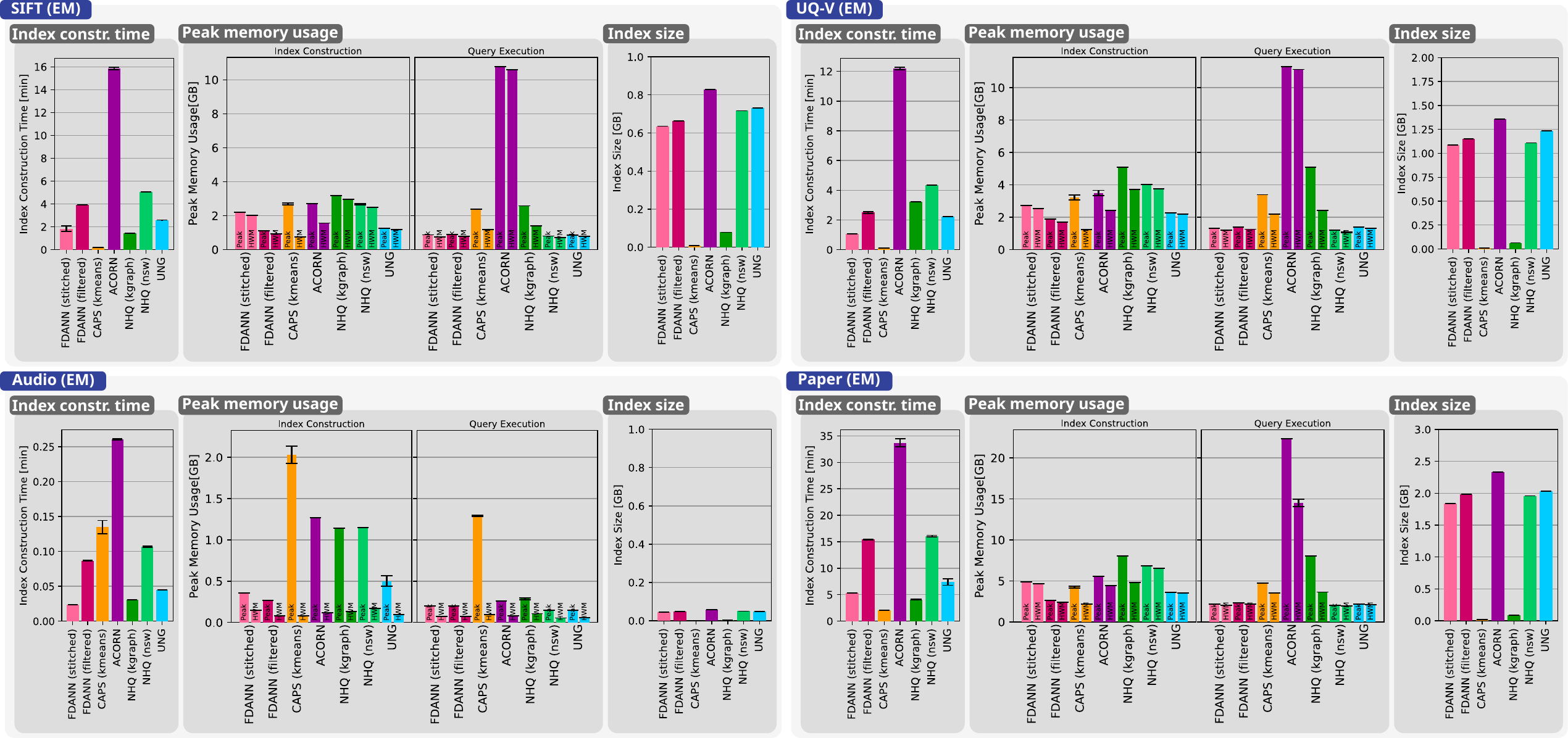}
\vspace{-2.5em}
\caption{\textbf{(\textsection \ref{sec:dataset}) Index construction time, peak memory usage, and index size} for EM filtering on four widely used datasets.}
\label{fig:bench-tti-mem-baselines}
\vspace{-1.0em}
\end{figure*}

%% file: fig/fig_dataset_properties.tex
\begin{figure*}[t]
\centering
\captionsetup{justification=centering}
\includegraphics[width=1.0\textwidth]{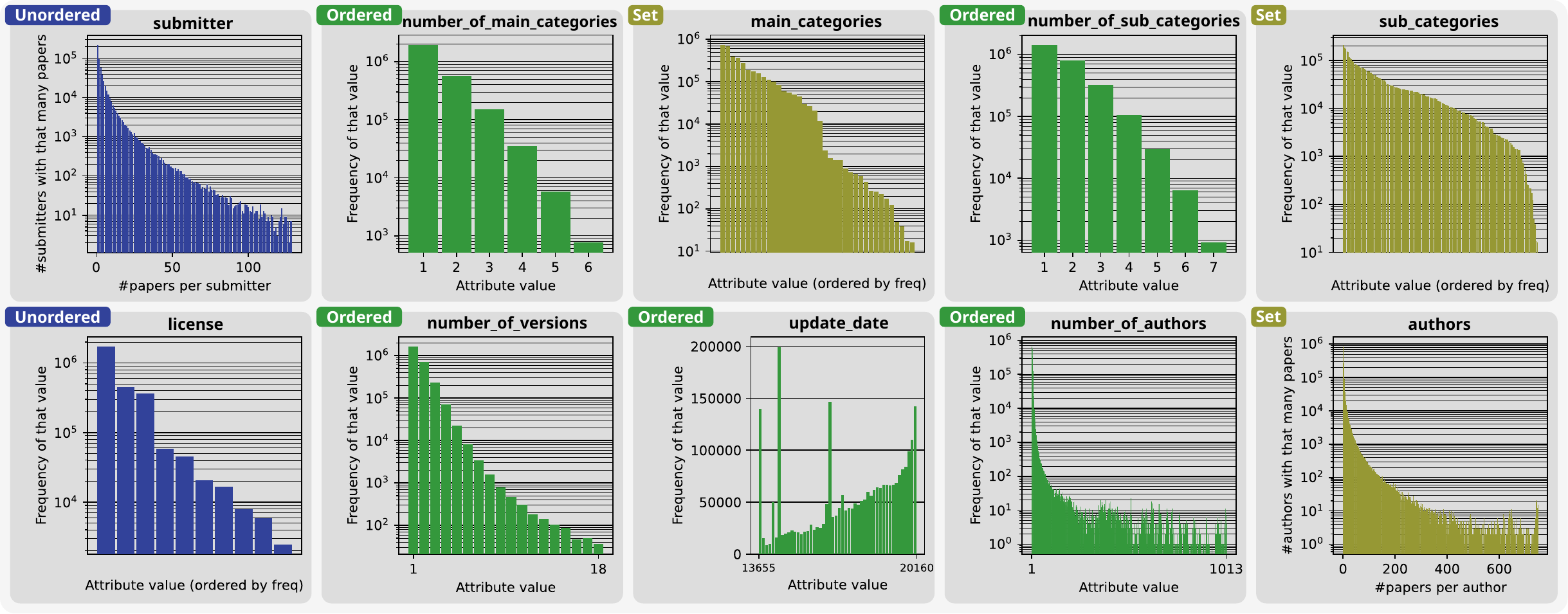}
\vspace{-2em}
\caption{\textbf{(\textsection \ref{sec:dataset}) Distribution of the unordered attributes (blue), ordered attributes (green), and set attributes (yellow) in the \texttt{arxiv-for-fanns-large} dataset.} We show inverse histograms for \textit{authors} and \textit{submitter} since they have many possible values and only few items with a given value. Outliers are omitted for clarity. The \textit{has\_comment} attribute is true for 74.216\% of items.}
\label{fig:dataset-properties}
\vspace{-0.5em}
\end{figure*}

%% file: fig/fig_query_properties.tex
\begin{figure}[H]
\centering
\captionsetup{justification=centering}
\includegraphics[width=1.0\columnwidth]{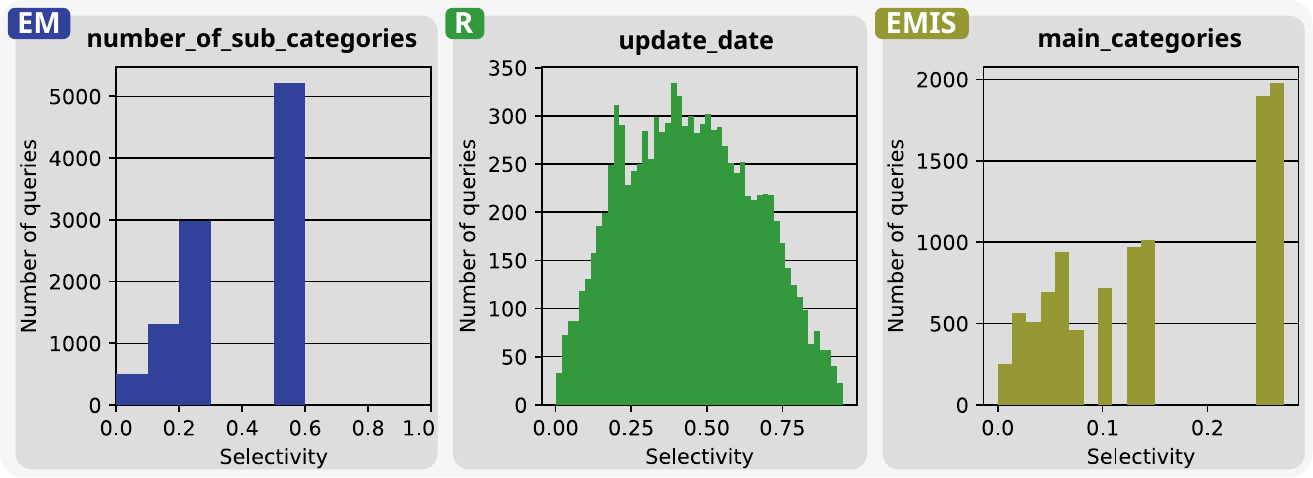}
\vspace{-2.25em}
\caption{\textbf{(\textsection \ref{sec:dataset}) Selectivity of the three query sets.}}
\label{fig:query-properties}
\end{figure}

%% file: 05_benchmark.tex
\section{Benchmarking FANNS}
\label{sec:bench}

We benchmark ACORN \cite{acorn}, CAPS \cite{caps}, two versions of FDANN \cite{fdann} (filtered and stitched), two versions of NHQ \cite{nhq} (nsw and kgraph), SeRF \cite{serf}, iRangeGraph \cite{irg}, DIGRA \cite{digra}, DSG \cite{dsg}, and UNG \cite{ung} on our \texttt{arxiv-for-fanns} dataset.
This selection of methods covers a wide range of filtering approaches, indexing techniques, and filter types.

\subsection{Collected Metrics}
\label{sec:bench-metrics}

To illustrate the trade-off between accuracy and query throughput, we focus on recall vs.~\gls{qps} plots as our primary evaluation metric.
We define recall@k as:

\vspace{-0.5em}
\begin{equation*}
\text{recall@}k = \frac{|\text{knn}_{\text{alg}} \cap \text{knn}_{\text{gt}}|}{k}
\end{equation*}

\noindent where $\text{knn}_{\text{alg}}$ denotes the set of $k$ nearest neighbors returned by the algorithm, and $\text{knn}_{\text{gt}}$ is the ground truth.
In addition, we report the index construction time, peak memory usage during both index construction and query execution, and the index size.

\subsection{Parameter Search}
\label{sec:bench-param}

Since all \gls{fanns} methods considered involve tunable parameters that influence index construction, we perform a dedicated parameter search for each combination of method, dataset, and filter type prior to benchmarking.
Given that some methods expose up to eight parameters, a full grid search is computationally infeasible and we therefore rely on a greedy search strategy.
For some combinations of method and dataset, even the greedy search takes multiple weeks to complete.
\Cref{alg:parameter-search} presents the pseudocode of our parameter search algorithm.

\input{alg/alg_parameter_search}

This algorithm takes as input a list of tunable parameters, their corresponding candidate values, and the index of each parameter's default value inside the list of candidate values.
Candidate and default values are selected based on the literature and publicly available implementations of each method.
The \texttt{get\_reward()} function performs two iterations of building the index with a given parameter configuration and querying it using $50$ randomly sampled queries from the dataset.
It computes the recall vs.~\gls{qps} curve averaged over both iterations and returns the highest QPS achieved at a recall of at least 0.95.
If a method does not reach a recall of 0.95, it returns the highest achieved recall instead.

To validate our parameter search strategy, we perform a full grid search for the ACORN algorithm (only 3 parameters) on the \texttt{Audio} dataset, which is much smaller than \texttt{arxiv-for-fanns}.
Both search strategies found parameters $M=24$ and $\beta=32$.
The greedy search settled for $\gamma = 15$ while the grid search found $\gamma = 12$.
\Cref{fig:parameter-search-evaluation} confirms that both parameter configurations yield nearly identical recall vs.~\gls{qps} curves, and the different parameters are likely caused by randomness in index construction and query execution stages.

\input{fig/fig_parameter_search_evaluation}

The parameters identified by our search strategy for the \texttt{SIFT}, \texttt{Audio}, \texttt{UQ-V}, and \texttt{Paper} datasets used in \Cref{sec:dataset} and for the \texttt{arxiv-for-fanns-medium} and \texttt{arxiv-for-fanns-large} datasets (see \Cref{sec:bench-medium,sec:bench-large}) are reported in \Cref{tab:bench-parameters}.
For the study varying the parameter $k$ (see \Cref{sec:bench-vary-k}), we use the parameters obtained for $k=10$, as the choice of $k$ does not affect index construction. 
For the study on varying query selectivity (see \Cref{sec:bench-selectivity}), we perform a separate parameter search for each selectivity value. 
The resulting parameters are available in our repository\footnote{\codeurl}.

\subsection{Benchmarking Methodology}
\label{sec:bench-method}

Once the parameter search is complete, we benchmark each method five times and report the mean and standard deviation for all measured metrics following scientific benchmarking practice \cite{bench-rules}.
Index construction is performed using all available hardware threads on the respective system (see \Cref{sec:bench-sw-hw}), while query execution is restricted to a single thread, following common practice~\cite{ung,caps,nhq}.
Unless stated otherwise, we search for $k=10$ nearest neighbors and report recall@10, which is a standard choice in the literature~\cite{nhq,hqann,fdann,acorn,ung}.
All benchmarks use Euclidean distance, as it is supported by all methods under evaluation.
\ifarx Since our transformer-based embeddings are normalized, \gls{knn} under Euclidean distance is equivalent to \gls{knn} under cosine distance. \fi
We exclude any preprocessing, such as sorting items by attribute value or transforming vectors into alternate formats when reporting index construction time to avoid making any assumptions on the format of the input data.

\input{tab/tab_bench_parameters}

\subsection{Software and Hardware Configuration}
\label{sec:bench-sw-hw}

We encapsulate our benchmarking infrastructure in a Docker container based on Ubuntu 20.04.6.
We use Python 3.8.10, \texttt{gcc} 9.4.0, and \texttt{g++} 9.4.0.
Benchmarks on the medium-scale dataset are run on an Intel Core i7-1165G7 CPU with 4 physical cores, 8 hardware threads, and 16 GB RAM, running Arch Linux with kernel version 6.15.5-arch1-1 and those on the large dataset run on a machine with 384 GB of RAM and two Intel Xeon Gold 6154 CPUs, each with 18 physical cores and 36 hardware threads, running CentOS Linux 8.

\input{fig/fig_bench_recall_vs_qps_medium} 			
\input{fig/fig_bench_tti_and_mem_medium}			

\ifarx
\subsection{Remarks on Algorithms}
\label{sec:bench-algos}

\paragraph{NHQ-kgraph}
We exclude the call to \texttt{optimize\_graph()} from query execution timing, as it is independent of the number of queries and its cost is amortized when processing large batches.

\paragraph{ACORN}
Algorithmically, ACORN only needs to check the filter condition for visited vertices during graph traversal.
However, the implementation we benchmark performs this check for all items prior to traversal.
Since prior work is inconsistent on whether this cost is included \cite{ung} or not \cite{acorn}, we report two recall vs. \gls{qps} curves for ACORN: one including the filtering overhead (labeled “upper bound”) and one excluding it (labeled “lower bound”).

\paragraph{SeRF, DIGRA}
Unlike most other methods, SeRF and DIGRA perform index construction and query execution within a single process, without persisting the index to disk.
Therefore, we report only the overall peak memory usage, and we omit the index size from our results.

\paragraph{FDANN}  
The code provided in the FDANN repository~\cite{fdann-code} stores two versions of the index, each approximately the size of the original dataset. We optimize the code to store only one index at a time, reducing the overall index size to half.

\paragraph{DSG, DIGRA}
Unlike most of the other methods, DSG and DIGRA do not parallelize index construction. 
Therefore, we report index construction time for a single thread in our benchmarking on \texttt{arxiv-for-fanns-medium} and we omit DSG and DIGRA from our benchmarking on \texttt{arxiv-for-fanns-large}.

\fi

\vspace{-0.5em}
\subsection{Results on \texttt{arxiv-for-fanns-medium}}
\label{sec:bench-medium}

\Cref{fig:bench-recall-medium} shows the recall vs.~\gls{qps} curves for the three filter types on the \texttt{arxiv-for-fanns-medium} dataset.
We observe that ACORN, the only evaluated method applicable to all three filter types, is generally outperformed by more specialized methods.
FDANN, which lagged slightly behind UNG on the established datasets (see \Cref{fig:bench-recall-baselines}), matches UNG's performance on our transformer-based embeddings with \gls{em} filters and even outperforms UNG with \gls{emis} filters.
NHQ (kgraph), a top performer on only two of the four established datasets, is among the best methods on our dataset.

\vspace{0.25em}
\gray{\textbf{Observation \rthree}:~} Specialized \gls{fanns} methods that only support specific filters outperform more versatile methods by up to $10\times$.
\vspace{0.25em}

\Cref{fig:bench-tti-medium} summarizes index construction time, peak memory usage, and index size on the \texttt{arxiv-for-fanns-medium} dataset.
Index construction time varies by more than an order of magnitude across methods, with DIGRA and DSG being the slowest since they are the only methods that do not parallelize index construction.
Except for NHQ (kgraph) and SeRF, which show noticeably higher peak memory usage, differences in memory consumption during index construction and query execution are moderate across methods.
The index built by most methods is approximately 1.6\,GB, corresponding to the size of the dataset.
Some methods produce very small indexes but require access to the original database vectors during query execution, resulting in the same effective index size.

\vspace{0.25em}
\gray{\textbf{Observation \rfour}:~} Index construction times vary by up to $16\times$ across methods but do not correlate with query performance.
\vspace{0.25em}

\input{fig/fig_bench_recall_vs_qps_large}

\input{fig/fig_bench_tti_and_mem_large}

\subsection{Results on \texttt{arxiv-for-fanns-large}}
\label{sec:bench-large}

\subsubsection{Challenges In Scaling Up the Dataset Size}
\label{sec:bench-large-challenges}

While $2.7$ million vectors may not sound particularly large, note that our transformer-based embeddings have $4{,}096$ dimensions, making our dataset over $85\times$ larger than the well-known SIFT1M dataset~\cite{sift-ds} which uses 128-dimensional embedding vectors.
We had to modify the source code of four out of the eleven methods we benchmarked to fix \texttt{int32} overflows during memory allocation (UNG~\cite{ung}, CAPS~\cite{caps}, and NHQ (KGraph)~\cite{nhq}) or to reduce the index size (FDANN (stitched) ~\cite{fdann}) to fit in the local persistent storage.
Because index construction for DSG and DIGRA is non-parallelized, takes over a day per run, and requires dozens of runs for parameter tuning, we could not benchmark these two methods on the large dataset.

\subsubsection{Results}
\label{sec:bench-large-results}

\Cref{fig:bench-recall-large} shows the recall vs.~\gls{qps} curves for the three filter types on the \texttt{arxiv-for-fanns-large} dataset.
Comparing these results to those on the medium-scale dataset in \Cref{fig:bench-recall-medium} provides insights into scalability.
ACORN, both NHQ variants, FDANN (filtered), SeRF, and iRangeGraph  maintain their relative performance, with throughput dropping by only a factor of about $4$ despite the dataset being $27\times$ larger.
This indicates good scalability, especially considering that both experiments used a single thread, albeit on different machines.
FDANN (stitched), CAPS, and UNG perform poorly on the large dataset and fail to exceed $25\%$ recall.
We cannot rule out that suboptimal parameter choices contribute to this performance drop as our greedy parameter search may have converged to local optima.
This highlights the difficulty of parameter tuning for \gls{fanns} methods on large datasets.

\vspace{0.25em}
\gray{\textbf{Observation \rfive}:~} Scaling to large datasets can cause recall to drop below $0.2$, rendering some methods useless. Adjustments of the implementation or parameter tuning strategies may be required.
\vspace{-0.5em}

\Cref{fig:bench-tti-large} reports index construction time, peak memory usage, and index size for all methods.
On the $27\times$ larger dataset, construction time increased by roughly an order of magnitude.
Since $9\times$ more threads were used, this suggests super-linear growth with dataset size or sublinear thread speedup.
CAPS remains the fastest to build, with NHQ and FDANN also being relatively fast.
UNG has by far the longest construction time, likely because it failed to reach the $0.95$ recall target and the search shifted toward more expensive configurations.
Memory usage scales roughly linearly with dataset size, and relative differences remain similar to the medium-scale case.
UNG is notable for compressing the index to about $25\%$ of the dataset with CAPS, iRangeGraph, and NHQ (kgraph) producing compact indexes but requiring access to the original vectors.

\vspace{0.25em}
\gray{\textbf{Observation \rsix}:~} Parameter tuning is difficult on large datasets, with indexes construction taking up to 6 hours for 2.7M items.
\vspace{-0.5em}

\input{fig/fig_bench_recall_vs_qps_vary_k}	

\ifarx
\input{fig/fig_bench_recall_vs_qps_vary_k_r}

\input{fig/fig_bench_recall_vs_qps_vary_k_emis}

\fi

\subsection{Analysis for Different Values of k}
\label{sec:bench-vary-k}

In \Cref{fig:bench-recall-vary-k}, we show recall vs.~\gls{qps} plots for \gls{em} filters on the \texttt{arxiv-for-fanns-medium} dataset with $k \in \{20,40,60,80,100\}$.
\ifarx
\Cref{fig:bench-recall-vary-k-r,fig:bench-recall-vary-k-emis} present the same study for \gls{r} and \gls{emis} filters, respectively. Index construction time, memory usage, and index size show almost no variation with $k$ and are therefore omitted.
\fi

\vspace{0.25em}
\gray{\textbf{Observation \rseven}:~} The relative performance of different \gls{fanns} methods remains consistent across different values of $k$.
\vspace{-0.5em}

\input{fig/fig_bench_tti_and_mem_selectivity_em}

\input{fig/fig_bench_recall_vs_qps_selectivity}

\input{fig/fig_bench_tti_and_mem_selectivity_r}

\input{fig/fig_bench_recall_vs_qps_selectivity_r}

\input{fig/fig_bench_tti_and_mem_selectivity_emis}

\input{fig/fig_bench_recall_vs_qps_selectivity_emis}

\vspace{-10em}
\subsection{Analysis for Different Query Selectivities}
\label{sec:bench-selectivity}

To explore the impact of query selectivity on \gls{fanns} performance, we benchmark a version of our \texttt{arxiv-for-fanns-medium} dataset where we generate synthetic attributes in order to control the query selectivity.
\Cref{fig:bench-recall-selectivity,fig:bench-recall-selectivity-r,fig:bench-recall-selectivity-emis} show recall vs.~\gls{qps} plots for \gls{em}, \gls{r}, and \gls{emis} filters with selectivities of $0.2$, $0.4$, $0.6$, and $0.8$. \Cref{fig:bench-tti-mem-selectivity-em,fig:bench-tti-mem-selectivity-r,fig:bench-tti-mem-selectivity-emis} report index construction time, peak memory usage, and index size for the same experiments.
We observe that NHQ (nsw) with EM filters exhibits substantial difficulty with low selectivity scenarios, while UNG with EMIS filters handles low selectivity scenarios slightly better than high selectivity ones.
There are also some variations in index construction time, with ACORN being slower to build for selectivity close to $0.5$ than for more extreme selectivities, and DIGRA being faster to build for lower selectivities.
Memory usage and index size are unaffected by selectivity.

\vspace{0.25em}
\gray{\textbf{Observation \reight}:~} The performance and resource usage of most \gls{fanns} methods is largely consistent across query selectivities; however, some methods exhibit significant performance degradation in low-selectivity scenarios.
\vspace{-0.5em}

%% file: alg/alg_parameter_search.tex
\vspace{-0.75em}
\begin{algorithm}[H]
\small
\caption{\textbf{(\textsection \ref{sec:bench-param}) Greedy Parameter Search}}
\label{alg:parameter-search}
\begin{algorithmic}[1]
\Require params, value\_lists, default\_indices
\State indices = default\_indices
\State P = \{p : value\_lists[p][indices[p]] for p in params\}
\State best\_reward = get\_reward(P)
\State do\_repeat = true
\While{do\_repeat}
	\State do\_repeat = false
	\ForAll{$\text{par} \in \text{params}$}
		\ForAll{$\text{change} \in [-1,1]$}
			\State P = \{p : value\_lists[p][indices[p]] for p in params\}
			\State P[par] = value\_lists[par][indices[par] + change]
			\State reward = get\_reward(P)
			\If{reward > best\_reward}
				\If{reward > (best\_reward * 1.01)}
					\State do\_repeat = true
				\EndIf
				\State best\_reward = reward
				\State indices[par] = indices[par] + change
			\EndIf
		\EndFor
	\EndFor
\EndWhile
\State 
\Return \{p : value\_lists[p][indices[p]] for p in params\}
\end{algorithmic}
\end{algorithm}
\vspace{-0.75em}

%% file: fig/fig_parameter_search_evaluation.tex
\begin{figure}[H]
\vspace{-1.0em}
\centering
\captionsetup{justification=centering}
\includegraphics[width=1.0\columnwidth]{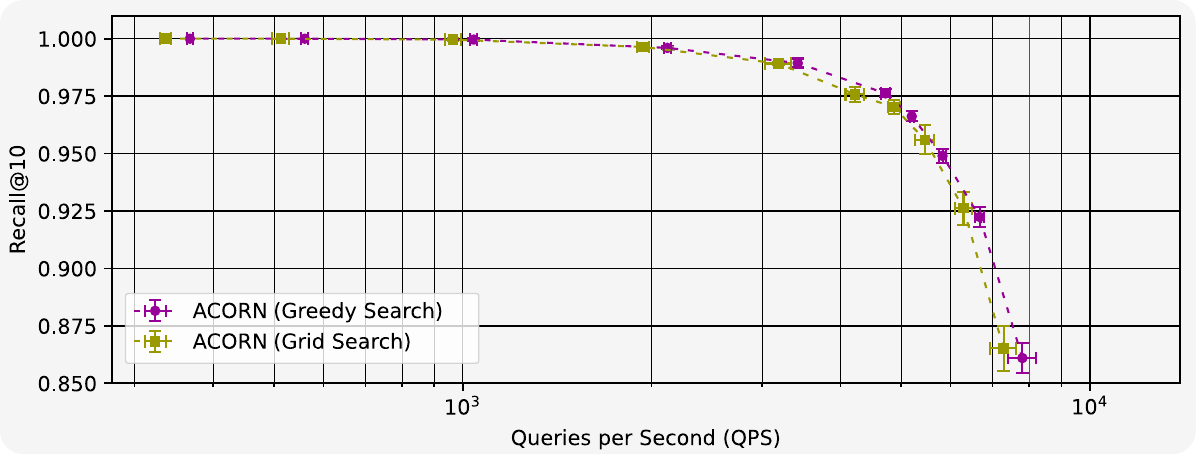}
\vspace{-2.5em}
\caption{\textbf{(\textsection \ref{sec:fanns-approaches}) Performance of ACORN on the \texttt{Audio} dataset with parameters from grid and greedy search.}}
\label{fig:parameter-search-evaluation}
\vspace{-1.0em}
\end{figure}

%% file: tab/tab_bench_parameters.tex
\balance
\begin{table*}[h]
\footnotesize
\setlength{\tabcolsep}{10pt}
\centering
\captionsetup{justification=centering}
\caption{\textbf{(\textsection \ref{sec:bench-param}) Parameters used for benchmarking} (found through parameter search, see \Cref{sec:bench-param}).}
\vspace{-1.5em}
\begin{tabular}{lll}
\specialrule{0.8pt}{0pt}{0pt}
\textbf{\makecell[l]{Method}}
&\textbf{\makecell[l]{Experiment}}
&\textbf{\makecell[l]{Parameters}}
\\
\specialrule{0.8pt}{0pt}{0pt}
\multirow{11}{*}{ACORN \cite{acorn}}	& all									& $efs \in \{10,15,20,25,30,50,100,250,500,750\}$											\\
										& \texttt{SIFT}, EM filter				& $M=32,M_\beta=16,\gamma=12$																\\
										& \texttt{Audio}, EM filter				& $M=24,M_\beta=32,\gamma=12$																\\
										& \texttt{UQ-V}, EM filter				& $M=24,M_\beta=32,\gamma=12$																\\
										& \texttt{Paper}, EM filter				& $M=32,M_\beta=24,\gamma=12$																\\
										& \texttt{arxiv-medium}, EM filter		& $M=16,M_\beta=24,\gamma=10$																\\
										& \texttt{arxiv-medium}, R filter		& $M=32,M_\beta=24,\gamma=12$																\\
										& \texttt{arxiv-medium}, EMIS filter	& $M=16,M_\beta=24,\gamma=15$																\\
										& \texttt{arxiv-large}, EM filter		& $M=32,M_\beta=32,\gamma=10$																\\
										& \texttt{arxiv-large}, R filter		& $M=32,M_\beta=16,\gamma=12$																\\
										& \texttt{arxiv-large}, EMIS filter		& $M=48,M_\beta=48,\gamma=15$																\\

\specialrule{0.2pt}{0pt}{0pt}
\multirow{7}{*}{CAPS (kmeans) \cite{caps}}& all									& $m \in \{200,400,1k,5k,10k,20k,40k,60k\}$													\\
										& \texttt{SIFT}, EM filter				& $B = 256$																					\\
										& \texttt{Audio}, EM filter				& $B = 256$																					\\
										& \texttt{UQ-V}, EM filter				& $B = 128$																					\\
										& \texttt{Paper}, EM filter				& $B = 1024$																				\\
										& \texttt{arxiv-medium}, EM filter		& $B = 128$																					\\
										& \texttt{arxiv-large}, EM filter		& $B = 512$																					\\
\specialrule{0.2pt}{0pt}{0pt}
\multirow{9}{*}{FDANN (stitched) \cite{fdann}} & all							& $L \in \{10,20,30,50,100,150,200,300,500,1k\}$											\\
										& \texttt{SIFT}, EM filter				& $R_\text{small}=32,L_\text{small}=80,R_\text{stitched}=48,\alpha=1.3$						\\
										& \texttt{Audio}, EM filter				& $R_\text{small}=16,L_\text{small}=80,R_\text{stitched}=96,\alpha=1.2$						\\
										& \texttt{UQ-V}, EM filter				& $R_\text{small}=32,L_\text{small}=80,R_\text{stitched}=64,\alpha=1.1$						\\
										& \texttt{Paper}, EM filter				& $R_\text{small}=32,L_\text{small}=100,R_\text{stitched}=128,\alpha=1.3$					\\
										& \texttt{arxiv-medium}, EM filter		& $R_\text{small}=32,L_\text{small}=80,R_\text{stitched}=48,\alpha=1.1$						\\
										& \texttt{arxiv-medium}, EMIS filter	& $R_\text{small}=32,L_\text{small}=100,R_\text{stitched}=48,\alpha=1.2$					\\
										& \texttt{arxiv-large}, EM filter		& $R_\text{small}=96,L_\text{small}=100,R_\text{stitched}=64,\alpha=1.4$					\\
										& \texttt{arxiv-large}, EMIS filter		& $R_\text{small}=64,L_\text{small}=80,R_\text{stitched}=64,\alpha=1.2$						\\
\specialrule{0.2pt}{0pt}{0pt}
\multirow{9}{*}{FDANN (filtered) \cite{fdann}} & all							& $L_\text{search} \in \{10,20,30,50,100,150,200,300,500,1k\}$								\\
										& \texttt{SIFT}, EM filter				& $R=48,L=100,\alpha=1.0$																	\\
										& \texttt{Audio}, EM filter				& $R=32,L=100,\alpha=1.2$																	\\
										& \texttt{UQ-V}, EM filter				& $R=96,L=60,\alpha=1.0$																	\\
										& \texttt{Paper}, EM filter				& $R=48,L=100,\alpha=1.3$																	\\
										& \texttt{arxiv-medium}, EM filter		& $R=48,L=100, \alpha=1.0$																	\\
										& \texttt{arxiv-medium}, EMIS filter	& $R=48,L=80,\alpha=1.2$																	\\
										& \texttt{arxiv-large}, EM filter		& $R=48,L=200,\alpha=1.2$																	\\
										& \texttt{arxiv-large}, EMIS filter		& $R=48,L=150,\alpha=1.2$																	\\
\specialrule{0.2pt}{0pt}{0pt}
\multirow{7}{*}{NHQ (kgraph) \cite{nhq}}& all									& $L_\text{search} \in \{10,25,50,75,100,150,200,300,400\}, \text{iter}=12, M=1.0$			\\
										& \texttt{SIFT}, EM filter 				& $K=100,L=100,S=15,R=300,\text{RANGE}=50,PL=400,B=0.4,\text{weight}=1M$ \hspace{2em}		\\
										& \texttt{Audio}, EM filter 			& $K=40,L=40,S=4,R=300,\text{RANGE}=70,PL=400,B=0.6,\text{weight}=10B$ \hspace{2em}			\\
										& \texttt{UQ-V}, EM filter 				& $K=80,L=80,S=10,R=200,\text{RANGE}=60,PL=300,B=0.5,\text{weight}=1M$ \hspace{2em}			\\
										& \texttt{Paper}, EM filter 			& $K=60,L=40,S=15,R=400,\text{RANGE}=20,PL=200,B=0.3,\text{weight}=100M$ \hspace{2em}		\\
										& \texttt{arxiv-medium}, EM filter 		& $K=100,L=80,S=15,R=200,\text{RANGE}=70,PL=200,B=0.5,\text{weight}=1M$ \hspace{2em}		\\
										& \texttt{arxiv-large}, EM filter 		& $K=80,L=60,S=10,R=200,\text{RANGE}=60,PL=300,B=0.6,\text{weight}=1M$ \hspace{2em}			\\
\specialrule{0.2pt}{0pt}{0pt}
\multirow{7}{*}{NHQ (nsw) \cite{nhq}}	& all									& $\text{ef\_search} \in \{50,100,150,200,300,500,1k,2k,4k\}$ 								\\
										& \texttt{SIFT}, EM filter 				& $M=50, \text{MaxM0}=50,\text{efConstruction}=200,\text{weight\_search}=1M$				\\
										& \texttt{Audio}, EM filter 			& $M=40, \text{MaxM0}=50,\text{efConstruction}=300,\text{weight\_search}=1M$				\\
										& \texttt{UQ-V}, EM filter 				& $M=20, \text{MaxM0}=40,\text{efConstruction}=300,\text{weight\_search}=1M$				\\
										& \texttt{Paper}, EM filter 			& $M=40, \text{MaxM0}=50,\text{efConstruction}=300,\text{weight\_search}=1M$				\\
										& \texttt{arxiv-medium}, EM filter 		& $M=40, \text{MaxM0}=40,\text{efConstruction}=300,\text{weight\_search}=1M$				\\
										& \texttt{arxiv-large}, EM filter 		& $M=40, \text{MaxM0}=60,\text{efConstruction}=150,\text{weight\_search}=1M$				\\
\specialrule{0.2pt}{0pt}{0pt}
\multirow{9}{*}{UNG \cite{ung}}			& all									& $L_\text{search} \in \{10,20,30,40,50,100,150,200,300,500,1k\}$ 							\\
										& \texttt{SIFT}, EM filter				& $\delta=6,R=32,L=100,\alpha=1.4,\sigma=12$												\\
										& \texttt{Audio}, EM filter				& $\delta=6,R=48,L=100,\alpha=1.4,\sigma=16$												\\
										& \texttt{UQ-V}, EM filter				& $\delta=6,R=48,L=100,\alpha=1.2,\sigma=16$												\\
										& \texttt{Paper}, EM filter				& $\delta=6,R=32,L=100,\alpha=1.4,\sigma=16$												\\
										& \texttt{arxiv-medium}, EM filter		& $\delta=2,R=24,L=80,\alpha=1.4,\sigma=16$													\\
										& \texttt{arxiv-medium}, EMIS filter	& $\delta=8,R=32,L=100,\alpha=1.4,\sigma=16$												\\
										& \texttt{arxiv-large}, EM filter		& $\delta=6,R=48,L=150,\alpha=1.6,\sigma=24$												\\
										& \texttt{arxiv-large}, EMIS filter		& $\delta=6,R=48,L=150,\alpha=1.4,\sigma=24$												\\
\specialrule{0.2pt}{0pt}{0pt}
\multirow{3}{*}{SeRF \cite{serf}}		& all									& $\text{ef\_search} \in \{4,8,16,32,64,128,256,512,1024\}$									\\
										& \texttt{arxiv-medium}, R filter		& $\text{index\_k}=100,\text{ef\_construction}=100,\text{ef\_max}=500$						\\
										& \texttt{arxiv-large}, R filter		& $\text{index\_k}=100,\text{ef\_construction}=100,\text{ef\_max}=600$						\\
\specialrule{0.2pt}{0pt}{0pt}
\multirow{3}{*}{iRangeGraph \cite{irg}}	& all									& $\text{ef\_search} \in \{1,2,3,4,5,6,8,10,15,20,30,50,100,200\}$							\\
										& \texttt{arxiv-medium}, R filter		& $M=32,\text{ef\_construction}=200$														\\
										& \texttt{arxiv-large}, R filter		& $M=24,\text{ef\_construction}=100$														\\
\specialrule{0.2pt}{0pt}{0pt}
\multirow{1}{*}{DIGRA \cite{digra}}		& \texttt{arxiv-medium}, R filter		& $\text{ef\_search} \in \{1,2,3,4,5,6,8,10,15,20,30,50,100,200\},M=24,\text{ef\_construction}=200$	\\
										
\specialrule{0.2pt}{0pt}{0pt}
\multirow{1}{*}{DSG \cite{dsg}}			& \texttt{arxiv-medium}, R filter		& $\text{ef\_search} \in \{4,8,16,32,64,128,256,512,1024\},M=24,\text{ef\_construction}=200,\text{ef\_max}=500,\alpha=1.0$	\\

\specialrule{0.8pt}{0pt}{0pt}
\end{tabular}
\vspace{-1.25em}
\label{tab:bench-parameters}
\end{table*}

%% file: fig/fig_bench_recall_vs_qps_medium.tex
\begin{figure*}[t]
\centering
\captionsetup{justification=centering}
\includegraphics[width=1.0\textwidth]{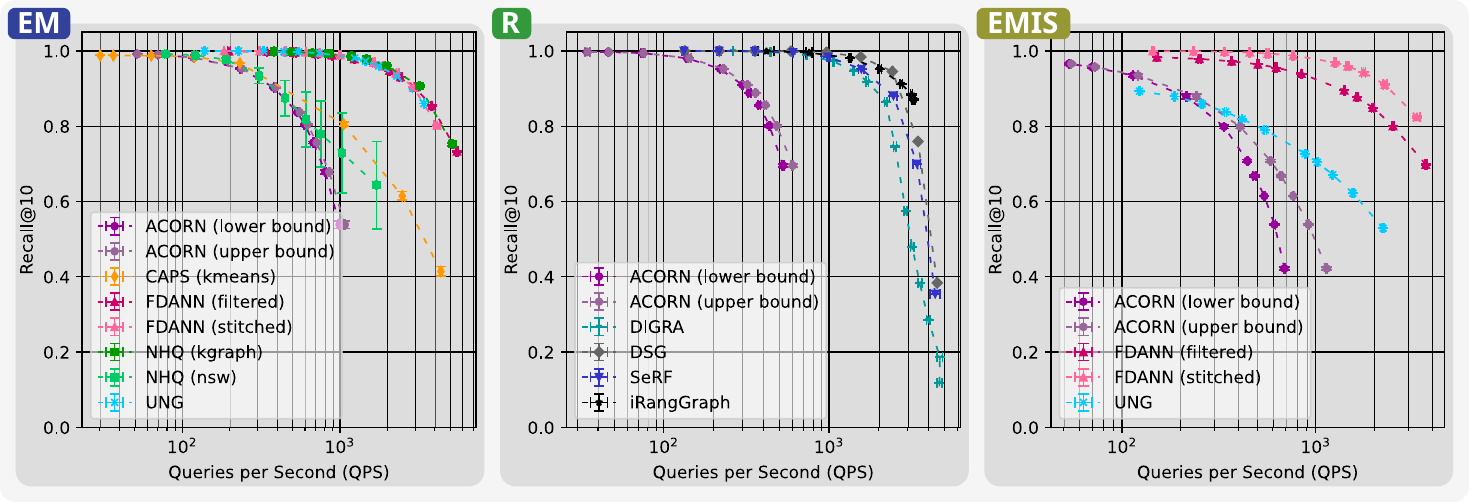}
\vspace{-2.5em}
\caption{\textbf{(\textsection \ref{sec:bench-medium}) Recall@10 vs. QPS plots} for the three filter types (EM, R, EMIS) on the \texttt{arxiv-for-fanns-medium} dataset.}
\label{fig:bench-recall-medium}
\end{figure*}

%% file: fig/fig_bench_tti_and_mem_medium.tex
\begin{figure*}[t]
\vspace{-1.0em}
\centering
\captionsetup{justification=centering}
\includegraphics[width=1.0\textwidth]{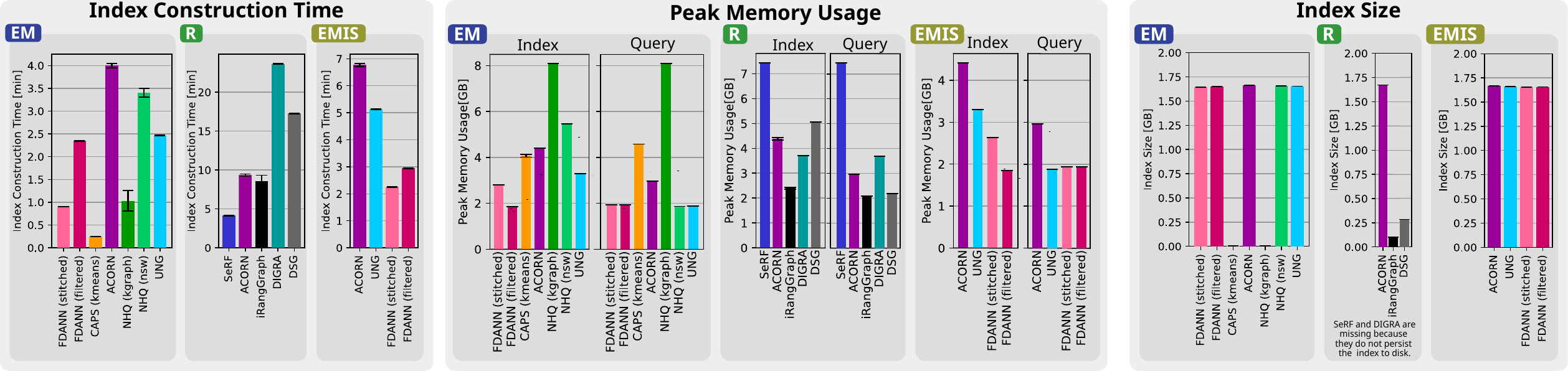}
\vspace{-2.5em}
	\caption{\textbf{(\textsection \ref{sec:bench-medium}) Index construction time, peak memory usage, and index size} on the \texttt{arxiv-for-fanns-medium} dataset.}
\label{fig:bench-tti-medium}
\vspace{-1.5em}
\end{figure*}

%% file: fig/fig_bench_recall_vs_qps_large.tex
\begin{figure*}[t]
\centering
\captionsetup{justification=centering}
\includegraphics[width=1.0\textwidth]{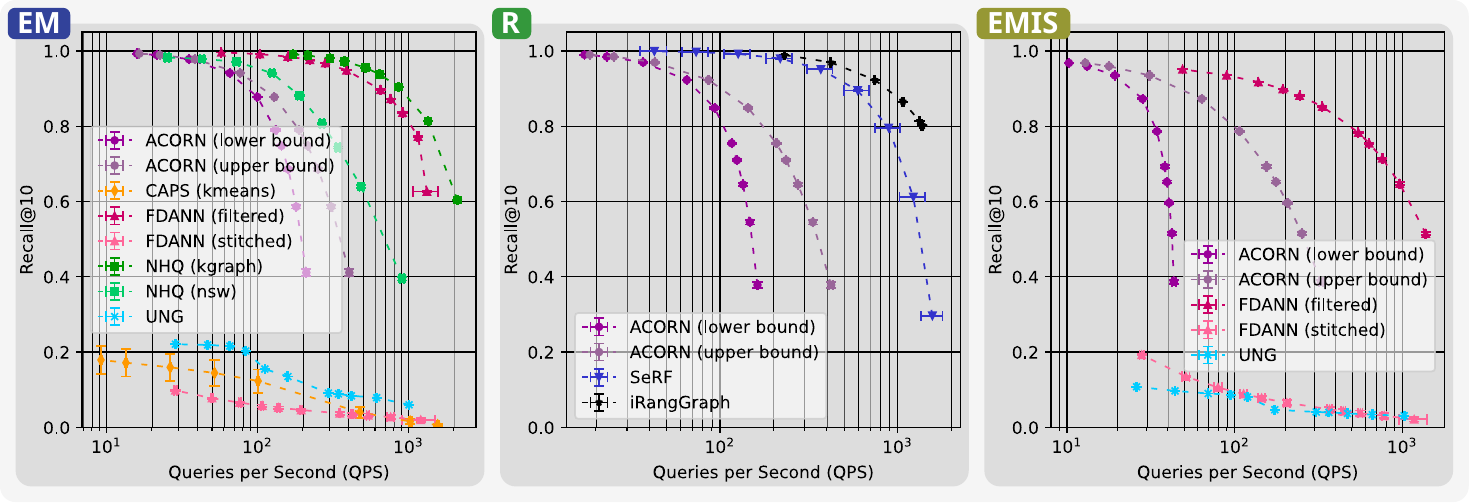}
\vspace{-2.5em}
\caption{\textbf{(\textsection \ref{sec:bench-large}) Recall@10 vs. QPS plots} for the three filter types (EM, R, EMIS) on the \texttt{arxiv-for-fanns-large} dataset.}
\label{fig:bench-recall-large}
\vspace{-0.5em}
\end{figure*}

%% file: fig/fig_bench_tti_and_mem_large.tex
\begin{figure*}[t]
\vspace{-0.5em}
\centering
\captionsetup{justification=centering}
\includegraphics[width=1.0\textwidth]{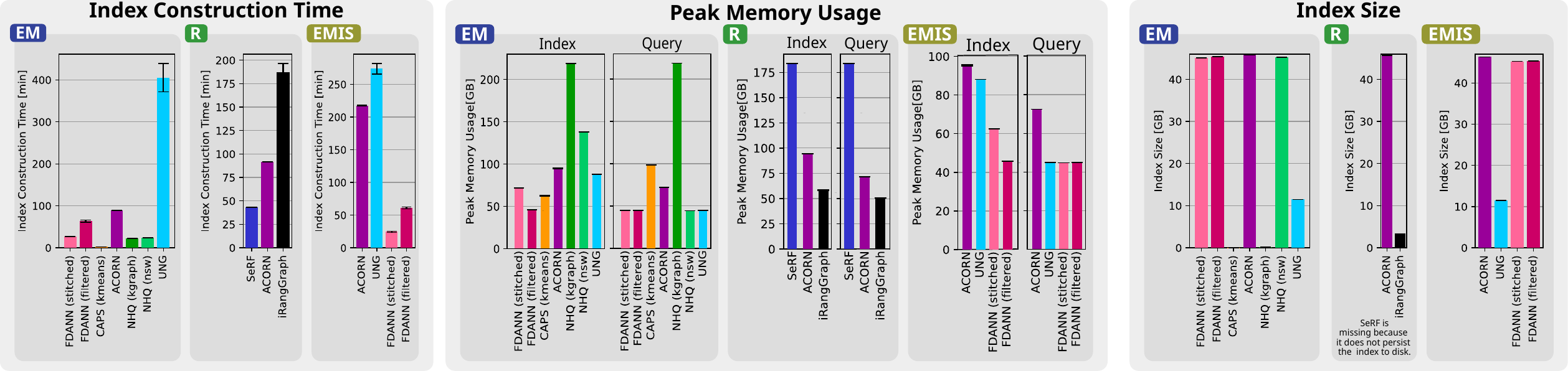}
\vspace{-2.5em}
\caption{\textbf{(\textsection \ref{sec:bench-large}) Index construction time, peak memory usage, and index size} on the \texttt{arxiv-for-fanns-large} dataset.}
\label{fig:bench-tti-large}
\vspace{-1.5em}
\end{figure*}

%% file: fig/fig_bench_recall_vs_qps_vary_k.tex
\begin{figure*}[t]
\centering
\captionsetup{justification=centering}
\includegraphics[width=0.98\textwidth]{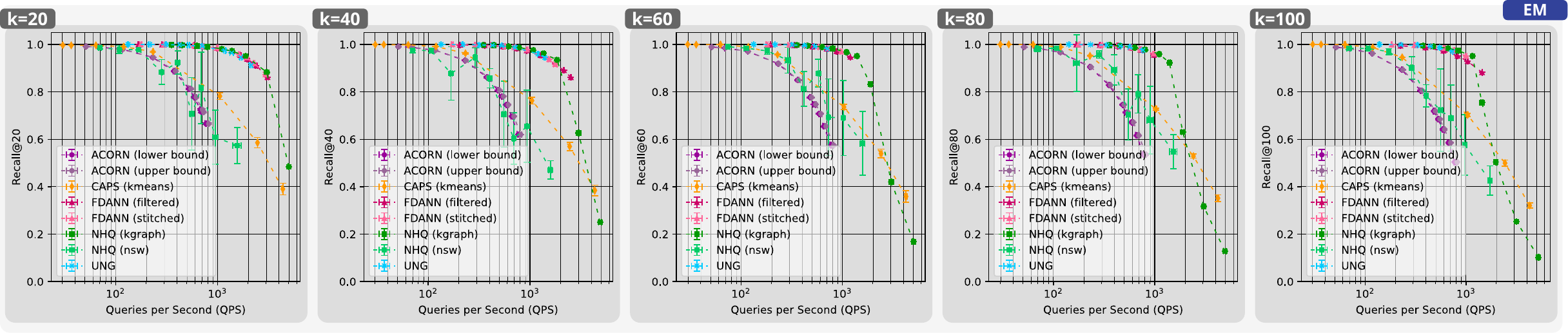}
\vspace{-1.25em}
\caption{\textbf{(\textsection \ref{sec:bench-vary-k}) Recall@k vs. QPS plots} for varying values of $k$ on the \texttt{arxiv-for-fanns-medium} dataset with EM filtering.}
\label{fig:bench-recall-vary-k}
\vspace{-0.25em}
\end{figure*}

%% file: fig/fig_bench_recall_vs_qps_vary_k_r.tex
\begin{figure*}[h]
\vspace{-1em}
\centering
\captionsetup{justification=centering}
\includegraphics[width=0.98\textwidth]{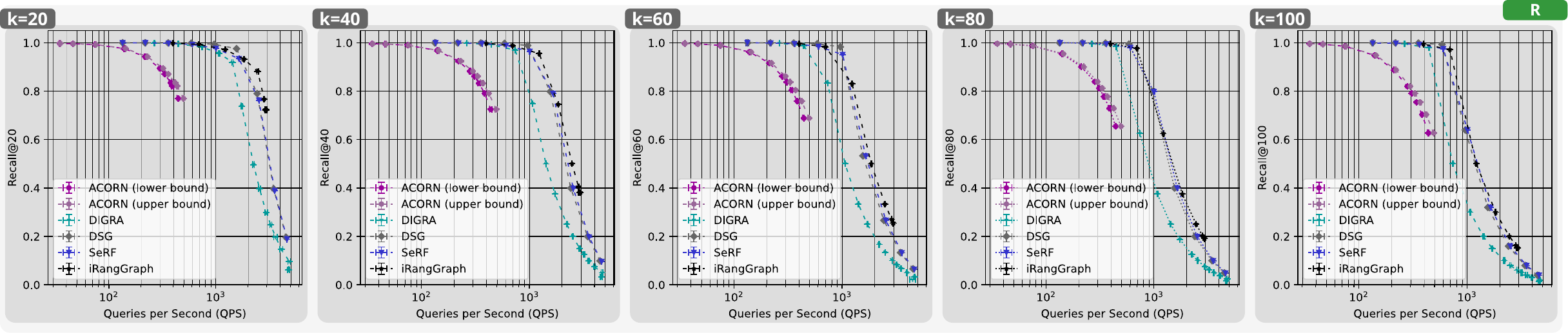}
\vspace{-1.25em}
\caption{\textbf{(\textsection \ref{sec:bench-vary-k}) Recall@k vs. QPS plots} for varying values of $k$ on the \texttt{arxiv-for-fanns-medium} dataset with R filtering.}
\label{fig:bench-recall-vary-k-r}
\vspace{-0.25em}
\end{figure*}

%% file: fig/fig_bench_recall_vs_qps_vary_k_emis.tex
\begin{figure*}[h]
\vspace{-1em}
\centering
\captionsetup{justification=centering}
\includegraphics[width=0.98\textwidth]{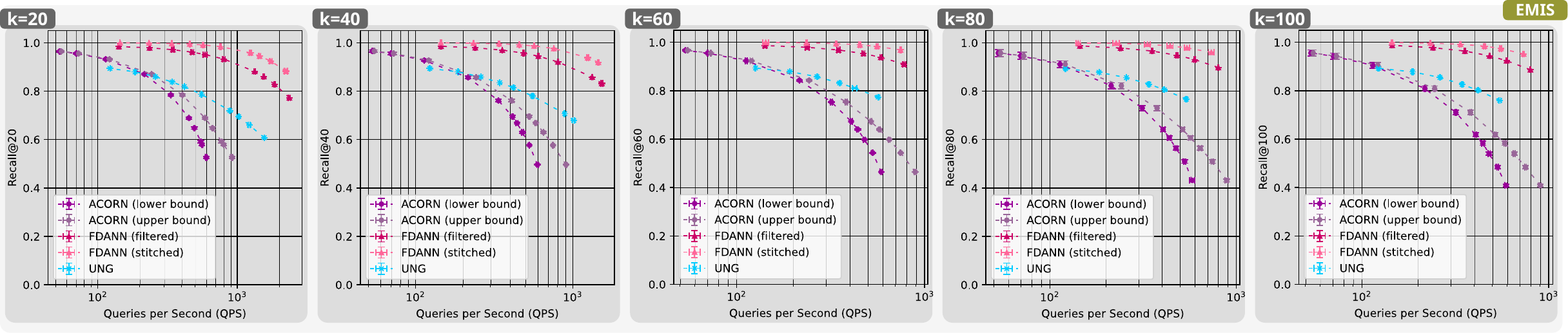}
\vspace{-1.25em}
\caption{\textbf{(\textsection \ref{sec:bench-vary-k}) Recall@k vs. QPS plots} for varying values of $k$ on the \texttt{arxiv-for-fanns-medium} dataset with EMIS filtering.}
\label{fig:bench-recall-vary-k-emis}
\vspace{-1.0em}
\end{figure*}

%% file: fig/fig_bench_tti_and_mem_selectivity_em.tex
\begin{figure*}[t]
\centering
\captionsetup{justification=centering}
\includegraphics[width=0.98\textwidth]{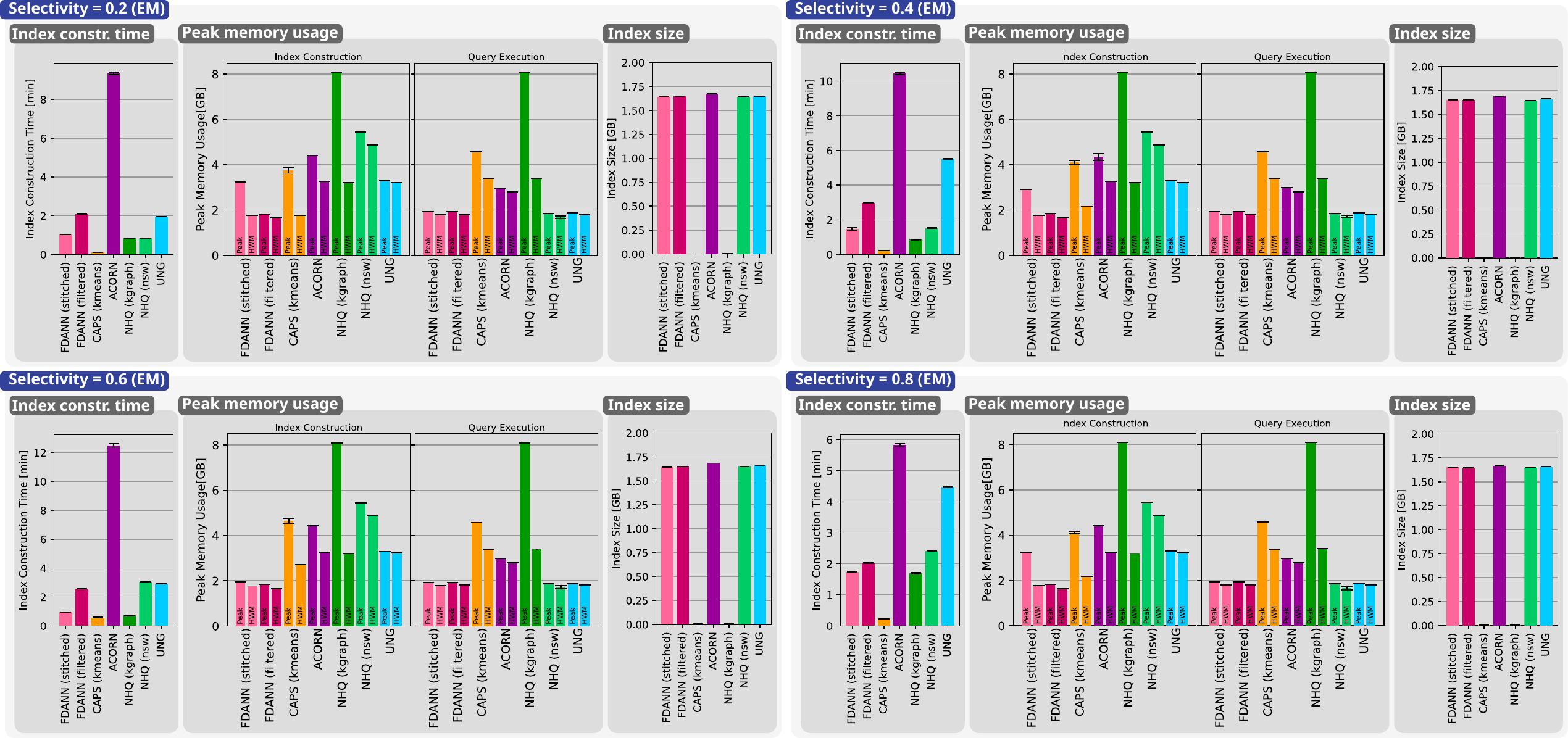}
\vspace{-1.0em}
\caption{\textbf{(\textsection \ref{sec:bench-selectivity}) Index construction time, peak memory usage, and index size} for EM filtering with varying selectivity.}
\label{fig:bench-tti-mem-selectivity-em}
\end{figure*}

%% file: fig/fig_bench_recall_vs_qps_selectivity.tex
\begin{figure*}[t]
\vspace{-0.5em}
\centering
\captionsetup{justification=centering}
\includegraphics[width=1.0\textwidth]{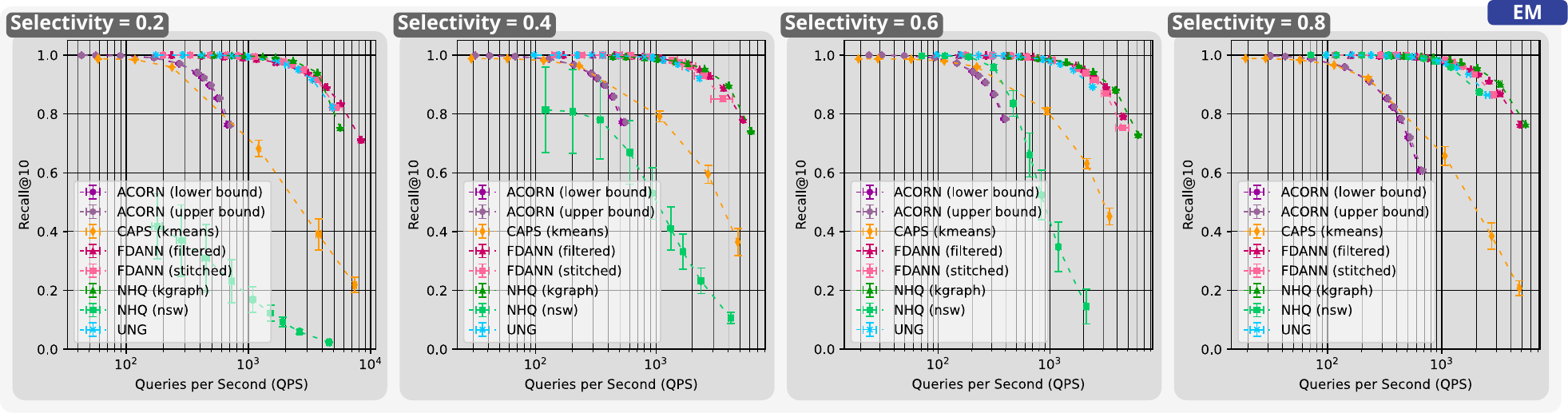}
\vspace{-2.5em}
\caption{\textbf{(\textsection \ref{sec:bench-selectivity}) Recall@10 vs. QPS plots} for EM filtering on \texttt{arxiv-for-fanns-medium} with different query selectivities.}
\label{fig:bench-recall-selectivity}
\end{figure*}

%% file: fig/fig_bench_tti_and_mem_selectivity_r.tex
\begin{figure*}[t]
\centering
\captionsetup{justification=centering}
\includegraphics[width=1.0\textwidth]{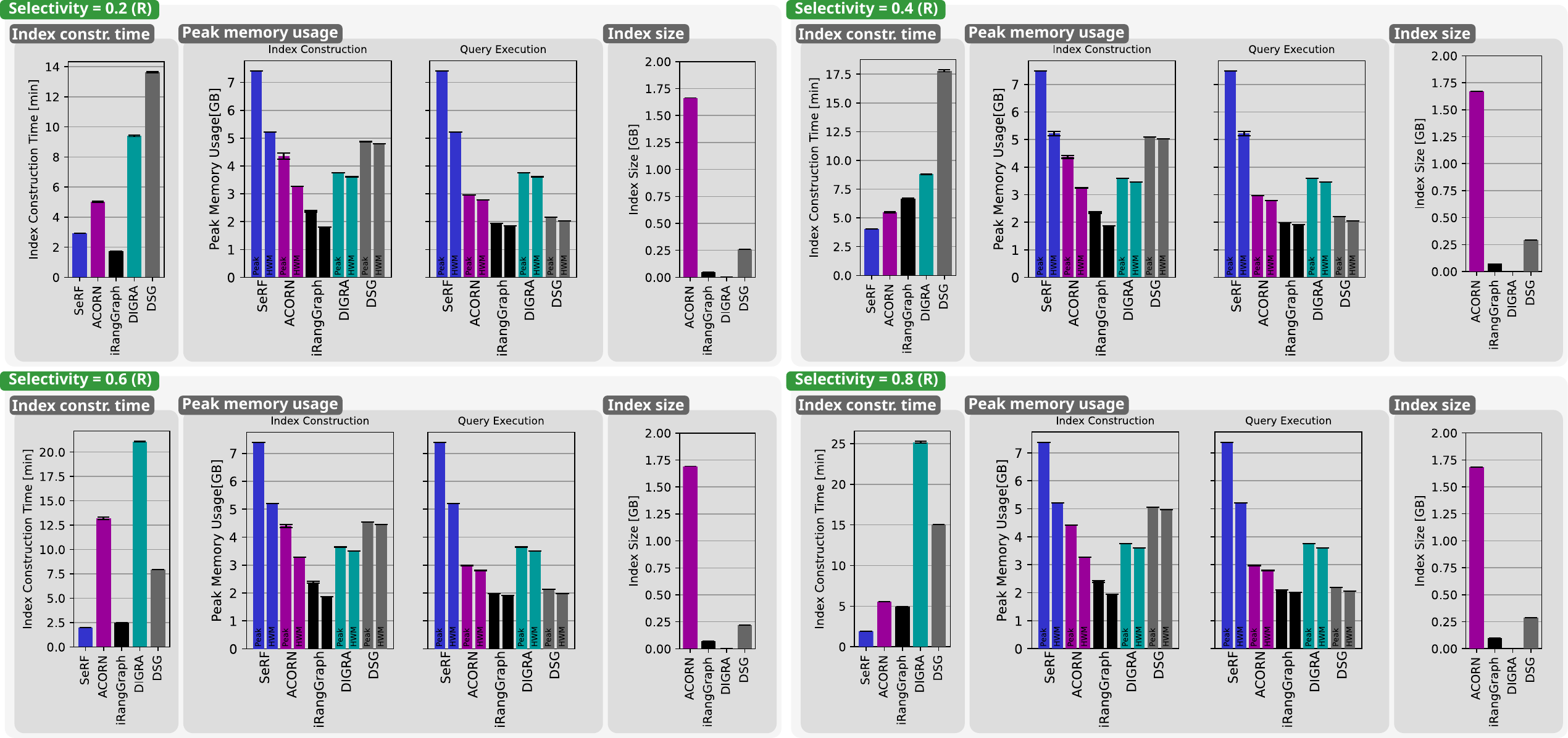}
\vspace{-2.5em}
\caption{\textbf{(\textsection \ref{sec:bench-selectivity}) Index construction time, peak memory usage, and index size} for R filtering with varying selectivity.}
\label{fig:bench-tti-mem-selectivity-r}
\end{figure*}

%% file: fig/fig_bench_recall_vs_qps_selectivity_r.tex
\begin{figure*}[t]
\centering
\captionsetup{justification=centering}
\includegraphics[width=1.0\textwidth]{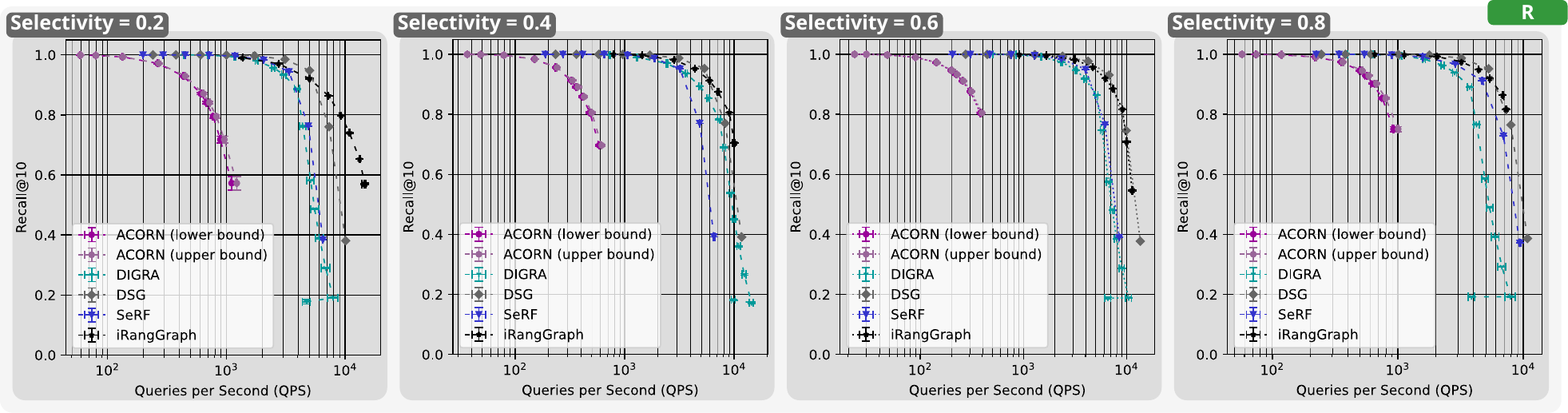}
\vspace{-2.5em}
\caption{\textbf{(\textsection \ref{sec:bench-selectivity}) Recall@10 vs. QPS plots} for R filtering on \texttt{arxiv-for-fanns-medium} with different query selectivities.}
\label{fig:bench-recall-selectivity-r}
\end{figure*}

%% file: fig/fig_bench_tti_and_mem_selectivity_emis.tex
\begin{figure*}[t]
\centering
\captionsetup{justification=centering}
\includegraphics[width=1.0\textwidth]{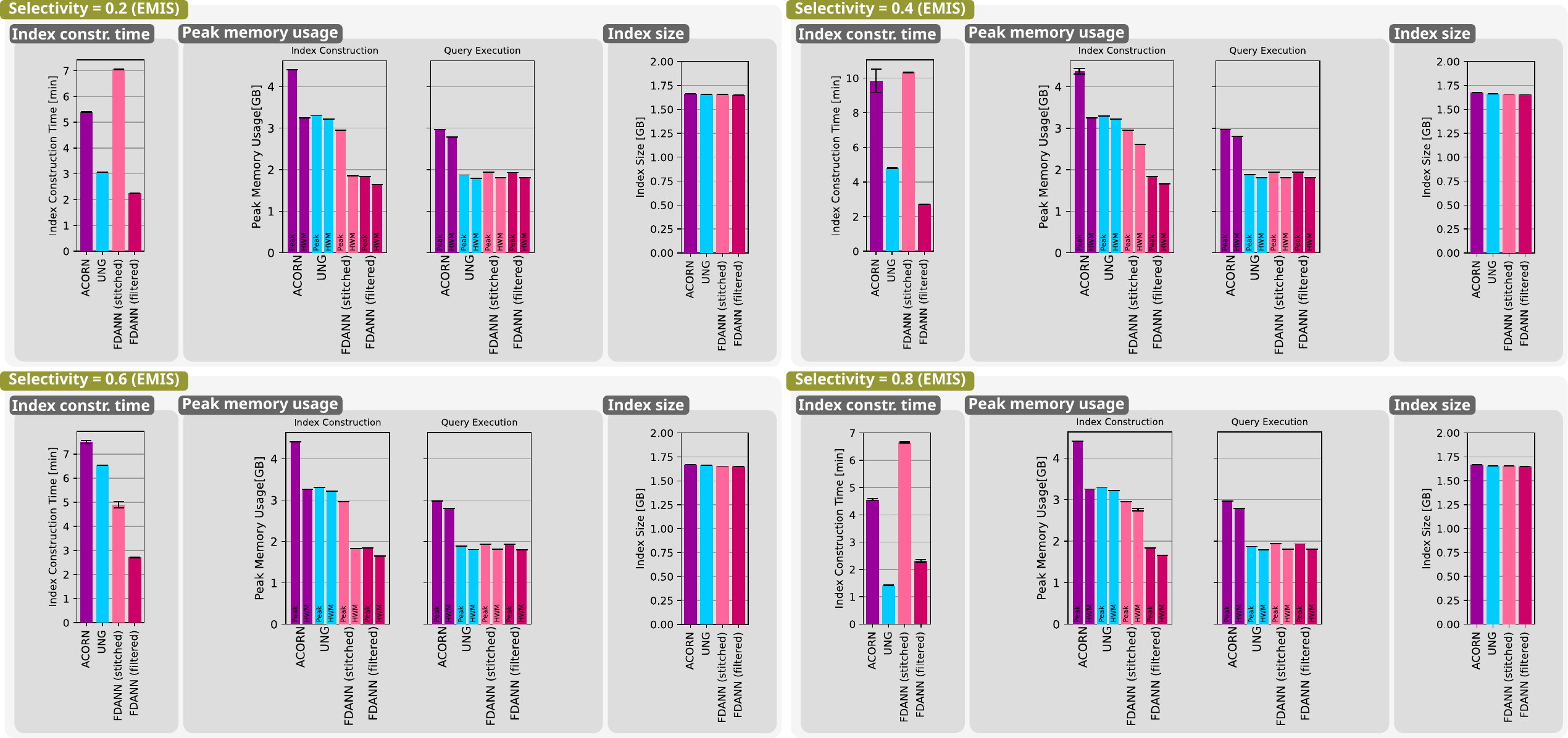}
\vspace{-2.5em}
\caption{\textbf{(\textsection \ref{sec:bench-selectivity}) Index construction time, peak memory usage, and index size} for EMIS filtering with varying selectivity.}
\label{fig:bench-tti-mem-selectivity-emis}
\vspace{-1.0em}
\end{figure*}

%% file: fig/fig_bench_recall_vs_qps_selectivity_emis.tex
\begin{figure*}[t]
\centering
\captionsetup{justification=centering}
\includegraphics[width=1.0\textwidth]{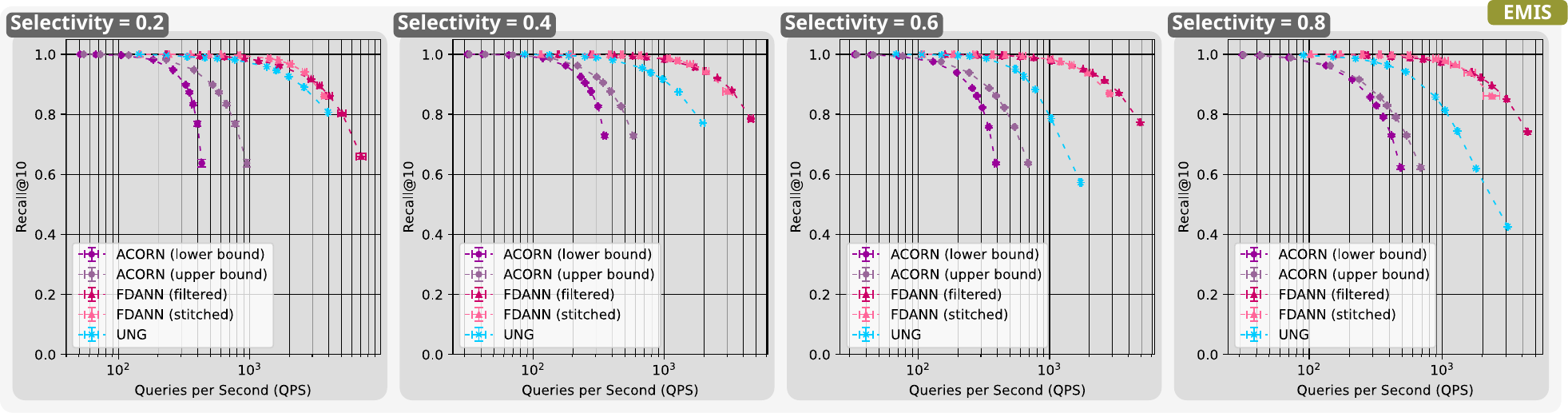}
\vspace{-2.5em}
\caption{\textbf{(\textsection \ref{sec:bench-selectivity}) Recall@10 vs. QPS plots} for EMIS filtering on \texttt{arxiv-for-fanns-medium} with different query selectivities.}
\label{fig:bench-recall-selectivity-emis}
\end{figure*}

%% file: 06_related_work.tex
\section{Related Work}
\label{sec:rw}

Interest in \gls{fanns} is growing, with new methods introduced each year, an ACM SIGMOD programming contest~\cite{sigmod-contest} dedicated to \gls{fanns}, and survey and benchmark preprints appearing around the same time as ours.  
Lin et al.~\cite{rl-fanns-survey} classify methods in a pruning-oriented framework, analyzing the combination of vector pruning (i.e., \gls{anns}) and scalar pruning (i.e., attribute filtering), while Shi et al.~\cite{rl-fanns-benchmark} benchmark selected methods on existing datasets, though, to the best of our knowledge, not on transformer-based embeddings.  
Our work complements these efforts by introducing a taxonomy along three dimensions: filtering approach, indexing technique, and filter types, and by contributing a transformer-based dataset together with benchmarks on both established and new datasets.  

Such transformer-based datasets that reflect emerging \gls{rag} workloads are currently missing in the \gls{anns} and \gls{fanns} literature.  
We generate our embedding vectors with the \texttt{stella\_en\_400M\_v5} model~\cite{stella-hf,stella}, widely adopted in \gls{llm} research~\cite{checkembed,rl_01}.  
Unlike our \texttt{arxiv-for-fanns} dataset, end-to-end \gls{rag} benchmarks such as the Massive Text Embedding Benchmark \cite{mteb} contain text passages rather than vectors, making them unsuitable for direct evaluation of \gls{fanns} methods.  

In the broader \gls{anns} field, surveys by Echihabi et al.~\cite{anns-survey-echihabi} and Han et al.~\cite{anns-survey-han} provide comprehensive overviews, and large-scale benchmarks~\cite{ann-benchmark,ann-benchmark-web,anns-graph-2} compare methods across datasets.  
To the best of our knowledge, these do not cover transformer-based embeddings; since our dataset also contains the ground truth for unfiltered \gls{anns}, it could be used to extend these prior works.  

%% file: 07_conclusion.tex
\section{Conclusion}
\label{sec:conclusion}

The rapid progress of embedding models for text, image, audio, and video has created strong demand for fast and accurate \gls{fanns} methods.  
Recent work has proposed diverse approaches supporting \gls{em}, \gls{r}, and \gls{emis} filtering.  
To structure this evolving field, we present a taxonomy classifying methods by filtering approach, indexing technique, and filter type, and we survey \gls{fanns} methods accordingly.  
By analyzing how \gls{fanns} methods are evaluated in literature, we identify a key limitation: the lack of open datasets with transformer-based embeddings and real-world attributes.  
Analyzing embeddings from images, audio, video, and text, we find that the embedding model strongly shapes vector characteristics and hence \gls{fanns} performance, which motivates the need for transformer-based datasets.

To address this gap, we release the \texttt{arxiv-for-fanns} dataset, which contains over 2.7 million 4096-dimensional vectors with 11 attributes and reflects realistic transformer-based workloads.
Using this dataset, we perform an in-depth benchmarking study of \gls{fanns} methods, analyzing their performance across different filter types, dataset scales, numbers of retrieved neighbors, and query selectivities.
By distilling our results into eight key observations, we provide practical guidance for selecting and configuring \gls{fanns} methods, and by publishing our dataset, we establish a benchmark that helps steer future research towards more efficient \gls{fanns} methods for state-of-the-art, transformer-based embedding vectors.

%% file: 08_acknowledgements.tex
\section{Acknowledgements}
\label{sec:acks}

We thank the Swiss National Supercomputing Centre (CSCS) for access to their Ault system, which we used to execute our benchmarks.
We gratefully acknowledge Polish high-performance computing infrastructure PLGrid (HPC Center: ACK Cyfronet AGH) for providing computer facilities and support within computational grants no. PLG/2024/017103 and PLG/2025/018259.
This work was supported by the ETH Future Computing Laboratory (EFCL), financed by a donation from Huawei Technologies.
It also received funding from the European Research Council \raisebox{-0.25em}{\includegraphics[height=1em]{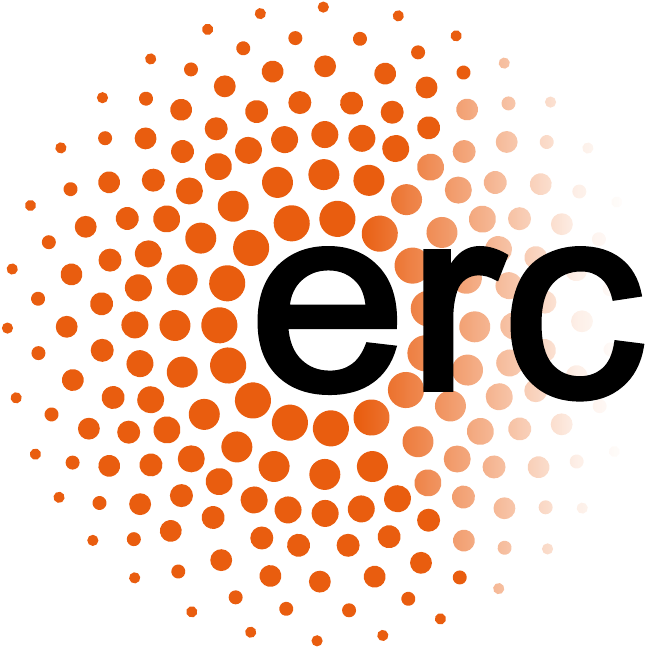}} (Project PSAP, No.~101002047) and from the European Union's HE research and innovation programme under the grant agreement No.~101070141 (Project GLACIATION).

%% file: paper.bbl

\begin{thebibliography}{164}


\ifx \showCODEN    \undefined \def \showCODEN     #1{\unskip}     \fi
\ifx \showISBNx    \undefined \def \showISBNx     #1{\unskip}     \fi
\ifx \showISBNxiii \undefined \def \showISBNxiii  #1{\unskip}     \fi
\ifx \showISSN     \undefined \def \showISSN      #1{\unskip}     \fi
\ifx \showLCCN     \undefined \def \showLCCN      #1{\unskip}     \fi
\ifx \shownote     \undefined \def \shownote      #1{#1}          \fi
\ifx \showarticletitle \undefined \def \showarticletitle #1{#1}   \fi
\ifx \showURL      \undefined \def \showURL       {\relax}        \fi
\providecommand\bibfield[2]{#2}
\providecommand\bibinfo[2]{#2}
\providecommand\natexlab[1]{#1}
\providecommand\showeprint[2][]{arXiv:#2}

\bibitem[tfa(2025)]%
        {tfanns-code}
 \bibinfo{year}{2025}\natexlab{}.
\newblock \bibinfo{title}{{Tag-Filtered ANNS}}.
\newblock
  \bibinfo{howpublished}{\url{https://github.com/SpaceIshtar/FilterGraph}}.
\newblock
\newblock
\shownote{Accessed: 2025-12-02}.


\bibitem[Abel(1984)]%
        {bptree}
\bibfield{author}{\bibinfo{person}{David~J Abel}.}
  \bibinfo{year}{1984}\natexlab{}.
\newblock \showarticletitle{A B+-tree structure for large quadtrees}.
\newblock \bibinfo{journal}{\emph{Computer Vision, Graphics, and Image
  Processing}} \bibinfo{volume}{27}, \bibinfo{number}{1}
  (\bibinfo{year}{1984}), \bibinfo{pages}{19--31}.
\newblock


\bibitem[Abu-El-Haija et~al\mbox{.}(2025)]%
        {youtube8m-ds}
\bibfield{author}{\bibinfo{person}{Sami Abu-El-Haija}, \bibinfo{person}{Anja
  Hauth}, \bibinfo{person}{Lu Jiang}, \bibinfo{person}{Nisarg Kothari},
  \bibinfo{person}{Joonseok Lee}, \bibinfo{person}{Hanhan Li},
  \bibinfo{person}{Paul Natsev}, \bibinfo{person}{Joe Ng},
  \bibinfo{person}{Sobhan~Naderi Parizi}, \bibinfo{person}{George Toderici},
  \bibinfo{person}{Balakrishnan Varadarajan}, \bibinfo{person}{Sudheendra
  Vijayanarasimhan}, {and} \bibinfo{person}{Shoou-I Yu}.}
  \bibinfo{year}{2025}\natexlab{}.
\newblock \bibinfo{title}{{YouTube 8M}}.
\newblock
  \bibinfo{howpublished}{\url{https://research.google.com/youtube8m/download.html}}.
\newblock
\newblock
\shownote{Accessed: 2025-03-06}.


\bibitem[Alibaba-NLP(2025)]%
        {gte-hf}
\bibfield{author}{\bibinfo{person}{Alibaba-NLP}.}
  \bibinfo{year}{2025}\natexlab{}.
\newblock \bibinfo{title}{{gte-Qwen2-7B-instruct}}.
\newblock
  \bibinfo{howpublished}{\url{https://huggingface.co/Alibaba-NLP/gte-Qwen2-7B-instruct}}.
\newblock
\newblock
\shownote{Accessed: 2025-01-22}.


\bibitem[Alipay(2025)]%
        {pase-code}
\bibfield{author}{\bibinfo{person}{Alipay}.} \bibinfo{year}{2025}\natexlab{}.
\newblock \bibinfo{title}{{Pase: PostgreSQL Ultra-High Dimensional Approximate
  Nearest Neighbor Search Extension}}.
\newblock \bibinfo{howpublished}{\url{https://github.com/alipay/PASE}}.
\newblock
\newblock
\shownote{Accessed: 2025-03-04}.


\bibitem[Andoni et~al\mbox{.}(2015)]%
        {anns-hash-1}
\bibfield{author}{\bibinfo{person}{Alexandr Andoni}, \bibinfo{person}{Piotr
  Indyk}, \bibinfo{person}{Thijs Laarhoven}, \bibinfo{person}{Ilya
  Razenshteyn}, {and} \bibinfo{person}{Ludwig Schmidt}.}
  \bibinfo{year}{2015}\natexlab{}.
\newblock \showarticletitle{Practical and optimal LSH for angular distance}.
\newblock \bibinfo{journal}{\emph{Advances in neural information processing
  systems}}  \bibinfo{volume}{28} (\bibinfo{year}{2015}).
\newblock


\bibitem[Andoni and Razenshteyn(2015)]%
        {anns-hash-2}
\bibfield{author}{\bibinfo{person}{Alexandr Andoni} {and} \bibinfo{person}{Ilya
  Razenshteyn}.} \bibinfo{year}{2015}\natexlab{}.
\newblock \showarticletitle{Optimal data-dependent hashing for approximate near
  neighbors}. In \bibinfo{booktitle}{\emph{Proceedings of the forty-seventh
  annual ACM symposium on Theory of computing}}. \bibinfo{pages}{793--801}.
\newblock


\bibitem[Andr{\'e} et~al\mbox{.}(2016)]%
        {anns-quant-3}
\bibfield{author}{\bibinfo{person}{Fabien Andr{\'e}},
  \bibinfo{person}{Anne-Marie Kermarrec}, {and} \bibinfo{person}{Nicolas
  Le~Scouarnec}.} \bibinfo{year}{2016}\natexlab{}.
\newblock \showarticletitle{Cache locality is not enough: High-performance
  nearest neighbor search with product quantization fast scan}. In
  \bibinfo{booktitle}{\emph{42nd International Conference on Very Large Data
  Bases}}, Vol.~\bibinfo{volume}{9}. \bibinfo{pages}{12}.
\newblock


\bibitem[Aoyama et~al\mbox{.}(2011)]%
        {dr-ng}
\bibfield{author}{\bibinfo{person}{Kazuo Aoyama}, \bibinfo{person}{Kazumi
  Saito}, \bibinfo{person}{Hiroshi Sawada}, {and} \bibinfo{person}{Naonori
  Ueda}.} \bibinfo{year}{2011}\natexlab{}.
\newblock \showarticletitle{Fast approximate similarity search based on
  degree-reduced neighborhood graphs}. In \bibinfo{booktitle}{\emph{Proceedings
  of the 17th ACM SIGKDD international conference on Knowledge discovery and
  data mining}}. \bibinfo{pages}{1055--1063}.
\newblock


\bibitem[at~Rutgers~University(2025a)]%
        {arkgraph-code}
\bibfield{author}{\bibinfo{person}{Data Curation~Lab at Rutgers~University}.}
  \bibinfo{year}{2025}\natexlab{a}.
\newblock \bibinfo{title}{{ARKGraph: All-Range Approximate K-Nearest-Neighbor
  Graph}}.
\newblock \bibinfo{howpublished}{\url{https://github.com/rutgers-db/ARKGraph}}.
\newblock
\newblock
\shownote{Accessed: 2025-02-25}.


\bibitem[at~Rutgers~University(2025b)]%
        {dsg-code}
\bibfield{author}{\bibinfo{person}{Data Curation~Lab at Rutgers~University}.}
  \bibinfo{year}{2025}\natexlab{b}.
\newblock \bibinfo{title}{{DynamicSegmentGraph}}.
\newblock
  \bibinfo{howpublished}{\url{https://github.com/rutgers-db/DynamicSegmentGraph/tree/release_version}}.
\newblock
\newblock
\shownote{Accessed: 2025-12-02}.


\bibitem[at~Rutgers~University(2025c)]%
        {serf-code}
\bibfield{author}{\bibinfo{person}{Data Curation~Lab at Rutgers~University}.}
  \bibinfo{year}{2025}\natexlab{c}.
\newblock \bibinfo{title}{{SeRF}}.
\newblock \bibinfo{howpublished}{\url{https://github.com/rutgers-db/SeRF}}.
\newblock
\newblock
\shownote{Accessed: 2025-02-24}.


\bibitem[at~The Chinese University~of Hong~Kong(2025a)]%
        {digra-code}
\bibfield{author}{\bibinfo{person}{Database~Group at~The Chinese University~of
  Hong~Kong}.} \bibinfo{year}{2025}\natexlab{a}.
\newblock \bibinfo{title}{{DIGRA}}.
\newblock \bibinfo{howpublished}{\url{https://github.com/CUHK-DBGroup/DIGRA}}.
\newblock
\newblock
\shownote{Accessed: 2025-12-02}.


\bibitem[at~The Chinese University~of Hong~Kong(2025b)]%
        {rpq-code}
\bibfield{author}{\bibinfo{person}{Database~Group at~The Chinese University~of
  Hong~Kong}.} \bibinfo{year}{2025}\natexlab{b}.
\newblock \bibinfo{title}{{WoW: A Window-to-Window Incremental Index for
  Range-Filtering Approximate Nearest Neighbor Search, SIGMOD 2026 (Round 2)}}.
\newblock
  \bibinfo{howpublished}{\url{https://github.com/CUHK-DBGroup/RangePQ}}.
\newblock
\newblock
\shownote{Accessed: 2025-12-04}.


\bibitem[Aumueller et~al\mbox{.}(2025)]%
        {ann-benchmark-web}
\bibfield{author}{\bibinfo{person}{Martin Aumueller}, \bibinfo{person}{Erik
  Bernhardsson}, {and} \bibinfo{person}{Alec Faitfull}.}
  \bibinfo{year}{2025}\natexlab{}.
\newblock \bibinfo{title}{{ANN Benchmarks}}.
\newblock \bibinfo{howpublished}{\url{https://ann-benchmarks.com/index.html}}.
\newblock
\newblock
\shownote{Accessed: 2025-01-22}.


\bibitem[Aum{\"u}ller et~al\mbox{.}(2020)]%
        {ann-benchmark}
\bibfield{author}{\bibinfo{person}{Martin Aum{\"u}ller}, \bibinfo{person}{Erik
  Bernhardsson}, {and} \bibinfo{person}{Alexander Faithfull}.}
  \bibinfo{year}{2020}\natexlab{}.
\newblock \showarticletitle{ANN-Benchmarks: A benchmarking tool for approximate
  nearest neighbor algorithms}.
\newblock \bibinfo{journal}{\emph{Information Systems}}  \bibinfo{volume}{87}
  (\bibinfo{year}{2020}), \bibinfo{pages}{101374}.
\newblock


\bibitem[Aytar et~al\mbox{.}(2016)]%
        {aud-embed-2}
\bibfield{author}{\bibinfo{person}{Yusuf Aytar}, \bibinfo{person}{Carl
  Vondrick}, {and} \bibinfo{person}{Antonio Torralba}.}
  \bibinfo{year}{2016}\natexlab{}.
\newblock \showarticletitle{Soundnet: Learning sound representations from
  unlabeled video}.
\newblock \bibinfo{journal}{\emph{Advances in neural information processing
  systems}}  \bibinfo{volume}{29} (\bibinfo{year}{2016}).
\newblock


\bibitem[Babenko and Lempitsky(2014a)]%
        {aq}
\bibfield{author}{\bibinfo{person}{Artem Babenko} {and} \bibinfo{person}{Victor
  Lempitsky}.} \bibinfo{year}{2014}\natexlab{a}.
\newblock \showarticletitle{Additive quantization for extreme vector
  compression}. In \bibinfo{booktitle}{\emph{Proceedings of the IEEE Conference
  on Computer Vision and Pattern Recognition}}. \bibinfo{pages}{931--938}.
\newblock


\bibitem[Babenko and Lempitsky(2014b)]%
        {anns-quant-1}
\bibfield{author}{\bibinfo{person}{Artem Babenko} {and} \bibinfo{person}{Victor
  Lempitsky}.} \bibinfo{year}{2014}\natexlab{b}.
\newblock \showarticletitle{The inverted multi-index}.
\newblock \bibinfo{journal}{\emph{IEEE transactions on pattern analysis and
  machine intelligence}} \bibinfo{volume}{37}, \bibinfo{number}{6}
  (\bibinfo{year}{2014}), \bibinfo{pages}{1247--1260}.
\newblock


\bibitem[Bayer and McCreight(1970)]%
        {btree}
\bibfield{author}{\bibinfo{person}{Rudolf Bayer} {and} \bibinfo{person}{Edward
  McCreight}.} \bibinfo{year}{1970}\natexlab{}.
\newblock \showarticletitle{Organization and maintenance of large ordered
  indices}. In \bibinfo{booktitle}{\emph{Proceedings of the 1970 ACM SIGFIDET
  (Now SIGMOD) Workshop on Data Description, Access and Control}}.
  \bibinfo{pages}{107--141}.
\newblock


\bibitem[Beis and Lowe(1997)]%
        {bbf}
\bibfield{author}{\bibinfo{person}{Jeffrey~S Beis} {and}
  \bibinfo{person}{David~G Lowe}.} \bibinfo{year}{1997}\natexlab{}.
\newblock \showarticletitle{Shape indexing using approximate nearest-neighbour
  search in high-dimensional spaces}. In \bibinfo{booktitle}{\emph{Proceedings
  of IEEE computer society conference on computer vision and pattern
  recognition}}. IEEE, \bibinfo{pages}{1000--1006}.
\newblock


\bibitem[Bentley(1975)]%
        {kdtree}
\bibfield{author}{\bibinfo{person}{Jon~Louis Bentley}.}
  \bibinfo{year}{1975}\natexlab{}.
\newblock \showarticletitle{Multidimensional binary search trees used for
  associative searching}.
\newblock \bibinfo{journal}{\emph{Commun. ACM}} \bibinfo{volume}{18},
  \bibinfo{number}{9} (\bibinfo{year}{1975}), \bibinfo{pages}{509--517}.
\newblock


\bibitem[Besta et~al\mbox{.}(2024)]%
        {checkembed}
\bibfield{author}{\bibinfo{person}{Maciej Besta}, \bibinfo{person}{Lorenzo
  Paleari}, \bibinfo{person}{Ales Kubicek}, \bibinfo{person}{Piotr Nyczyk},
  \bibinfo{person}{Robert Gerstenberger}, \bibinfo{person}{Patrick Iff},
  \bibinfo{person}{Tomasz Lehmann}, \bibinfo{person}{Hubert Niewiadomski},
  {and} \bibinfo{person}{Torsten Hoefler}.} \bibinfo{year}{2024}\natexlab{}.
\newblock \showarticletitle{Checkembed: Effective verification of llm solutions
  to open-ended tasks}.
\newblock \bibinfo{journal}{\emph{arXiv preprint arXiv:2406.02524}}
  (\bibinfo{year}{2024}).
\newblock


\bibitem[Beygelzimer et~al\mbox{.}(2006)]%
        {covertree}
\bibfield{author}{\bibinfo{person}{Alina Beygelzimer}, \bibinfo{person}{Sham
  Kakade}, {and} \bibinfo{person}{John Langford}.}
  \bibinfo{year}{2006}\natexlab{}.
\newblock \showarticletitle{Cover trees for nearest neighbor}. In
  \bibinfo{booktitle}{\emph{Proceedings of the 23rd international conference on
  Machine learning}}. \bibinfo{pages}{97--104}.
\newblock


\bibitem[Cai(2025)]%
        {ung-code}
\bibfield{author}{\bibinfo{person}{Yuzheng Cai}.}
  \bibinfo{year}{2025}\natexlab{}.
\newblock \bibinfo{title}{{Unified Navigating Graph Algorithm for Filtered
  Approximate Nearest Neighbor Search.}}
\newblock
  \bibinfo{howpublished}{\url{https://github.com/YZ-Cai/Unified-Navigating-Graph}}.
\newblock
\newblock
\shownote{Accessed: 2025-04-23}.


\bibitem[Cai et~al\mbox{.}(2024)]%
        {ung}
\bibfield{author}{\bibinfo{person}{Yuzheng Cai}, \bibinfo{person}{Jiayang Shi},
  \bibinfo{person}{Yizhuo Chen}, {and} \bibinfo{person}{Weiguo Zheng}.}
  \bibinfo{year}{2024}\natexlab{}.
\newblock \showarticletitle{Navigating Labels and Vectors: A Unified Approach
  to Filtered Approximate Nearest Neighbor Search}.
\newblock \bibinfo{journal}{\emph{Proceedings of the ACM on Management of
  Data}} \bibinfo{volume}{2}, \bibinfo{number}{6} (\bibinfo{year}{2024}),
  \bibinfo{pages}{1--27}.
\newblock


\bibitem[Chen et~al\mbox{.}(2024)]%
        {bge}
\bibfield{author}{\bibinfo{person}{Jianlv Chen}, \bibinfo{person}{Shitao Xiao},
  \bibinfo{person}{Peitian Zhang}, \bibinfo{person}{Kun Luo},
  \bibinfo{person}{Defu Lian}, {and} \bibinfo{person}{Zheng Liu}.}
  \bibinfo{year}{2024}\natexlab{}.
\newblock \showarticletitle{Bge m3-embedding: Multi-lingual,
  multi-functionality, multi-granularity text embeddings through self-knowledge
  distillation}.
\newblock \bibinfo{journal}{\emph{arXiv preprint arXiv:2402.03216}}
  (\bibinfo{year}{2024}).
\newblock


\bibitem[Chen et~al\mbox{.}(2021)]%
        {spann}
\bibfield{author}{\bibinfo{person}{Qi Chen}, \bibinfo{person}{Bing Zhao},
  \bibinfo{person}{Haidong Wang}, \bibinfo{person}{Mingqin Li},
  \bibinfo{person}{Chuanjie Liu}, \bibinfo{person}{Zengzhong Li},
  \bibinfo{person}{Mao Yang}, {and} \bibinfo{person}{Jingdong Wang}.}
  \bibinfo{year}{2021}\natexlab{}.
\newblock \showarticletitle{Spann: Highly-efficient billion-scale approximate
  nearest neighborhood search}.
\newblock \bibinfo{journal}{\emph{Advances in Neural Information Processing
  Systems}}  \bibinfo{volume}{34} (\bibinfo{year}{2021}),
  \bibinfo{pages}{5199--5212}.
\newblock


\bibitem[Ciaccia et~al\mbox{.}(1997)]%
        {mtree}
\bibfield{author}{\bibinfo{person}{Paolo Ciaccia}, \bibinfo{person}{Marco
  Patella}, \bibinfo{person}{Pavel Zezula}, {et~al\mbox{.}}}
  \bibinfo{year}{1997}\natexlab{}.
\newblock \showarticletitle{M-tree: An efficient access method for similarity
  search in metric spaces}. In \bibinfo{booktitle}{\emph{Vldb}},
  Vol.~\bibinfo{volume}{97}. Citeseer, \bibinfo{pages}{426--435}.
\newblock


\bibitem[Contest(2025)]%
        {sigmod-contest}
\bibfield{author}{\bibinfo{person}{ACM SIGMOD 2024~Programming Contest}.}
  \bibinfo{year}{2025}\natexlab{}.
\newblock \bibinfo{title}{{ACM SIGMOD Programming Contest 2024}}.
\newblock
  \bibinfo{howpublished}{\url{https://dbgroup.cs.tsinghua.edu.cn/sigmod2024/index.shtml}}.
\newblock
\newblock
\shownote{Accessed: 2025-10-02}.


\bibitem[{Cornell University} et~al\mbox{.}(2025)]%
        {arxiv-ds}
\bibfield{author}{\bibinfo{person}{{Cornell University}},
  \bibinfo{person}{Devrishi}, \bibinfo{person}{Joe Tricot},
  \bibinfo{person}{Brian Maltzan}, \bibinfo{person}{Shamsi Brinn}, {and}
  \bibinfo{person}{Timo Bozsolik}.} \bibinfo{year}{2025}\natexlab{}.
\newblock \bibinfo{title}{{arXiv Dataset}}.
\newblock
  \bibinfo{howpublished}{\url{https://www.kaggle.com/datasets/Cornell-University/arxiv}}.
\newblock
\newblock
\shownote{Accessed: 2025-03-13}.


\bibitem[Cramer et~al\mbox{.}(2019)]%
        {aud-embed-1}
\bibfield{author}{\bibinfo{person}{Aurora~Linh Cramer},
  \bibinfo{person}{Ho-Hsiang Wu}, \bibinfo{person}{Justin Salamon}, {and}
  \bibinfo{person}{Juan~Pablo Bello}.} \bibinfo{year}{2019}\natexlab{}.
\newblock \showarticletitle{Look, listen, and learn more: Design choices for
  deep audio embeddings}. In \bibinfo{booktitle}{\emph{ICASSP 2019-2019 IEEE
  International Conference on Acoustics, Speech and Signal Processing
  (ICASSP)}}. IEEE, \bibinfo{pages}{3852--3856}.
\newblock


\bibitem[Dasgupta and Freund(2008)]%
        {rp-tree}
\bibfield{author}{\bibinfo{person}{Sanjoy Dasgupta} {and} \bibinfo{person}{Yoav
  Freund}.} \bibinfo{year}{2008}\natexlab{}.
\newblock \showarticletitle{Random projection trees and low dimensional
  manifolds}. In \bibinfo{booktitle}{\emph{Proceedings of the fortieth annual
  ACM symposium on Theory of computing}}. \bibinfo{pages}{537--546}.
\newblock


\bibitem[Datar et~al\mbox{.}(2004)]%
        {lsh-2}
\bibfield{author}{\bibinfo{person}{Mayur Datar}, \bibinfo{person}{Nicole
  Immorlica}, \bibinfo{person}{Piotr Indyk}, {and} \bibinfo{person}{Vahab~S
  Mirrokni}.} \bibinfo{year}{2004}\natexlab{}.
\newblock \showarticletitle{Locality-sensitive hashing scheme based on p-stable
  distributions}. In \bibinfo{booktitle}{\emph{Proceedings of the twentieth
  annual symposium on Computational geometry}}. \bibinfo{pages}{253--262}.
\newblock


\bibitem[Datasets(2025)]%
        {wit-ds}
\bibfield{author}{\bibinfo{person}{Google~Research Datasets}.}
  \bibinfo{year}{2025}\natexlab{}.
\newblock \bibinfo{title}{{WIT : Wikipedia-based Image Text Dataset}}.
\newblock
  \bibinfo{howpublished}{\url{https://github.com/google-research-datasets/wit}}.
\newblock
\newblock
\shownote{Accessed: 2025-03-06}.


\bibitem[DBGroup(2025)]%
        {unify-code}
\bibfield{author}{\bibinfo{person}{SJTU DBGroup}.}
  \bibinfo{year}{2025}\natexlab{}.
\newblock \bibinfo{title}{{UNIFY - Unified Index for Range Filtered Approximate
  Nearest Neighbors Search}}.
\newblock \bibinfo{howpublished}{\url{https://github.com/sjtu-dbgroup/UNIFY}}.
\newblock
\newblock
\shownote{Accessed: 2025-02-20}.


\bibitem[De~Berg(2000)]%
        {seg-tree}
\bibfield{author}{\bibinfo{person}{Mark De~Berg}.}
  \bibinfo{year}{2000}\natexlab{}.
\newblock \bibinfo{booktitle}{\emph{Computational geometry: algorithms and
  applications}}.
\newblock \bibinfo{publisher}{Springer Science \& Business Media}.
\newblock


\bibitem[Desai et~al\mbox{.}(2025)]%
        {redcaps-ds}
\bibfield{author}{\bibinfo{person}{Karan Desai}, \bibinfo{person}{Gaurav Kaul},
  \bibinfo{person}{Zubin Aysola}, {and} \bibinfo{person}{Justin Johnson}.}
  \bibinfo{year}{2025}\natexlab{}.
\newblock \bibinfo{title}{{RedCaps: Web-curated image-text data created by the
  people, for the people}}.
\newblock \bibinfo{howpublished}{\url{https://redcaps.xyz/}}.
\newblock
\newblock
\shownote{Accessed: 2025-03-06}.


\bibitem[Dmitry~Baranchuk(2025)]%
        {deep1b-ds}
\bibfield{author}{\bibinfo{person}{Artem~Babenko Dmitry~Baranchuk}.}
  \bibinfo{year}{2025}\natexlab{}.
\newblock \bibinfo{title}{{Benchmarks for Billion-Scale Similarity Search}}.
\newblock
  \bibinfo{howpublished}{\url{https://research.yandex.com/blog/benchmarks-for-billion-scale-similarity-search}}.
\newblock
\newblock
\shownote{Accessed: 2025-03-06}.


\bibitem[Dong et~al\mbox{.}(2011)]%
        {knn-graph}
\bibfield{author}{\bibinfo{person}{Wei Dong}, \bibinfo{person}{Charikar Moses},
  {and} \bibinfo{person}{Kai Li}.} \bibinfo{year}{2011}\natexlab{}.
\newblock \showarticletitle{Efficient k-nearest neighbor graph construction for
  generic similarity measures}. In \bibinfo{booktitle}{\emph{Proceedings of the
  20th international conference on World wide web}}. \bibinfo{pages}{577--586}.
\newblock


\bibitem[Douze et~al\mbox{.}(2024)]%
        {faiss}
\bibfield{author}{\bibinfo{person}{Matthijs Douze}, \bibinfo{person}{Alexandr
  Guzhva}, \bibinfo{person}{Chengqi Deng}, \bibinfo{person}{Jeff Johnson},
  \bibinfo{person}{Gergely Szilvasy}, \bibinfo{person}{Pierre-Emmanuel
  Mazaré}, \bibinfo{person}{Maria Lomeli}, \bibinfo{person}{Lucas Hosseini},
  {and} \bibinfo{person}{Hervé Jégou}.} \bibinfo{year}{2024}\natexlab{}.
\newblock \showarticletitle{The Faiss library}.
\newblock  (\bibinfo{year}{2024}).
\newblock
\showeprint[arxiv]{2401.08281}~[cs.LG]


\bibitem[Echihabi et~al\mbox{.}(2021)]%
        {anns-survey-echihabi}
\bibfield{author}{\bibinfo{person}{Karima Echihabi}, \bibinfo{person}{Kostas
  Zoumpatianos}, {and} \bibinfo{person}{Themis Palpanas}.}
  \bibinfo{year}{2021}\natexlab{}.
\newblock \showarticletitle{New trends in high-d vector similarity search:
  al-driven, progressive, and distributed}.
\newblock \bibinfo{journal}{\emph{Proceedings of the VLDB Endowment}}
  \bibinfo{volume}{14}, \bibinfo{number}{12} (\bibinfo{year}{2021}),
  \bibinfo{pages}{3198--3201}.
\newblock


\bibitem[Engels(2025)]%
        {bwst-code}
\bibfield{author}{\bibinfo{person}{Josh Engels}.}
  \bibinfo{year}{2025}\natexlab{}.
\newblock \bibinfo{title}{{RangeFilteredANN}}.
\newblock
  \bibinfo{howpublished}{\url{https://github.com/JoshEngels/RangeFilteredANN}}.
\newblock
\newblock
\shownote{Accessed: 2025-02-25}.


\bibitem[Engels et~al\mbox{.}(2024)]%
        {bwst}
\bibfield{author}{\bibinfo{person}{Joshua Engels}, \bibinfo{person}{Benjamin
  Landrum}, \bibinfo{person}{Shangdi Yu}, \bibinfo{person}{Laxman Dhulipala},
  {and} \bibinfo{person}{Julian Shun}.} \bibinfo{year}{2024}\natexlab{}.
\newblock \showarticletitle{Approximate Nearest Neighbor Search with Window
  Filters}.
\newblock \bibinfo{journal}{\emph{arXiv preprint arXiv:2402.00943}}
  (\bibinfo{year}{2024}).
\newblock


\bibitem[Farrall(2025)]%
        {words-ds}
\bibfield{author}{\bibinfo{person}{Eric~Yang Farrall}.}
  \bibinfo{year}{2025}\natexlab{}.
\newblock \bibinfo{title}{{Word Embeddings}}.
\newblock
  \bibinfo{howpublished}{\url{https://huggingface.co/datasets/efarrall/word_embeddings}}.
\newblock
\newblock
\shownote{Accessed: 2025-04-23}.


\bibitem[Foster et~al\mbox{.}(2025)]%
        {grng}
\bibfield{author}{\bibinfo{person}{Cole Foster}, \bibinfo{person}{Berk
  Sevilmis}, {and} \bibinfo{person}{Benjamin Kimia}.}
  \bibinfo{year}{2025}\natexlab{}.
\newblock \showarticletitle{Generalized relative neighborhood graph (GRNG) for
  similarity search}.
\newblock \bibinfo{journal}{\emph{Pattern Recognition Letters}}
  \bibinfo{volume}{188} (\bibinfo{year}{2025}), \bibinfo{pages}{103--110}.
\newblock


\bibitem[Fu et~al\mbox{.}(2019)]%
        {anns-graph-6}
\bibfield{author}{\bibinfo{person}{Cong Fu}, \bibinfo{person}{Changxu Wang},
  {and} \bibinfo{person}{Deng Cai}.} \bibinfo{year}{2019}\natexlab{}.
\newblock \showarticletitle{Satellite system graph: Towards the efficiency
  up-boundary of graph-based approximate nearest neighbor search}.
\newblock \bibinfo{journal}{\emph{CoRR}} (\bibinfo{year}{2019}).
\newblock


\bibitem[Fu et~al\mbox{.}(2021)]%
        {anns-graph-3}
\bibfield{author}{\bibinfo{person}{Cong Fu}, \bibinfo{person}{Changxu Wang},
  {and} \bibinfo{person}{Deng Cai}.} \bibinfo{year}{2021}\natexlab{}.
\newblock \showarticletitle{High dimensional similarity search with satellite
  system graph: Efficiency, scalability, and unindexed query compatibility}.
\newblock \bibinfo{journal}{\emph{IEEE Transactions on Pattern Analysis and
  Machine Intelligence}} \bibinfo{volume}{44}, \bibinfo{number}{8}
  (\bibinfo{year}{2021}), \bibinfo{pages}{4139--4150}.
\newblock


\bibitem[Fu et~al\mbox{.}(2017)]%
        {nsg}
\bibfield{author}{\bibinfo{person}{Cong Fu}, \bibinfo{person}{Chao Xiang},
  \bibinfo{person}{Changxu Wang}, {and} \bibinfo{person}{Deng Cai}.}
  \bibinfo{year}{2017}\natexlab{}.
\newblock \showarticletitle{Fast approximate nearest neighbor search with the
  navigating spreading-out graph}.
\newblock \bibinfo{journal}{\emph{arXiv preprint arXiv:1707.00143}}
  (\bibinfo{year}{2017}).
\newblock


\bibitem[Fu(2025)]%
        {nhq-code}
\bibfield{author}{\bibinfo{person}{Yujian Fu}.}
  \bibinfo{year}{2025}\natexlab{}.
\newblock \bibinfo{title}{{NHQ: Native Hybrid Query Framework for Vector
  Similarity Search with Attribute Constraint}}.
\newblock \bibinfo{howpublished}{\url{https://github.com/YujianFu97/NHQ}}.
\newblock
\newblock
\shownote{Accessed: 2025-02-27}.


\bibitem[Gan et~al\mbox{.}(2012)]%
        {anns-hash-8}
\bibfield{author}{\bibinfo{person}{Junhao Gan}, \bibinfo{person}{Jianlin Feng},
  \bibinfo{person}{Qiong Fang}, {and} \bibinfo{person}{Wilfred Ng}.}
  \bibinfo{year}{2012}\natexlab{}.
\newblock \showarticletitle{Locality-sensitive hashing scheme based on dynamic
  collision counting}. In \bibinfo{booktitle}{\emph{Proceedings of the 2012 ACM
  SIGMOD international conference on management of data}}.
  \bibinfo{pages}{541--552}.
\newblock


\bibitem[Gao and Long(2024)]%
        {rabitq}
\bibfield{author}{\bibinfo{person}{Jianyang Gao} {and} \bibinfo{person}{Cheng
  Long}.} \bibinfo{year}{2024}\natexlab{}.
\newblock \showarticletitle{RaBitQ: quantizing high-dimensional vectors with a
  theoretical error bound for approximate nearest neighbor search}.
\newblock \bibinfo{journal}{\emph{Proceedings of the ACM on Management of
  Data}} \bibinfo{volume}{2}, \bibinfo{number}{3} (\bibinfo{year}{2024}),
  \bibinfo{pages}{1--27}.
\newblock


\bibitem[Gao et~al\mbox{.}(2023)]%
        {rag-survey}
\bibfield{author}{\bibinfo{person}{Yunfan Gao}, \bibinfo{person}{Yun Xiong},
  \bibinfo{person}{Xinyu Gao}, \bibinfo{person}{Kangxiang Jia},
  \bibinfo{person}{Jinliu Pan}, \bibinfo{person}{Yuxi Bi}, \bibinfo{person}{Yi
  Dai}, \bibinfo{person}{Jiawei Sun}, \bibinfo{person}{Haofen Wang}, {and}
  \bibinfo{person}{Haofen Wang}.} \bibinfo{year}{2023}\natexlab{}.
\newblock \showarticletitle{Retrieval-augmented generation for large language
  models: A survey}.
\newblock \bibinfo{journal}{\emph{arXiv preprint arXiv:2312.10997}}
  \bibinfo{volume}{2} (\bibinfo{year}{2023}).
\newblock


\bibitem[Ge et~al\mbox{.}(2013)]%
        {opq}
\bibfield{author}{\bibinfo{person}{Tiezheng Ge}, \bibinfo{person}{Kaiming He},
  \bibinfo{person}{Qifa Ke}, {and} \bibinfo{person}{Jian Sun}.}
  \bibinfo{year}{2013}\natexlab{}.
\newblock \showarticletitle{Optimized product quantization for approximate
  nearest neighbor search}. In \bibinfo{booktitle}{\emph{Proceedings of the
  IEEE conference on computer vision and pattern recognition}}.
  \bibinfo{pages}{2946--2953}.
\newblock


\bibitem[Gionis et~al\mbox{.}(1999)]%
        {anns-hash-3}
\bibfield{author}{\bibinfo{person}{Aristides Gionis}, \bibinfo{person}{Piotr
  Indyk}, \bibinfo{person}{Rajeev Motwani}, {et~al\mbox{.}}}
  \bibinfo{year}{1999}\natexlab{}.
\newblock \showarticletitle{Similarity search in high dimensions via hashing}.
  In \bibinfo{booktitle}{\emph{Vldb}}, Vol.~\bibinfo{volume}{99}.
  \bibinfo{pages}{518--529}.
\newblock


\bibitem[Gollapudi et~al\mbox{.}(2023)]%
        {fdann}
\bibfield{author}{\bibinfo{person}{Siddharth Gollapudi}, \bibinfo{person}{Neel
  Karia}, \bibinfo{person}{Varun Sivashankar}, \bibinfo{person}{Ravishankar
  Krishnaswamy}, \bibinfo{person}{Nikit Begwani}, \bibinfo{person}{Swapnil
  Raz}, \bibinfo{person}{Yiyong Lin}, \bibinfo{person}{Yin Zhang},
  \bibinfo{person}{Neelam Mahapatro}, \bibinfo{person}{Premkumar Srinivasan},
  {et~al\mbox{.}}} \bibinfo{year}{2023}\natexlab{}.
\newblock \showarticletitle{Filtered-diskann: Graph algorithms for approximate
  nearest neighbor search with filters}. In
  \bibinfo{booktitle}{\emph{Proceedings of the ACM Web Conference 2023}}.
  \bibinfo{pages}{3406--3416}.
\newblock


\bibitem[Gong et~al\mbox{.}(2020)]%
        {idec}
\bibfield{author}{\bibinfo{person}{Long Gong}, \bibinfo{person}{Huayi Wang},
  \bibinfo{person}{Mitsunori Ogihara}, {and} \bibinfo{person}{Jun Xu}.}
  \bibinfo{year}{2020}\natexlab{}.
\newblock \showarticletitle{iDEC: indexable distance estimating codes for
  approximate nearest neighbor search}.
\newblock \bibinfo{journal}{\emph{Proceedings of the VLDB Endowment}}
  \bibinfo{volume}{13}, \bibinfo{number}{9} (\bibinfo{year}{2020}).
\newblock


\bibitem[Grother et~al\mbox{.}(2019)]%
        {face-rec-1}
\bibfield{author}{\bibinfo{person}{Patrick Grother}, \bibinfo{person}{Patrick
  Grother}, \bibinfo{person}{Mei Ngan}, {and} \bibinfo{person}{Kayee Hanaoka}.}
  \bibinfo{year}{2019}\natexlab{}.
\newblock \bibinfo{title}{Face recognition vendor test (frvt) part 2:
  Identification}.
\newblock


\bibitem[Group(2025)]%
        {postgres}
\bibfield{author}{\bibinfo{person}{The PostgreSQL Global~Development Group}.}
  \bibinfo{year}{2025}\natexlab{}.
\newblock \bibinfo{title}{{PostgreSQL}}.
\newblock \bibinfo{howpublished}{\url{https://www.postgresql.org/}}.
\newblock
\newblock
\shownote{Accessed: 2025-01-22}.


\bibitem[Guestrin Lab~at Stanford~University(2025)]%
        {acorn-code}
\bibfield{author}{\bibinfo{person}{Department of Computer~Science Guestrin
  Lab~at Stanford~University}.} \bibinfo{year}{2025}\natexlab{}.
\newblock \bibinfo{title}{{ACORN}}.
\newblock \bibinfo{howpublished}{\url{https://github.com/guestrin-lab/ACORN}}.
\newblock
\newblock
\shownote{Accessed: 2025-02-25}.


\bibitem[Guo et~al\mbox{.}(2016)]%
        {bapq}
\bibfield{author}{\bibinfo{person}{Qin-Zhen Guo}, \bibinfo{person}{Zhi Zeng},
  \bibinfo{person}{Shuwu Zhang}, \bibinfo{person}{Guixuan Zhang}, {and}
  \bibinfo{person}{Yuan Zhang}.} \bibinfo{year}{2016}\natexlab{}.
\newblock \showarticletitle{Adaptive bit allocation product quantization}.
\newblock \bibinfo{journal}{\emph{Neurocomputing}}  \bibinfo{volume}{171}
  (\bibinfo{year}{2016}), \bibinfo{pages}{866--877}.
\newblock


\bibitem[Gupta(2025)]%
        {caps-code}
\bibfield{author}{\bibinfo{person}{Gaurav Gupta}.}
  \bibinfo{year}{2025}\natexlab{}.
\newblock \bibinfo{title}{{CAPS}}.
\newblock
  \bibinfo{howpublished}{\url{https://github.com/gaurav16gupta/constrainedANN}}.
\newblock
\newblock
\shownote{Accessed: 2025-02-26}.


\bibitem[Gupta et~al\mbox{.}(2023)]%
        {caps}
\bibfield{author}{\bibinfo{person}{Gaurav Gupta}, \bibinfo{person}{Jonah Yi},
  \bibinfo{person}{Benjamin Coleman}, \bibinfo{person}{Chen Luo},
  \bibinfo{person}{Vihan Lakshman}, {and} \bibinfo{person}{Anshumali
  Shrivastava}.} \bibinfo{year}{2023}\natexlab{}.
\newblock \showarticletitle{CAPS: A Practical Partition Index for Filtered
  Similarity Search}.
\newblock \bibinfo{journal}{\emph{arXiv preprint arXiv:2308.15014}}
  (\bibinfo{year}{2023}).
\newblock


\bibitem[Guttman(1984)]%
        {rtree}
\bibfield{author}{\bibinfo{person}{Antonin Guttman}.}
  \bibinfo{year}{1984}\natexlab{}.
\newblock \showarticletitle{R-trees: A dynamic index structure for spatial
  searching}. In \bibinfo{booktitle}{\emph{Proceedings of the 1984 ACM SIGMOD
  international conference on Management of data}}. \bibinfo{pages}{47--57}.
\newblock


\bibitem[Han et~al\mbox{.}(2023)]%
        {anns-survey-han}
\bibfield{author}{\bibinfo{person}{Yikun Han}, \bibinfo{person}{Chunjiang Liu},
  {and} \bibinfo{person}{Pengfei Wang}.} \bibinfo{year}{2023}\natexlab{}.
\newblock \showarticletitle{A comprehensive survey on vector database: Storage
  and retrieval technique, challenge}.
\newblock \bibinfo{journal}{\emph{arXiv preprint arXiv:2310.11703}}
  (\bibinfo{year}{2023}).
\newblock


\bibitem[Harwood and Drummond(2016)]%
        {anns-graph-4}
\bibfield{author}{\bibinfo{person}{Ben Harwood} {and} \bibinfo{person}{Tom
  Drummond}.} \bibinfo{year}{2016}\natexlab{}.
\newblock \showarticletitle{Fanng: Fast approximate nearest neighbour graphs}.
  In \bibinfo{booktitle}{\emph{Proceedings of the IEEE Conference on Computer
  Vision and Pattern Recognition}}. \bibinfo{pages}{5713--5722}.
\newblock


\bibitem[Hoefler and Belli(2015)]%
        {bench-rules}
\bibfield{author}{\bibinfo{person}{Torsten Hoefler} {and}
  \bibinfo{person}{Roberto Belli}.} \bibinfo{year}{2015}\natexlab{}.
\newblock \showarticletitle{Scientific benchmarking of parallel computing
  systems: twelve ways to tell the masses when reporting performance results}.
  In \bibinfo{booktitle}{\emph{Proceedings of the international conference for
  high performance computing, networking, storage and analysis}}.
  \bibinfo{pages}{1--12}.
\newblock


\bibitem[Houle and Nett(2014)]%
        {anns-tree-1}
\bibfield{author}{\bibinfo{person}{Michael~E Houle} {and}
  \bibinfo{person}{Michael Nett}.} \bibinfo{year}{2014}\natexlab{}.
\newblock \showarticletitle{Rank-based similarity search: Reducing the
  dimensional dependence}.
\newblock \bibinfo{journal}{\emph{IEEE transactions on pattern analysis and
  machine intelligence}} \bibinfo{volume}{37}, \bibinfo{number}{1}
  (\bibinfo{year}{2014}), \bibinfo{pages}{136--150}.
\newblock


\bibitem[Huang et~al\mbox{.}(2015)]%
        {anns-hash-9}
\bibfield{author}{\bibinfo{person}{Qiang Huang}, \bibinfo{person}{Jianlin
  Feng}, \bibinfo{person}{Yikai Zhang}, \bibinfo{person}{Qiong Fang}, {and}
  \bibinfo{person}{Wilfred Ng}.} \bibinfo{year}{2015}\natexlab{}.
\newblock \showarticletitle{Query-aware locality-sensitive hashing for
  approximate nearest neighbor search}.
\newblock \bibinfo{journal}{\emph{Proceedings of the VLDB Endowment}}
  \bibinfo{volume}{9}, \bibinfo{number}{1} (\bibinfo{year}{2015}),
  \bibinfo{pages}{1--12}.
\newblock


\bibitem[Hurst et~al\mbox{.}(2024)]%
        {gpt4o}
\bibfield{author}{\bibinfo{person}{Aaron Hurst}, \bibinfo{person}{Adam Lerer},
  \bibinfo{person}{Adam~P Goucher}, \bibinfo{person}{Adam Perelman},
  \bibinfo{person}{Aditya Ramesh}, \bibinfo{person}{Aidan Clark},
  \bibinfo{person}{AJ Ostrow}, \bibinfo{person}{Akila Welihinda},
  \bibinfo{person}{Alan Hayes}, \bibinfo{person}{Alec Radford},
  {et~al\mbox{.}}} \bibinfo{year}{2024}\natexlab{}.
\newblock \showarticletitle{Gpt-4o system card}.
\newblock \bibinfo{journal}{\emph{arXiv preprint arXiv:2410.21276}}
  (\bibinfo{year}{2024}).
\newblock


\bibitem[Indyk and Motwani(1998)]%
        {lsh-1}
\bibfield{author}{\bibinfo{person}{Piotr Indyk} {and} \bibinfo{person}{Rajeev
  Motwani}.} \bibinfo{year}{1998}\natexlab{}.
\newblock \showarticletitle{Approximate nearest neighbors: towards removing the
  curse of dimensionality}. In \bibinfo{booktitle}{\emph{Proceedings of the
  thirtieth annual ACM symposium on Theory of computing}}.
  \bibinfo{pages}{604--613}.
\newblock


\bibitem[Iwasaki(2016)]%
        {anns-graph-5}
\bibfield{author}{\bibinfo{person}{Masajiro Iwasaki}.}
  \bibinfo{year}{2016}\natexlab{}.
\newblock \showarticletitle{Pruned bi-directed k-nearest neighbor graph for
  proximity search}. In \bibinfo{booktitle}{\emph{International Conference on
  Similarity Search and Applications}}. Springer, \bibinfo{pages}{20--33}.
\newblock


\bibitem[Jafari et~al\mbox{.}(2020)]%
        {mmlsh}
\bibfield{author}{\bibinfo{person}{Omid Jafari}, \bibinfo{person}{Parth
  Nagarkar}, {and} \bibinfo{person}{Jonathan Monta{\~n}o}.}
  \bibinfo{year}{2020}\natexlab{}.
\newblock \showarticletitle{mmlsh: A practical and efficient technique for
  processing approximate nearest neighbor queries on multimedia data}. In
  \bibinfo{booktitle}{\emph{International Conference on Similarity Search and
  Applications}}. Springer, \bibinfo{pages}{47--61}.
\newblock


\bibitem[Jayaram~Subramanya et~al\mbox{.}(2019)]%
        {diskann}
\bibfield{author}{\bibinfo{person}{Suhas Jayaram~Subramanya},
  \bibinfo{person}{Fnu Devvrit}, \bibinfo{person}{Harsha~Vardhan Simhadri},
  \bibinfo{person}{Ravishankar Krishnawamy}, {and} \bibinfo{person}{Rohan
  Kadekodi}.} \bibinfo{year}{2019}\natexlab{}.
\newblock \showarticletitle{Diskann: Fast accurate billion-point nearest
  neighbor search on a single node}.
\newblock \bibinfo{journal}{\emph{Advances in neural information processing
  Systems}}  \bibinfo{volume}{32} (\bibinfo{year}{2019}).
\newblock


\bibitem[Jegou et~al\mbox{.}(2010)]%
        {pq}
\bibfield{author}{\bibinfo{person}{Herve Jegou}, \bibinfo{person}{Matthijs
  Douze}, {and} \bibinfo{person}{Cordelia Schmid}.}
  \bibinfo{year}{2010}\natexlab{}.
\newblock \showarticletitle{Product quantization for nearest neighbor search}.
\newblock \bibinfo{journal}{\emph{IEEE transactions on pattern analysis and
  machine intelligence}} \bibinfo{volume}{33}, \bibinfo{number}{1}
  (\bibinfo{year}{2010}), \bibinfo{pages}{117--128}.
\newblock


\bibitem[Jiang et~al\mbox{.}(2025)]%
        {digra}
\bibfield{author}{\bibinfo{person}{Mengxu Jiang}, \bibinfo{person}{Zhi Yang},
  \bibinfo{person}{Fangyuan Zhang}, \bibinfo{person}{Guanhao Hou},
  \bibinfo{person}{Jieming Shi}, \bibinfo{person}{Wenchao Zhou},
  \bibinfo{person}{Feifei Li}, {and} \bibinfo{person}{Sibo Wang}.}
  \bibinfo{year}{2025}\natexlab{}.
\newblock \showarticletitle{DIGRA: A Dynamic Graph Indexing for Approximate
  Nearest Neighbor Search with Range Filter}.
\newblock \bibinfo{journal}{\emph{Proceedings of the ACM on Management of
  Data}} \bibinfo{volume}{3}, \bibinfo{number}{3} (\bibinfo{year}{2025}),
  \bibinfo{pages}{1--26}.
\newblock


\bibitem[Kalantidis and Avrithis(2014)]%
        {lopq}
\bibfield{author}{\bibinfo{person}{Yannis Kalantidis} {and}
  \bibinfo{person}{Yannis Avrithis}.} \bibinfo{year}{2014}\natexlab{}.
\newblock \showarticletitle{Locally optimized product quantization for
  approximate nearest neighbor search}. In
  \bibinfo{booktitle}{\emph{Proceedings of the IEEE conference on computer
  vision and pattern recognition}}. \bibinfo{pages}{2321--2328}.
\newblock


\bibitem[Kim et~al\mbox{.}(2025)]%
        {rl_01}
\bibfield{author}{\bibinfo{person}{Sejong Kim}, \bibinfo{person}{Hyunseo Song},
  \bibinfo{person}{Hyunwoo Seo}, {and} \bibinfo{person}{Hyunjun Kim}.}
  \bibinfo{year}{2025}\natexlab{}.
\newblock \showarticletitle{Optimizing retrieval strategies for financial
  question answering documents in retrieval-augmented generation systems}.
\newblock \bibinfo{journal}{\emph{arXiv preprint arXiv:2503.15191}}
  (\bibinfo{year}{2025}).
\newblock


\bibitem[Koenigstein et~al\mbox{.}(2012)]%
        {rec-sys-2}
\bibfield{author}{\bibinfo{person}{Noam Koenigstein},
  \bibinfo{person}{Parikshit Ram}, {and} \bibinfo{person}{Yuval Shavitt}.}
  \bibinfo{year}{2012}\natexlab{}.
\newblock \showarticletitle{Efficient retrieval of recommendations in a matrix
  factorization framework}. In \bibinfo{booktitle}{\emph{Proceedings of the
  21st ACM international conference on Information and knowledge management}}.
  \bibinfo{pages}{535--544}.
\newblock


\bibitem[Laurent~Amsaleg(2025)]%
        {sift-ds}
\bibfield{author}{\bibinfo{person}{Hervé~Jégou Laurent~Amsaleg}.}
  \bibinfo{year}{2025}\natexlab{}.
\newblock \bibinfo{title}{{Datasets for approximate nearest neighbor search}}.
\newblock \bibinfo{howpublished}{\url{http://corpus-texmex.irisa.fr/}}.
\newblock
\newblock
\shownote{Accessed: 2025-03-06}.


\bibitem[LeCun(2025)]%
        {mnist-ds}
\bibfield{author}{\bibinfo{person}{Yann LeCun}.}
  \bibinfo{year}{2025}\natexlab{}.
\newblock \bibinfo{title}{{mnist}}.
\newblock
  \bibinfo{howpublished}{\url{https://huggingface.co/datasets/ylecun/mnist/viewer?views\%5B\%5D=train}}.
\newblock
\newblock
\shownote{Accessed: 2025-03-06}.


\bibitem[Lee et~al\mbox{.}(2024)]%
        {nvembed}
\bibfield{author}{\bibinfo{person}{Chankyu Lee}, \bibinfo{person}{Rajarshi
  Roy}, \bibinfo{person}{Mengyao Xu}, \bibinfo{person}{Jonathan Raiman},
  \bibinfo{person}{Mohammad Shoeybi}, \bibinfo{person}{Bryan Catanzaro}, {and}
  \bibinfo{person}{Wei Ping}.} \bibinfo{year}{2024}\natexlab{}.
\newblock \showarticletitle{NV-Embed: Improved Techniques for Training LLMs as
  Generalist Embedding Models}.
\newblock \bibinfo{journal}{\emph{arXiv preprint arXiv:2405.17428}}
  (\bibinfo{year}{2024}).
\newblock


\bibitem[Lei(2025)]%
        {lens-hf}
\bibfield{author}{\bibinfo{person}{Yibin Lei}.}
  \bibinfo{year}{2025}\natexlab{}.
\newblock \bibinfo{title}{{LENS-d8000}}.
\newblock
  \bibinfo{howpublished}{\url{https://huggingface.co/yibinlei/LENS-d8000}}.
\newblock
\newblock
\shownote{Accessed: 2025-01-22}.


\bibitem[Lei et~al\mbox{.}(2025)]%
        {lens}
\bibfield{author}{\bibinfo{person}{Yibin Lei}, \bibinfo{person}{Tao Shen},
  \bibinfo{person}{Yu Cao}, {and} \bibinfo{person}{Andrew Yates}.}
  \bibinfo{year}{2025}\natexlab{}.
\newblock \showarticletitle{Enhancing Lexicon-Based Text Embeddings with Large
  Language Models}.
\newblock \bibinfo{journal}{\emph{arXiv preprint arXiv:2501.09749}}
  (\bibinfo{year}{2025}).
\newblock


\bibitem[Lerer et~al\mbox{.}(2019)]%
        {biggraph}
\bibfield{author}{\bibinfo{person}{Adam Lerer}, \bibinfo{person}{Ledell Wu},
  \bibinfo{person}{Jiajun Shen}, \bibinfo{person}{Timothee Lacroix},
  \bibinfo{person}{Luca Wehrstedt}, \bibinfo{person}{Abhijit Bose}, {and}
  \bibinfo{person}{Alex Peysakhovich}.} \bibinfo{year}{2019}\natexlab{}.
\newblock \showarticletitle{Pytorch-biggraph: A large scale graph embedding
  system}.
\newblock \bibinfo{journal}{\emph{Proceedings of Machine Learning and Systems}}
   \bibinfo{volume}{1} (\bibinfo{year}{2019}), \bibinfo{pages}{120--131}.
\newblock


\bibitem[Li et~al\mbox{.}(2018)]%
        {ecom-1}
\bibfield{author}{\bibinfo{person}{Jie Li}, \bibinfo{person}{Haifeng Liu},
  \bibinfo{person}{Chuanghua Gui}, \bibinfo{person}{Jianyu Chen},
  \bibinfo{person}{Zhenyuan Ni}, \bibinfo{person}{Ning Wang}, {and}
  \bibinfo{person}{Yuan Chen}.} \bibinfo{year}{2018}\natexlab{}.
\newblock \showarticletitle{The design and implementation of a real time visual
  search system on JD E-commerce platform}. In
  \bibinfo{booktitle}{\emph{Proceedings of the 19th International Middleware
  Conference Industry}}. \bibinfo{pages}{9--16}.
\newblock


\bibitem[Li et~al\mbox{.}(2020)]%
        {anns-hash-4}
\bibfield{author}{\bibinfo{person}{Mingjie Li}, \bibinfo{person}{Ying Zhang},
  \bibinfo{person}{Yifang Sun}, \bibinfo{person}{Wei Wang},
  \bibinfo{person}{Ivor~W Tsang}, {and} \bibinfo{person}{Xuemin Lin}.}
  \bibinfo{year}{2020}\natexlab{}.
\newblock \showarticletitle{I/O efficient approximate nearest neighbour search
  based on learned functions}. In \bibinfo{booktitle}{\emph{2020 IEEE 36th
  international conference on data engineering (ICDE)}}. IEEE,
  \bibinfo{pages}{289--300}.
\newblock


\bibitem[Li et~al\mbox{.}(2023)]%
        {gte}
\bibfield{author}{\bibinfo{person}{Zehan Li}, \bibinfo{person}{Xin Zhang},
  \bibinfo{person}{Yanzhao Zhang}, \bibinfo{person}{Dingkun Long},
  \bibinfo{person}{Pengjun Xie}, {and} \bibinfo{person}{Meishan Zhang}.}
  \bibinfo{year}{2023}\natexlab{}.
\newblock \showarticletitle{Towards general text embeddings with multi-stage
  contrastive learning}.
\newblock \bibinfo{journal}{\emph{arXiv preprint arXiv:2308.03281}}
  (\bibinfo{year}{2023}).
\newblock


\bibitem[Liang et~al\mbox{.}(2024)]%
        {unify}
\bibfield{author}{\bibinfo{person}{Anqi Liang}, \bibinfo{person}{Pengcheng
  Zhang}, \bibinfo{person}{Bin Yao}, \bibinfo{person}{Zhongpu Chen},
  \bibinfo{person}{Yitong Song}, {and} \bibinfo{person}{Guangxu Cheng}.}
  \bibinfo{year}{2024}\natexlab{}.
\newblock \showarticletitle{UNIFY: Unified Index for Range Filtered Approximate
  Nearest Neighbors Search}.
\newblock \bibinfo{journal}{\emph{arXiv preprint arXiv:2412.02448}}
  (\bibinfo{year}{2024}).
\newblock


\bibitem[Lin et~al\mbox{.}(2025)]%
        {rl-fanns-survey}
\bibfield{author}{\bibinfo{person}{Yanjun Lin}, \bibinfo{person}{Kai Zhang},
  \bibinfo{person}{Zhenying He}, \bibinfo{person}{Yinan Jing}, {and}
  \bibinfo{person}{X~Sean Wang}.} \bibinfo{year}{2025}\natexlab{}.
\newblock \showarticletitle{Survey of Filtered Approximate Nearest Neighbor
  Search over the Vector-Scalar Hybrid Data}.
\newblock \bibinfo{journal}{\emph{arXiv preprint arXiv:2505.06501}}
  (\bibinfo{year}{2025}).
\newblock


\bibitem[Liu et~al\mbox{.}(2016b)]%
        {dh}
\bibfield{author}{\bibinfo{person}{Haomiao Liu}, \bibinfo{person}{Ruiping
  Wang}, \bibinfo{person}{Shiguang Shan}, {and} \bibinfo{person}{Xilin Chen}.}
  \bibinfo{year}{2016}\natexlab{b}.
\newblock \showarticletitle{Deep supervised hashing for fast image retrieval}.
  In \bibinfo{booktitle}{\emph{Proceedings of the IEEE conference on computer
  vision and pattern recognition}}. \bibinfo{pages}{2064--2072}.
\newblock


\bibitem[Liu et~al\mbox{.}(2021)]%
        {ei-lsh}
\bibfield{author}{\bibinfo{person}{Wanqi Liu}, \bibinfo{person}{Hanchen Wang},
  \bibinfo{person}{Ying Zhang}, \bibinfo{person}{Wei Wang}, \bibinfo{person}{Lu
  Qin}, {and} \bibinfo{person}{Xuemin Lin}.} \bibinfo{year}{2021}\natexlab{}.
\newblock \showarticletitle{EI-LSH: An early-termination driven I/O efficient
  incremental c-approximate nearest neighbor search}.
\newblock \bibinfo{journal}{\emph{The VLDB Journal}}  \bibinfo{volume}{30}
  (\bibinfo{year}{2021}), \bibinfo{pages}{215--235}.
\newblock


\bibitem[Liu et~al\mbox{.}(2016a)]%
        {re-ident-1}
\bibfield{author}{\bibinfo{person}{Xinchen Liu}, \bibinfo{person}{Wu Liu},
  \bibinfo{person}{Huadong Ma}, {and} \bibinfo{person}{Huiyuan Fu}.}
  \bibinfo{year}{2016}\natexlab{a}.
\newblock \showarticletitle{Large-scale vehicle re-identification in urban
  surveillance videos}. In \bibinfo{booktitle}{\emph{2016 IEEE international
  conference on multimedia and expo (ICME)}}. IEEE, \bibinfo{pages}{1--6}.
\newblock


\bibitem[Liu et~al\mbox{.}(2024)]%
        {seresnext}
\bibfield{author}{\bibinfo{person}{Yi Liu}, \bibinfo{person}{Minghui Wang},
  {and} \bibinfo{person}{Changxin Li}.} \bibinfo{year}{2024}\natexlab{}.
\newblock \showarticletitle{Research on High-Accuracy Indoor Visual Positioning
  Technology Using an Optimized SE-ResNeXt Architecture}. In
  \bibinfo{booktitle}{\emph{Proceedings of the 2024 7th International
  Conference on Signal Processing and Machine Learning}}.
  \bibinfo{pages}{313--320}.
\newblock


\bibitem[Lu and Kudo(2020)]%
        {r2lsh}
\bibfield{author}{\bibinfo{person}{Kejing Lu} {and} \bibinfo{person}{Mineichi
  Kudo}.} \bibinfo{year}{2020}\natexlab{}.
\newblock \showarticletitle{R2LSH: A nearest neighbor search scheme based on
  two-dimensional projected spaces}. In \bibinfo{booktitle}{\emph{2020 IEEE
  36th International Conference on Data Engineering (ICDE)}}. IEEE,
  \bibinfo{pages}{1045--1056}.
\newblock


\bibitem[Lu et~al\mbox{.}(2020)]%
        {anns-tree-2}
\bibfield{author}{\bibinfo{person}{Kejing Lu}, \bibinfo{person}{Hongya Wang},
  \bibinfo{person}{Wei Wang}, {and} \bibinfo{person}{Mineichi Kudo}.}
  \bibinfo{year}{2020}\natexlab{}.
\newblock \showarticletitle{VHP: approximate nearest neighbor search via
  virtual hypersphere partitioning}.
\newblock \bibinfo{journal}{\emph{Proceedings of the VLDB Endowment}}
  \bibinfo{volume}{13}, \bibinfo{number}{9} (\bibinfo{year}{2020}),
  \bibinfo{pages}{1443--1455}.
\newblock


\bibitem[Luo et~al\mbox{.}(2025)]%
        {tfanns}
\bibfield{author}{\bibinfo{person}{Jiarui Luo}, \bibinfo{person}{Miao Qiao},
  \bibinfo{person}{Chaoji Zuo}, {and} \bibinfo{person}{Dong Deng}.}
  \bibinfo{year}{2025}\natexlab{}.
\newblock \showarticletitle{Tag-Filtered Approximate Nearest Neighbor Search}.
  In \bibinfo{booktitle}{\emph{2025 IEEE 41st International Conference on Data
  Engineering (ICDE)}}. IEEE, \bibinfo{pages}{3642--3654}.
\newblock


\bibitem[Lv et~al\mbox{.}(2017)]%
        {anns-hash-5}
\bibfield{author}{\bibinfo{person}{Qin Lv}, \bibinfo{person}{William
  Josephson}, \bibinfo{person}{Zhe Wang}, \bibinfo{person}{Moses Charikar},
  {and} \bibinfo{person}{Kai Li}.} \bibinfo{year}{2017}\natexlab{}.
\newblock \showarticletitle{Intelligent probing for locality sensitive hashing:
  Multi-probe LSH and beyond}.
\newblock \bibinfo{journal}{\emph{Proceedings of the VLDB Endowment}}
  (\bibinfo{year}{2017}).
\newblock


\bibitem[Malkov et~al\mbox{.}(2014)]%
        {nsw}
\bibfield{author}{\bibinfo{person}{Yury Malkov}, \bibinfo{person}{Alexander
  Ponomarenko}, \bibinfo{person}{Andrey Logvinov}, {and}
  \bibinfo{person}{Vladimir Krylov}.} \bibinfo{year}{2014}\natexlab{}.
\newblock \showarticletitle{Approximate nearest neighbor algorithm based on
  navigable small world graphs}.
\newblock \bibinfo{journal}{\emph{Information Systems}}  \bibinfo{volume}{45}
  (\bibinfo{year}{2014}), \bibinfo{pages}{61--68}.
\newblock


\bibitem[Malkov and Yashunin(2018)]%
        {hnsw}
\bibfield{author}{\bibinfo{person}{Yu~A Malkov} {and} \bibinfo{person}{Dmitry~A
  Yashunin}.} \bibinfo{year}{2018}\natexlab{}.
\newblock \showarticletitle{Efficient and robust approximate nearest neighbor
  search using hierarchical navigable small world graphs}.
\newblock \bibinfo{journal}{\emph{IEEE transactions on pattern analysis and
  machine intelligence}} \bibinfo{volume}{42}, \bibinfo{number}{4}
  (\bibinfo{year}{2018}), \bibinfo{pages}{824--836}.
\newblock


\bibitem[Marimont and Shapiro(1979)]%
        {curse-of-dim}
\bibfield{author}{\bibinfo{person}{Rosalind~B Marimont} {and}
  \bibinfo{person}{Marvin~B Shapiro}.} \bibinfo{year}{1979}\natexlab{}.
\newblock \showarticletitle{Nearest neighbour searches and the curse of
  dimensionality}.
\newblock \bibinfo{journal}{\emph{IMA Journal of Applied Mathematics}}
  \bibinfo{volume}{24}, \bibinfo{number}{1} (\bibinfo{year}{1979}),
  \bibinfo{pages}{59--70}.
\newblock


\bibitem[Matsui(2025)]%
        {rii-code}
\bibfield{author}{\bibinfo{person}{Yusuke Matsui}.}
  \bibinfo{year}{2025}\natexlab{}.
\newblock \bibinfo{title}{{Reconfigurable Inverted Index (Rii): IVFPQ-based
  fast and memory efficient approximate nearest neighbor search method with a
  subset-search functionality.}}
\newblock \bibinfo{howpublished}{\url{https://github.com/matsui528/rii}}.
\newblock
\newblock
\shownote{Accessed: 2025-04-23}.


\bibitem[Matsui et~al\mbox{.}(2018)]%
        {rii}
\bibfield{author}{\bibinfo{person}{Yusuke Matsui}, \bibinfo{person}{Ryota
  Hinami}, {and} \bibinfo{person}{Shin'ichi Satoh}.}
  \bibinfo{year}{2018}\natexlab{}.
\newblock \showarticletitle{Reconfigurable inverted index}. In
  \bibinfo{booktitle}{\emph{Proceedings of the 26th ACM international
  conference on Multimedia}}. \bibinfo{pages}{1715--1723}.
\newblock


\bibitem[Matsui et~al\mbox{.}(2015)]%
        {anns-quant-4}
\bibfield{author}{\bibinfo{person}{Yusuke Matsui}, \bibinfo{person}{Toshihiko
  Yamasaki}, {and} \bibinfo{person}{Kiyoharu Aizawa}.}
  \bibinfo{year}{2015}\natexlab{}.
\newblock \showarticletitle{Pqtable: Fast exact asymmetric distance neighbor
  search for product quantization using hash tables}. In
  \bibinfo{booktitle}{\emph{Proceedings of the IEEE International Conference on
  Computer Vision}}. \bibinfo{pages}{1940--1948}.
\newblock


\bibitem[Microsoft(2025a)]%
        {fdann-code}
\bibfield{author}{\bibinfo{person}{Microsoft}.}
  \bibinfo{year}{2025}\natexlab{a}.
\newblock \bibinfo{title}{{DiskANN}}.
\newblock \bibinfo{howpublished}{\url{https://github.com/microsoft/DiskANN}}.
\newblock
\newblock
\shownote{Accessed: 2025-02-26}.


\bibitem[Microsoft(2025b)]%
        {vbase-code}
\bibfield{author}{\bibinfo{person}{Microsoft}.}
  \bibinfo{year}{2025}\natexlab{b}.
\newblock \bibinfo{title}{{https://github.com/microsoft/MSVBASE}}.
\newblock \bibinfo{howpublished}{\url{https://github.com/microsoft/MSVBASE}}.
\newblock
\newblock
\shownote{Accessed: 2025-02-26}.


\bibitem[Mohoney et~al\mbox{.}(2023)]%
        {hqi}
\bibfield{author}{\bibinfo{person}{Jason Mohoney}, \bibinfo{person}{Anil
  Pacaci}, \bibinfo{person}{Shihabur~Rahman Chowdhury}, \bibinfo{person}{Ali
  Mousavi}, \bibinfo{person}{Ihab~F Ilyas}, \bibinfo{person}{Umar~Farooq
  Minhas}, \bibinfo{person}{Jeffrey Pound}, {and} \bibinfo{person}{Theodoros
  Rekatsinas}.} \bibinfo{year}{2023}\natexlab{}.
\newblock \showarticletitle{High-throughput vector similarity search in
  knowledge graphs}.
\newblock \bibinfo{journal}{\emph{Proceedings of the ACM on Management of
  Data}} \bibinfo{volume}{1}, \bibinfo{number}{2} (\bibinfo{year}{2023}),
  \bibinfo{pages}{1--25}.
\newblock


\bibitem[Muennighoff et~al\mbox{.}(2022)]%
        {mteb}
\bibfield{author}{\bibinfo{person}{Niklas Muennighoff},
  \bibinfo{person}{Nouamane Tazi}, \bibinfo{person}{Lo{\"\i}c Magne}, {and}
  \bibinfo{person}{Nils Reimers}.} \bibinfo{year}{2022}\natexlab{}.
\newblock \showarticletitle{MTEB: Massive text embedding benchmark}.
\newblock \bibinfo{journal}{\emph{arXiv preprint arXiv:2210.07316}}
  (\bibinfo{year}{2022}).
\newblock


\bibitem[Muja and Lowe(2014)]%
        {anns-tree-3}
\bibfield{author}{\bibinfo{person}{Marius Muja} {and} \bibinfo{person}{David~G
  Lowe}.} \bibinfo{year}{2014}\natexlab{}.
\newblock \showarticletitle{Scalable nearest neighbor algorithms for high
  dimensional data}.
\newblock \bibinfo{journal}{\emph{IEEE transactions on pattern analysis and
  machine intelligence}} \bibinfo{volume}{36}, \bibinfo{number}{11}
  (\bibinfo{year}{2014}), \bibinfo{pages}{2227--2240}.
\newblock


\bibitem[NovaSearch(2025)]%
        {stella-hf}
\bibfield{author}{\bibinfo{person}{NovaSearch}.}
  \bibinfo{year}{2025}\natexlab{}.
\newblock \bibinfo{title}{{Stella\_em\_400M\_v5}}.
\newblock
  \bibinfo{howpublished}{\url{https://huggingface.co/NovaSearch/stella_en_400M_v5}}.
\newblock
\newblock
\shownote{Accessed: 2025-03-11}.


\bibitem[NVIDIA(2025)]%
        {nvembed-hf}
\bibfield{author}{\bibinfo{person}{NVIDIA}.} \bibinfo{year}{2025}\natexlab{}.
\newblock \bibinfo{title}{{NV-Embed-v2}}.
\newblock
  \bibinfo{howpublished}{\url{https://huggingface.co/nvidia/NV-Embed-v2}}.
\newblock
\newblock
\shownote{Accessed: 2025-01-22}.


\bibitem[of~Artificial~Intelligence(2025)]%
        {bge-hf}
\bibfield{author}{\bibinfo{person}{Beijing~Academy of
  Artificial~Intelligence}.} \bibinfo{year}{2025}\natexlab{}.
\newblock \bibinfo{title}{{bge-en-icl}}.
\newblock \bibinfo{howpublished}{\url{https://huggingface.co/BAAI/bge-en-icl}}.
\newblock
\newblock
\shownote{Accessed: 2025-01-22}.


\bibitem[Omohundro(1989)]%
        {balltree}
\bibfield{author}{\bibinfo{person}{Stephen~M Omohundro}.}
  \bibinfo{year}{1989}\natexlab{}.
\newblock \showarticletitle{Five balltree construction algorithms}.
\newblock  (\bibinfo{year}{1989}).
\newblock


\bibitem[Park et~al\mbox{.}(2015)]%
        {anns-hash-6}
\bibfield{author}{\bibinfo{person}{Yongjoo Park}, \bibinfo{person}{Michael
  Cafarella}, {and} \bibinfo{person}{Barzan Mozafari}.}
  \bibinfo{year}{2015}\natexlab{}.
\newblock \showarticletitle{Neighbor-sensitive hashing}.
\newblock \bibinfo{journal}{\emph{Proceedings of the VLDB Endowment}}
  \bibinfo{volume}{9}, \bibinfo{number}{3} (\bibinfo{year}{2015}),
  \bibinfo{pages}{144--155}.
\newblock


\bibitem[Patel et~al\mbox{.}(2024)]%
        {acorn}
\bibfield{author}{\bibinfo{person}{Liana Patel}, \bibinfo{person}{Peter Kraft},
  \bibinfo{person}{Carlos Guestrin}, {and} \bibinfo{person}{Matei Zaharia}.}
  \bibinfo{year}{2024}\natexlab{}.
\newblock \showarticletitle{ACORN: Performant and Predicate-Agnostic Search
  Over Vector Embeddings and Structured Data}.
\newblock \bibinfo{journal}{\emph{Proceedings of the ACM on Management of
  Data}} \bibinfo{volume}{2}, \bibinfo{number}{3} (\bibinfo{year}{2024}),
  \bibinfo{pages}{1--27}.
\newblock


\bibitem[Peng et~al\mbox{.}(2025)]%
        {dsg}
\bibfield{author}{\bibinfo{person}{Zhencan Peng}, \bibinfo{person}{Miao Qiao},
  \bibinfo{person}{Wenchao Zhou}, \bibinfo{person}{Feifei Li}, {and}
  \bibinfo{person}{Dong Deng}.} \bibinfo{year}{2025}\natexlab{}.
\newblock \showarticletitle{Dynamic Range-Filtering Approximate Nearest
  Neighbor Search}.
\newblock \bibinfo{journal}{\emph{Proceedings of the VLDB Endowment}}
  \bibinfo{volume}{18}, \bibinfo{number}{10} (\bibinfo{year}{2025}),
  \bibinfo{pages}{3256--3268}.
\newblock


\bibitem[Pennington et~al\mbox{.}(2025)]%
        {glove-ds}
\bibfield{author}{\bibinfo{person}{Jeffrey Pennington},
  \bibinfo{person}{Richard Socher}, {and} \bibinfo{person}{Christopher~D.
  Manning}.} \bibinfo{year}{2025}\natexlab{}.
\newblock \bibinfo{title}{{glove100\_angular}}.
\newblock
  \bibinfo{howpublished}{\url{https://www.tensorflow.org/datasets/catalog/glove100_angular}}.
\newblock
\newblock
\shownote{Accessed: 2025-03-06}.


\bibitem[Pinecone(2025)]%
        {pinecone}
\bibfield{author}{\bibinfo{person}{Pinecone}.} \bibinfo{year}{2025}\natexlab{}.
\newblock \bibinfo{title}{{Pinecone: The vector database to build knowledgeable
  AI}}.
\newblock \bibinfo{howpublished}{\url{https://www.pinecone.io/}}.
\newblock
\newblock
\shownote{Accessed: 2025-03-11}.


\bibitem[Project(2025)]%
        {milvus-code}
\bibfield{author}{\bibinfo{person}{The~Milvus Project}.}
  \bibinfo{year}{2025}\natexlab{}.
\newblock \bibinfo{title}{{Milvus}}.
\newblock \bibinfo{howpublished}{\url{https://github.com/milvus-io/milvus}}.
\newblock
\newblock
\shownote{Accessed: 2025-02-27}.


\bibitem[Rekabsaz et~al\mbox{.}(2025)]%
        {tripclick-ds}
\bibfield{author}{\bibinfo{person}{Navid Rekabsaz}, \bibinfo{person}{Oleg
  Lesota}, \bibinfo{person}{Markus Schedl}, \bibinfo{person}{Jon Brassey},
  {and} \bibinfo{person}{Carsten Eickhoff}.} \bibinfo{year}{2025}\natexlab{}.
\newblock \bibinfo{title}{{TripClick}}.
\newblock
  \bibinfo{howpublished}{\url{https://tripdatabase.github.io/tripclick/}}.
\newblock
\newblock
\shownote{Accessed: 2025-07-28}.


\bibitem[Research(2025)]%
        {faiss-github}
\bibfield{author}{\bibinfo{person}{Meta Research}.}
  \bibinfo{year}{2025}\natexlab{}.
\newblock \bibinfo{title}{{Faiss}}.
\newblock
  \bibinfo{howpublished}{\url{https://github.com/facebookresearch/faiss}}.
\newblock
\newblock
\shownote{Accessed: 2025-01-17}.


\bibitem[Richardson et~al\mbox{.}(2015)]%
        {voice-rec-1}
\bibfield{author}{\bibinfo{person}{Fred Richardson}, \bibinfo{person}{Douglas
  Reynolds}, {and} \bibinfo{person}{Najim Dehak}.}
  \bibinfo{year}{2015}\natexlab{}.
\newblock \showarticletitle{A unified deep neural network for speaker and
  language recognition}.
\newblock \bibinfo{journal}{\emph{arXiv preprint arXiv:1504.00923}}
  (\bibinfo{year}{2015}).
\newblock


\bibitem[Schuhmann(2025)]%
        {laion-ds}
\bibfield{author}{\bibinfo{person}{Christoph Schuhmann}.}
  \bibinfo{year}{2025}\natexlab{}.
\newblock \bibinfo{title}{{LAION-400-MILLION OPEN DATASET}}.
\newblock
  \bibinfo{howpublished}{\url{https://laion.ai/blog/laion-400-open-dataset/}}.
\newblock
\newblock
\shownote{Accessed: 2025-03-06}.


\bibitem[Shapkin et~al\mbox{.}(2023)]%
        {rag-2}
\bibfield{author}{\bibinfo{person}{Anton Shapkin}, \bibinfo{person}{Denis
  Litvinov}, \bibinfo{person}{Yaroslav Zharov}, \bibinfo{person}{Egor
  Bogomolov}, \bibinfo{person}{Timur Galimzyanov}, {and}
  \bibinfo{person}{Timofey Bryksin}.} \bibinfo{year}{2023}\natexlab{}.
\newblock \showarticletitle{Dynamic Retrieval-Augmented Generation}.
\newblock \bibinfo{journal}{\emph{arXiv preprint arXiv:2312.08976}}
  (\bibinfo{year}{2023}).
\newblock


\bibitem[Shi et~al\mbox{.}(2025)]%
        {rl-fanns-benchmark}
\bibfield{author}{\bibinfo{person}{Jiayang Shi}, \bibinfo{person}{Yuzheng Cai},
  {and} \bibinfo{person}{Weiguo Zheng}.} \bibinfo{year}{2025}\natexlab{}.
\newblock \showarticletitle{Filtered Approximate Nearest Neighbor Search: A
  Unified Benchmark and Systematic Experimental Study [Experiment, Analysis \&
  Benchmark]}.
\newblock \bibinfo{journal}{\emph{arXiv preprint arXiv:2509.07789}}
  (\bibinfo{year}{2025}).
\newblock


\bibitem[Silpa-Anan and Hartley(2008)]%
        {anns-tree-4}
\bibfield{author}{\bibinfo{person}{Chanop Silpa-Anan} {and}
  \bibinfo{person}{Richard Hartley}.} \bibinfo{year}{2008}\natexlab{}.
\newblock \showarticletitle{Optimised KD-trees for fast image descriptor
  matching}. In \bibinfo{booktitle}{\emph{2008 IEEE conference on computer
  vision and pattern recognition}}. IEEE, \bibinfo{pages}{1--8}.
\newblock


\bibitem[Simhadri et~al\mbox{.}(2025)]%
        {bigann-ds}
\bibfield{author}{\bibinfo{person}{Harsha~Vardhan Simhadri},
  \bibinfo{person}{George Williams}, \bibinfo{person}{Martin Aumüller},
  \bibinfo{person}{Artem Babenko}, \bibinfo{person}{Dmitry Baranchuk},
  \bibinfo{person}{Qi Chen}, \bibinfo{person}{Matthijs Douze},
  \bibinfo{person}{Lucas Hosseini}, \bibinfo{person}{Ravishankar Krishnaswamy},
  \bibinfo{person}{Gopal Srinivasa}, \bibinfo{person}{Suhas~Jayaram
  Subramanya}, {and} \bibinfo{person}{Jingdong Wang}.}
  \bibinfo{year}{2025}\natexlab{}.
\newblock \bibinfo{title}{{Billion-Scale Approximate Nearest Neighbor Search
  Challenge: NeurIPS'21 competition track}}.
\newblock
  \bibinfo{howpublished}{\url{https://big-ann-benchmarks.com/neurips21.html}}.
\newblock
\newblock
\shownote{Accessed: 2025-03-06}.


\bibitem[Singh et~al\mbox{.}(2021)]%
        {freshdann}
\bibfield{author}{\bibinfo{person}{Aditi Singh}, \bibinfo{person}{Suhas~Jayaram
  Subramanya}, \bibinfo{person}{Ravishankar Krishnaswamy}, {and}
  \bibinfo{person}{Harsha~Vardhan Simhadri}.} \bibinfo{year}{2021}\natexlab{}.
\newblock \showarticletitle{Freshdiskann: A fast and accurate graph-based ann
  index for streaming similarity search}.
\newblock \bibinfo{journal}{\emph{arXiv preprint arXiv:2105.09613}}
  (\bibinfo{year}{2021}).
\newblock


\bibitem[Sivic and Zisserman(2003a)]%
        {vid-embed-1}
\bibfield{author}{\bibinfo{person}{Sivic} {and} \bibinfo{person}{Zisserman}.}
  \bibinfo{year}{2003}\natexlab{a}.
\newblock \showarticletitle{Video Google: A text retrieval approach to object
  matching in videos}. In \bibinfo{booktitle}{\emph{Proceedings ninth IEEE
  international conference on computer vision}}. IEEE,
  \bibinfo{pages}{1470--1477}.
\newblock


\bibitem[Sivic and Zisserman(2003b)]%
        {ivf}
\bibfield{author}{\bibinfo{person}{Sivic} {and} \bibinfo{person}{Zisserman}.}
  \bibinfo{year}{2003}\natexlab{b}.
\newblock \showarticletitle{Video Google: A text retrieval approach to object
  matching in videos}. In \bibinfo{booktitle}{\emph{Proceedings ninth IEEE
  international conference on computer vision}}. IEEE,
  \bibinfo{pages}{1470--1477}.
\newblock


\bibitem[Suchal and N{\'a}vrat(2010)]%
        {rec-sys-1}
\bibfield{author}{\bibinfo{person}{J{\'a}n Suchal} {and} \bibinfo{person}{Pavol
  N{\'a}vrat}.} \bibinfo{year}{2010}\natexlab{}.
\newblock \showarticletitle{Full text search engine as scalable k-nearest
  neighbor recommendation system}. In \bibinfo{booktitle}{\emph{Artificial
  Intelligence in Theory and Practice III: Third IFIP TC 12 International
  Conference on Artificial Intelligence, IFIP AI 2010, Held as Part of WCC
  2010, Brisbane, Australia, September 20-23, 2010. Proceedings 3}}. Springer,
  \bibinfo{pages}{165--173}.
\newblock


\bibitem[Sundaram et~al\mbox{.}(2013)]%
        {anns-hash-7}
\bibfield{author}{\bibinfo{person}{Narayanan Sundaram}, \bibinfo{person}{Aizana
  Turmukhametova}, \bibinfo{person}{Nadathur Satish}, \bibinfo{person}{Todd
  Mostak}, \bibinfo{person}{Piotr Indyk}, \bibinfo{person}{Samuel Madden},
  {and} \bibinfo{person}{Pradeep Dubey}.} \bibinfo{year}{2013}\natexlab{}.
\newblock \showarticletitle{Streaming similarity search over one billion tweets
  using parallel locality-sensitive hashing}.
\newblock \bibinfo{journal}{\emph{Proceedings of the VLDB Endowment}}
  \bibinfo{volume}{6}, \bibinfo{number}{14} (\bibinfo{year}{2013}),
  \bibinfo{pages}{1930--1941}.
\newblock


\bibitem[Szegedy et~al\mbox{.}(2015)]%
        {googlenet}
\bibfield{author}{\bibinfo{person}{Christian Szegedy}, \bibinfo{person}{Wei
  Liu}, \bibinfo{person}{Yangqing Jia}, \bibinfo{person}{Pierre Sermanet},
  \bibinfo{person}{Scott Reed}, \bibinfo{person}{Dragomir Anguelov},
  \bibinfo{person}{Dumitru Erhan}, \bibinfo{person}{Vincent Vanhoucke}, {and}
  \bibinfo{person}{Andrew Rabinovich}.} \bibinfo{year}{2015}\natexlab{}.
\newblock \showarticletitle{Going deeper with convolutions}. In
  \bibinfo{booktitle}{\emph{Proceedings of the IEEE conference on computer
  vision and pattern recognition}}. \bibinfo{pages}{1--9}.
\newblock


\bibitem[Tavallali et~al\mbox{.}(2021)]%
        {kmtree}
\bibfield{author}{\bibinfo{person}{Pooya Tavallali}, \bibinfo{person}{Peyman
  Tavallali}, {and} \bibinfo{person}{Mukesh Singhal}.}
  \bibinfo{year}{2021}\natexlab{}.
\newblock \showarticletitle{K-means tree: an optimal clustering tree for
  unsupervised learning}.
\newblock \bibinfo{journal}{\emph{The journal of supercomputing}}
  \bibinfo{volume}{77}, \bibinfo{number}{5} (\bibinfo{year}{2021}),
  \bibinfo{pages}{5239--5266}.
\newblock


\bibitem[Trevor(2025)]%
        {mtg-ds}
\bibfield{author}{\bibinfo{person}{Trevor}.} \bibinfo{year}{2025}\natexlab{}.
\newblock \bibinfo{title}{{Mtg Scryfall Cropped Art Embeddings}}.
\newblock
  \bibinfo{howpublished}{\url{https://huggingface.co/datasets/TrevorJS/mtg-scryfall-cropped-art-embeddings-open-clip-ViT-SO400M-14-SigLIP-384}}.
\newblock
\newblock
\shownote{Accessed: 2025-04-23}.


\bibitem[Vectara(2025)]%
        {vectara}
\bibfield{author}{\bibinfo{person}{Vectara}.} \bibinfo{year}{2025}\natexlab{}.
\newblock \bibinfo{title}{{Vectara: The AI Agent and Assistant platform for
  enterprises}}.
\newblock \bibinfo{howpublished}{\url{https://www.vectara.com/}}.
\newblock
\newblock
\shownote{Accessed: 2025-03-11}.


\bibitem[Vespa(2025)]%
        {vespa}
\bibfield{author}{\bibinfo{person}{Vespa}.} \bibinfo{year}{2025}\natexlab{}.
\newblock \bibinfo{title}{{Vespa: We Make AI Work}}.
\newblock \bibinfo{howpublished}{\url{https://vespa.ai/}}.
\newblock
\newblock
\shownote{Accessed: 2025-03-11}.


\bibitem[Wang and Li(2012)]%
        {ng}
\bibfield{author}{\bibinfo{person}{Jingdong Wang} {and}
  \bibinfo{person}{Shipeng Li}.} \bibinfo{year}{2012}\natexlab{}.
\newblock \showarticletitle{Query-driven iterated neighborhood graph search for
  large scale indexing}. In \bibinfo{booktitle}{\emph{Proceedings of the 20th
  ACM international conference on Multimedia}}. \bibinfo{pages}{179--188}.
\newblock


\bibitem[Wang et~al\mbox{.}(2021d)]%
        {milvus}
\bibfield{author}{\bibinfo{person}{Jianguo Wang}, \bibinfo{person}{Xiaomeng
  Yi}, \bibinfo{person}{Rentong Guo}, \bibinfo{person}{Hai Jin},
  \bibinfo{person}{Peng Xu}, \bibinfo{person}{Shengjun Li},
  \bibinfo{person}{Xiangyu Wang}, \bibinfo{person}{Xiangzhou Guo},
  \bibinfo{person}{Chengming Li}, \bibinfo{person}{Xiaohai Xu},
  {et~al\mbox{.}}} \bibinfo{year}{2021}\natexlab{d}.
\newblock \showarticletitle{Milvus: A purpose-built vector data management
  system}. In \bibinfo{booktitle}{\emph{Proceedings of the 2021 International
  Conference on Management of Data}}. \bibinfo{pages}{2614--2627}.
\newblock


\bibitem[Wang et~al\mbox{.}(2022)]%
        {nhq-2}
\bibfield{author}{\bibinfo{person}{Mengzhao Wang}, \bibinfo{person}{Lingwei
  Lv}, \bibinfo{person}{Xiaoliang Xu}, \bibinfo{person}{Yuxiang Wang},
  \bibinfo{person}{Qiang Yue}, {and} \bibinfo{person}{Jiongkang Ni}.}
  \bibinfo{year}{2022}\natexlab{}.
\newblock \showarticletitle{Navigable proximity graph-driven native hybrid
  queries with structured and unstructured constraints}.
\newblock \bibinfo{journal}{\emph{arXiv preprint arXiv:2203.13601}}
  (\bibinfo{year}{2022}).
\newblock


\bibitem[Wang et~al\mbox{.}(2024)]%
        {nhq}
\bibfield{author}{\bibinfo{person}{Mengzhao Wang}, \bibinfo{person}{Lingwei
  Lv}, \bibinfo{person}{Xiaoliang Xu}, \bibinfo{person}{Yuxiang Wang},
  \bibinfo{person}{Qiang Yue}, {and} \bibinfo{person}{Jiongkang Ni}.}
  \bibinfo{year}{2024}\natexlab{}.
\newblock \showarticletitle{An efficient and robust framework for approximate
  nearest neighbor search with attribute constraint}.
\newblock \bibinfo{journal}{\emph{Advances in Neural Information Processing
  Systems}}  \bibinfo{volume}{36} (\bibinfo{year}{2024}).
\newblock


\bibitem[Wang et~al\mbox{.}(2021a)]%
        {anns-quant-2}
\bibfield{author}{\bibinfo{person}{Mengzhao Wang}, \bibinfo{person}{Xiaoliang
  Xu}, \bibinfo{person}{Qiang Yue}, {and} \bibinfo{person}{Yuxiang Wang}.}
  \bibinfo{year}{2021}\natexlab{a}.
\newblock \showarticletitle{A comprehensive survey and experimental comparison
  of graph-based approximate nearest neighbor search}.
\newblock \bibinfo{journal}{\emph{arXiv preprint arXiv:2101.12631}}
  (\bibinfo{year}{2021}).
\newblock


\bibitem[Wang et~al\mbox{.}(2021b)]%
        {551}
\bibfield{author}{\bibinfo{person}{Mengzhao Wang}, \bibinfo{person}{Xiaoliang
  Xu}, \bibinfo{person}{Qiang Yue}, {and} \bibinfo{person}{Yuxiang Wang}.}
  \bibinfo{year}{2021}\natexlab{b}.
\newblock \showarticletitle{A comprehensive survey and experimental comparison
  of graph-based approximate nearest neighbor search}.
\newblock \bibinfo{journal}{\emph{arXiv preprint arXiv:2101.12631}}
  (\bibinfo{year}{2021}).
\newblock


\bibitem[Wang et~al\mbox{.}(2021c)]%
        {anns-graph-2}
\bibfield{author}{\bibinfo{person}{Mengzhao Wang}, \bibinfo{person}{Xiaoliang
  Xu}, \bibinfo{person}{Qiang Yue}, {and} \bibinfo{person}{Yuxiang Wang}.}
  \bibinfo{year}{2021}\natexlab{c}.
\newblock \showarticletitle{A comprehensive survey and experimental comparison
  of graph-based approximate nearest neighbor search}.
\newblock \bibinfo{journal}{\emph{arXiv preprint arXiv:2101.12631}}
  (\bibinfo{year}{2021}).
\newblock


\bibitem[Wang and Deng(2020)]%
        {anns-quant-5}
\bibfield{author}{\bibinfo{person}{Runhui Wang} {and} \bibinfo{person}{Dong
  Deng}.} \bibinfo{year}{2020}\natexlab{}.
\newblock \showarticletitle{DeltaPQ: lossless product quantization code
  compression for high dimensional similarity search}.
\newblock \bibinfo{journal}{\emph{Proceedings of the VLDB Endowment}}
  \bibinfo{volume}{13}, \bibinfo{number}{13} (\bibinfo{year}{2020}),
  \bibinfo{pages}{3603--3616}.
\newblock


\bibitem[Wang et~al\mbox{.}(2025)]%
        {wow}
\bibfield{author}{\bibinfo{person}{Ziqi Wang}, \bibinfo{person}{Jingzhe Zhang},
  {and} \bibinfo{person}{Wei Hu}.} \bibinfo{year}{2025}\natexlab{}.
\newblock \showarticletitle{WoW: A Window-to-Window Incremental Index for
  Range-Filtering Approximate Nearest Neighbor Search}.
\newblock \bibinfo{journal}{\emph{arXiv preprint arXiv:2508.18617}}
  (\bibinfo{year}{2025}).
\newblock


\bibitem[Wei et~al\mbox{.}(2020)]%
        {adbv}
\bibfield{author}{\bibinfo{person}{Chuangxian Wei}, \bibinfo{person}{Bin Wu},
  \bibinfo{person}{Sheng Wang}, \bibinfo{person}{Renjie Lou},
  \bibinfo{person}{Chaoqun Zhan}, \bibinfo{person}{Feifei Li}, {and}
  \bibinfo{person}{Yuanzhe Cai}.} \bibinfo{year}{2020}\natexlab{}.
\newblock \showarticletitle{AnalyticDB-V: a hybrid analytical engine towards
  query fusion for structured and unstructured data}.
\newblock \bibinfo{journal}{\emph{Proceedings of the VLDB Endowment}}
  \bibinfo{volume}{13}, \bibinfo{number}{12} (\bibinfo{year}{2020}),
  \bibinfo{pages}{3152--3165}.
\newblock


\bibitem[Weiss et~al\mbox{.}(2008)]%
        {sh}
\bibfield{author}{\bibinfo{person}{Yair Weiss}, \bibinfo{person}{Antonio
  Torralba}, {and} \bibinfo{person}{Rob Fergus}.}
  \bibinfo{year}{2008}\natexlab{}.
\newblock \showarticletitle{Spectral hashing}.
\newblock \bibinfo{journal}{\emph{Advances in neural information processing
  systems}}  \bibinfo{volume}{21} (\bibinfo{year}{2008}).
\newblock


\bibitem[Xu et~al\mbox{.}(2021)]%
        {rag-1}
\bibfield{author}{\bibinfo{person}{J Xu}, \bibinfo{person}{A Szlam}, {and}
  \bibinfo{person}{J Weston}.} \bibinfo{year}{2021}\natexlab{}.
\newblock \showarticletitle{Beyond goldfish memory: Long-term open-domain
  conversation. arXiv 2021}.
\newblock \bibinfo{journal}{\emph{arXiv preprint arXiv:2107.07567}}
  (\bibinfo{year}{2021}).
\newblock


\bibitem[Xu et~al\mbox{.}(2020)]%
        {mansw}
\bibfield{author}{\bibinfo{person}{Xiaoliang Xu}, \bibinfo{person}{Chang Li},
  \bibinfo{person}{Yuxiang Wang}, {and} \bibinfo{person}{Yixing Xia}.}
  \bibinfo{year}{2020}\natexlab{}.
\newblock \showarticletitle{Multiattribute approximate nearest neighbor search
  based on navigable small world graph}.
\newblock \bibinfo{journal}{\emph{Concurrency and Computation: Practice and
  Experience}} \bibinfo{volume}{32}, \bibinfo{number}{24}
  (\bibinfo{year}{2020}), \bibinfo{pages}{e5970}.
\newblock


\bibitem[Xu(2025)]%
        {irg-code}
\bibfield{author}{\bibinfo{person}{Yuexuan Xu}.}
  \bibinfo{year}{2025}\natexlab{}.
\newblock \bibinfo{title}{{iRangeGraph}}.
\newblock
  \bibinfo{howpublished}{\url{https://github.com/YuexuanXu7/iRangeGraph}}.
\newblock
\newblock
\shownote{Accessed: 2025-02-21}.


\bibitem[Xu et~al\mbox{.}(2024)]%
        {irg}
\bibfield{author}{\bibinfo{person}{Yuexuan Xu}, \bibinfo{person}{Jianyang Gao},
  \bibinfo{person}{Yutong Gou}, \bibinfo{person}{Cheng Long}, {and}
  \bibinfo{person}{Christian~S Jensen}.} \bibinfo{year}{2024}\natexlab{}.
\newblock \showarticletitle{iRangeGraph: Improvising Range-dedicated Graphs for
  Range-filtering Nearest Neighbor Search}.
\newblock \bibinfo{journal}{\emph{Proceedings of the ACM on Management of
  Data}} \bibinfo{volume}{2}, \bibinfo{number}{6} (\bibinfo{year}{2024}),
  \bibinfo{pages}{1--26}.
\newblock


\bibitem[Yang et~al\mbox{.}(2020b)]%
        {pase}
\bibfield{author}{\bibinfo{person}{Wen Yang}, \bibinfo{person}{Tao Li},
  \bibinfo{person}{Gai Fang}, {and} \bibinfo{person}{Hong Wei}.}
  \bibinfo{year}{2020}\natexlab{b}.
\newblock \showarticletitle{Pase: Postgresql ultra-high-dimensional approximate
  nearest neighbor search extension}. In \bibinfo{booktitle}{\emph{Proceedings
  of the 2020 ACM SIGMOD international conference on management of data}}.
  \bibinfo{pages}{2241--2253}.
\newblock


\bibitem[Yang et~al\mbox{.}(2020c)]%
        {hqann}
\bibfield{author}{\bibinfo{person}{Wen Yang}, \bibinfo{person}{Tao Li},
  \bibinfo{person}{Gai Fang}, {and} \bibinfo{person}{Hong Wei}.}
  \bibinfo{year}{2020}\natexlab{c}.
\newblock \showarticletitle{Pase: Postgresql ultra-high-dimensional approximate
  nearest neighbor search extension}. In \bibinfo{booktitle}{\emph{Proceedings
  of the 2020 ACM SIGMOD international conference on management of data}}.
  \bibinfo{pages}{2241--2253}.
\newblock


\bibitem[Yang et~al\mbox{.}(2020a)]%
        {qdtree}
\bibfield{author}{\bibinfo{person}{Zongheng Yang}, \bibinfo{person}{Badrish
  Chandramouli}, \bibinfo{person}{Chi Wang}, \bibinfo{person}{Johannes Gehrke},
  \bibinfo{person}{Yinan Li}, \bibinfo{person}{Umar~Farooq Minhas},
  \bibinfo{person}{Per-{\AA}ke Larson}, \bibinfo{person}{Donald Kossmann},
  {and} \bibinfo{person}{Rajeev Acharya}.} \bibinfo{year}{2020}\natexlab{a}.
\newblock \showarticletitle{Qd-tree: Learning data layouts for big data
  analytics}. In \bibinfo{booktitle}{\emph{Proceedings of the 2020 ACM SIGMOD
  international conference on management of data}}. \bibinfo{pages}{193--208}.
\newblock


\bibitem[Zhan et~al\mbox{.}(2019)]%
        {adb}
\bibfield{author}{\bibinfo{person}{Chaoqun Zhan}, \bibinfo{person}{Maomeng Su},
  \bibinfo{person}{Chuangxian Wei}, \bibinfo{person}{Xiaoqiang Peng},
  \bibinfo{person}{Liang Lin}, \bibinfo{person}{Sheng Wang},
  \bibinfo{person}{Zhe Chen}, \bibinfo{person}{Feifei Li}, \bibinfo{person}{Yue
  Pan}, \bibinfo{person}{Fang Zheng}, {et~al\mbox{.}}}
  \bibinfo{year}{2019}\natexlab{}.
\newblock \showarticletitle{AnalyticDB: real-time OLAP database system at
  Alibaba cloud}.
\newblock \bibinfo{journal}{\emph{Proceedings of the VLDB Endowment}}
  \bibinfo{volume}{12}, \bibinfo{number}{12} (\bibinfo{year}{2019}),
  \bibinfo{pages}{2059--2070}.
\newblock


\bibitem[Zhang et~al\mbox{.}(2024)]%
        {stella}
\bibfield{author}{\bibinfo{person}{Dun Zhang}, \bibinfo{person}{Jiacheng Li},
  \bibinfo{person}{Ziyang Zeng}, {and} \bibinfo{person}{Fulong Wang}.}
  \bibinfo{year}{2024}\natexlab{}.
\newblock \showarticletitle{Jasper and Stella: distillation of SOTA embedding
  models}.
\newblock \bibinfo{journal}{\emph{arXiv preprint arXiv:2412.19048}}
  (\bibinfo{year}{2024}).
\newblock


\bibitem[Zhang et~al\mbox{.}(2025)]%
        {rpq}
\bibfield{author}{\bibinfo{person}{Fangyuan Zhang}, \bibinfo{person}{Mengxu
  Jiang}, \bibinfo{person}{Guanhao Hou}, \bibinfo{person}{Jieming Shi},
  \bibinfo{person}{Hua Fan}, \bibinfo{person}{Wenchao Zhou},
  \bibinfo{person}{Feifei Li}, {and} \bibinfo{person}{Sibo Wang}.}
  \bibinfo{year}{2025}\natexlab{}.
\newblock \showarticletitle{Efficient Dynamic Indexing for Range Filtered
  Approximate Nearest Neighbor Search}.
\newblock \bibinfo{journal}{\emph{Proceedings of the ACM on Management of
  Data}} \bibinfo{volume}{3}, \bibinfo{number}{3} (\bibinfo{year}{2025}),
  \bibinfo{pages}{1--26}.
\newblock


\bibitem[Zhang et~al\mbox{.}(2023)]%
        {vbase}
\bibfield{author}{\bibinfo{person}{Qianxi Zhang}, \bibinfo{person}{Shuotao Xu},
  \bibinfo{person}{Qi Chen}, \bibinfo{person}{Guoxin Sui},
  \bibinfo{person}{Jiadong Xie}, \bibinfo{person}{Zhizhen Cai},
  \bibinfo{person}{Yaoqi Chen}, \bibinfo{person}{Yinxuan He},
  \bibinfo{person}{Yuqing Yang}, \bibinfo{person}{Fan Yang}, {et~al\mbox{.}}}
  \bibinfo{year}{2023}\natexlab{}.
\newblock \showarticletitle{$\{$VBASE$\}$: Unifying Online Vector Similarity
  Search and Relational Queries via Relaxed Monotonicity}. In
  \bibinfo{booktitle}{\emph{17th USENIX Symposium on Operating Systems Design
  and Implementation (OSDI 23)}}. \bibinfo{pages}{377--395}.
\newblock


\bibitem[Zhao et~al\mbox{.}(2020)]%
        {anns-graph-1}
\bibfield{author}{\bibinfo{person}{Weijie Zhao}, \bibinfo{person}{Shulong Tan},
  {and} \bibinfo{person}{Ping Li}.} \bibinfo{year}{2020}\natexlab{}.
\newblock \showarticletitle{Song: Approximate nearest neighbor search on gpu}.
  In \bibinfo{booktitle}{\emph{2020 IEEE 36th International Conference on Data
  Engineering (ICDE)}}. IEEE, \bibinfo{pages}{1033--1044}.
\newblock


\bibitem[Zhao et~al\mbox{.}(2022)]%
        {airship}
\bibfield{author}{\bibinfo{person}{Weijie Zhao}, \bibinfo{person}{Shulong Tan},
  {and} \bibinfo{person}{Ping Li}.} \bibinfo{year}{2022}\natexlab{}.
\newblock \showarticletitle{Constrained approximate similarity search on
  proximity graph}.
\newblock \bibinfo{journal}{\emph{arXiv preprint arXiv:2210.14958}}
  (\bibinfo{year}{2022}).
\newblock


\bibitem[Zheng et~al\mbox{.}(2020)]%
        {pm-lsh}
\bibfield{author}{\bibinfo{person}{Bolong Zheng}, \bibinfo{person}{Zhao Xi},
  \bibinfo{person}{Lianggui Weng}, \bibinfo{person}{Nguyen Quoc~Viet Hung},
  \bibinfo{person}{Hang Liu}, {and} \bibinfo{person}{Christian~S Jensen}.}
  \bibinfo{year}{2020}\natexlab{}.
\newblock \showarticletitle{PM-LSH: A fast and accurate LSH framework for
  high-dimensional approximate NN search}.
\newblock \bibinfo{journal}{\emph{Proceedings of the VLDB Endowment}}
  \bibinfo{volume}{13}, \bibinfo{number}{5} (\bibinfo{year}{2020}),
  \bibinfo{pages}{643--655}.
\newblock


\bibitem[Zuo and Deng(2023)]%
        {arkgraph}
\bibfield{author}{\bibinfo{person}{Chaoji Zuo} {and} \bibinfo{person}{Dong
  Deng}.} \bibinfo{year}{2023}\natexlab{}.
\newblock \showarticletitle{ARKGraph: All-Range Approximate K-Nearest-Neighbor
  Graph}.
\newblock \bibinfo{journal}{\emph{Proceedings of the VLDB Endowment}}
  \bibinfo{volume}{16}, \bibinfo{number}{10} (\bibinfo{year}{2023}),
  \bibinfo{pages}{2645--2658}.
\newblock


\bibitem[Zuo et~al\mbox{.}(2024)]%
        {serf}
\bibfield{author}{\bibinfo{person}{Chaoji Zuo}, \bibinfo{person}{Miao Qiao},
  \bibinfo{person}{Wenchao Zhou}, \bibinfo{person}{Feifei Li}, {and}
  \bibinfo{person}{Dong Deng}.} \bibinfo{year}{2024}\natexlab{}.
\newblock \showarticletitle{SeRF: Segment Graph for Range-Filtering Approximate
  Nearest Neighbor Search}.
\newblock \bibinfo{journal}{\emph{Proceedings of the ACM on Management of
  Data}} \bibinfo{volume}{2}, \bibinfo{number}{1} (\bibinfo{year}{2024}),
  \bibinfo{pages}{1--26}.
\newblock


\end{thebibliography}
